\documentclass[final,3p,fleqn,11pt]{elsarticle}

\usepackage{amssymb}
\usepackage{multicol}
\usepackage[svgnames]{xcolor}

\usepackage{booktabs}
\usepackage{array}
\newcommand{\PreserveBackslash}[1]{\let\temp=\\#1\let\\=\temp}
\newcolumntype{C}[1]{>{\PreserveBackslash\centering}p{#1}}
\newcolumntype{R}[1]{>{\PreserveBackslash\raggedleft}p{#1}}
\newcolumntype{L}[1]{>{\PreserveBackslash\raggedright}p{#1}}

\usepackage{enumitem}
\usepackage{amsmath}
\usepackage{amsfonts}
\usepackage{amsthm}
\usepackage{mathtools}
\usepackage{subfig}
\usepackage{lineno}
\usepackage[hidelinks]{hyperref}
\usepackage{textcomp}
\usepackage{algorithm,algpseudocode,algorithmicx}
\usepackage{pifont}
\usepackage{tabularx}

\usepackage{physics}
\usepackage{MnSymbol}

\usepackage{float}
\usepackage{array,multirow}
\usepackage{color}
\usepackage{tikz}
\usetikzlibrary{patterns}
\usetikzlibrary{spy}
\usetikzlibrary{spy,decorations.fractals,shapes.misc}

\usetikzlibrary{backgrounds}
\usepackage{pgfplots}
\usepgfplotslibrary{patchplots}
\usepgfplotslibrary{fillbetween}
\usepgfplotslibrary{groupplots}
\usepgfplotslibrary{polar}
\usepgfplotslibrary{smithchart}
\usepgfplotslibrary{statistics}
\usepgfplotslibrary{dateplot}
\usepgfplotslibrary{ternary}
\usetikzlibrary{arrows.meta, positioning, shapes.geometric}

\pgfplotsset{compat=1.16}
\newsavebox\Axis

\definecolor{lightgray}{gray}{0.80}

\usetikzlibrary{decorations.pathreplacing}
\usepackage[listings,theorems,skins,breakable]{tcolorbox}
\tcbset{skin=enhanced}
\newtcolorbox{lbracebox}[1][Word]{%
   frame hidden,enlarge left by=2cm,width=\linewidth-2cm,%
  overlay unbroken = {\draw [decorate,decoration={brace,amplitude=10pt},]%
                     (frame.south west)-- (frame.north west)
                    node [black,midway,left,xshift=-.6cm] {#1};},%
}

\newcolumntype{Y}{>{\centering\arraybackslash}X} 
\newcolumntype{Z}{>{\raggedright\arraybackslash}X}

\newcolumntype{A}{>{\raggedright\arraybackslash\hsize=0.4\hsize}X}
\newcolumntype{B}{>{\centering\arraybackslash\hsize=0.3\hsize}X}
\newcolumntype{C}{>{\centering\arraybackslash\hsize=0.3\hsize}X}

\definecolor{grey1}{rgb}{0.5, 0.5, 0.5}
\definecolor{green1}{rgb}{0.1412, 0.5294, 0.1294} 
\definecolor{blue1}{rgb}{0.26, 0.41, 0.88} 
\definecolor{red1}{rgb}{0.8600, 0.0800, 0.2400}
\definecolor{yellow1}{rgb}{1.0, 0.76, 0.15}
\definecolor{purple1}{rgb}{0.4940, 0.1840, 0.5560}
\definecolor{lightblue1}{rgb}{0.3010, 0.7450, 0.9330}
\definecolor{bordeaux1}{rgb}{0.6350, 0.0780, 0.1840}
\definecolor{brown1}{rgb}{0.65, 0.16, 0.16}
\definecolor{pink1}{rgb}{1.0, 0.08, 0.58}

\definecolor{green2}{rgb}{0.467, 0.675, 0.188} 
\definecolor{blue2}{rgb}{0, 0.447, 0.741} 

\definecolor{burntorange}{rgb}{0.851, 0.3255, 0.098}

\usepackage{color, soul}

\DeclareRobustCommand{\reviewerI}[1]{{\sethlcolor{pink}\hl{#1}}}
\DeclareRobustCommand{\reviewerII}[1]{{\sethlcolor{yellow}\hl{#1}}}

\soulregister\reviewerI{1}
\soulregister\reviewerII{1}

\soulregister\ref7
\soulregister\pageref7
\soulregister\eqref7
\soulregister\cite7

\makeatletter
\renewcommand*\env@matrix[1][*\c@MaxMatrixCols c]{%
  \hskip -\arraycolsep
  \let\@ifnextchar\new@ifnextchar
  \array{#1}}
\makeatother

\theoremstyle{plain}
\newtheorem{theorem}{Theorem}[section]

\newtheorem{remark}[theorem]{Remark}

\theoremstyle{definition}

\newcommand{\vect}[1]{\boldsymbol{#1}} 									
\newcommand{\mat}[1]{\mathbf{#1}} 											
\newcommand{\domain}{\Omega}														

\newcommand{\map}{{\Theta}} 
\newcommand{\refconfig}{\vect{\Phi}}
\newcommand{\curconfig}{\vect{\varphi}}

\newcommand{\datafieldset}{\mathcal{D}}
\newcommand{\dataset}{\mathfrak{D}}

\newcommand{\mdofs}{M}																	
\newcommand{\numDataPts}{n_D}

\newcommand{\arclen}{\xi}

\newcommand{\numBeam}{n_B}

\newcommand{\xF}{x}																	
\newcommand{\yF}{y}																	
\newcommand{\lF}{z}																	

\newcommand{\xSpace}{X}																	
\newcommand{\ySpace}{Y}																	
\newcommand{\lSpace}{Z}																	

\newcommand{\ytilde}{\tilde{y}}																	

\newcommand{\yhat}{\hat{y}}

\newcommand{\qV}{\vect{q}}                            
\newcommand{\qhat}{\hat{\vect{q}}}

\newcommand{\eV}{\vect{e}}
\newcommand{\sV}{\vect{s}}

\newcommand{\shat}{\hat{\vect{s}}}

\newcommand{\eTilV}{\vect{\tilde{e}}}
\newcommand{\sTilV}{\vect{\tilde{s}}}

\newcommand{\gammaV}{\vect{\gamma}}
\newcommand{\omegaV}{\vect{\omega}}

\newcommand{\chiV}{\vect{\chi}}
\newcommand{\lambV}{\vect{\lambda}}
\newcommand{\muV}{\vect{\mu}}
\newcommand{\nuV}{\vect{\nu}}

\newcommand{\etaV}{\vect{\eta}}
\newcommand{\tauV}{\vect{\tau}}

\newcommand{\hV}{\vect{h}}
\newcommand{\idset}{\Xi}               

\newcommand{\globObjFunc}{\text{dist}_G}      
\newcommand{\eleObjFunc}{\text{dist}_E}      

\newcommand{\globObjFuncP}{J_G}      

\newcommand{\penFunc}{P}
\newcommand{\penFac}{\kappa}
\newcommand{\penSca}{\alpha}
\newcommand{\penEps}{\varepsilon}

\newcommand{\loadFac}{\zeta}

\DeclareMathOperator*{\argmin}{arg\,min}

\allowdisplaybreaks

\modulolinenumbers[5]

\begin{document}

\begin{frontmatter}

\title{A data-driven solving strategy based on a greedy optimization algorithm for the analysis of nonlinear beam structures}

\author[address1]{Thi-Hoa Nguyen \corref{cor1}}
\ead{hoa.nguyen@uib.no}

\author[address1]{Bruno A. Roccia}
\ead{bruno.roccia@uib.no}

\author[address1]{Cristian G. Gebhardt}
\ead{cristian.gebhardt@uib.no}

\cortext[cor1]{Corresponding author}

\address[address1]{Geophysical Institute and Bergen Offshore Wind Centre, University of Bergen, Norway}

\begin{abstract}
  In the last decade, data-driven computational mechanics (DDCM) has emerged as a novel paradigm in computational mechanics, enabling the direct use of constitutive data -- such as stress--strain pairs obtained from experiments, without relying on ad-hoc material models and thereby avoiding information loss. 
  In this work, we extend our data-driven solving strategy GO-ADM, which combines a greedy optimization algorithm with the alternating direction method (ADM), to the structural analysis of geometrically exact beams formulated using director-based kinematics.
  We discuss a data initialization strategy for nonlinear systems based on a conventional finite element analysis of the same structure using a prescribed constitutive model. 
  The resulting discrete stress and strain fields, possibly obtained under multiple loading scenarios, may also be employed as artificial datasets for the subsequent data-driven computations. 
  Furthermore, we investigate the thermomechanical consistency of both the dataset and the discrete solution, and propose a weak enforcement of this consistency in the latter via a penalty approach.
  Numerical examples involving single- and multi-member structures demonstrate that the proposed penalty term leads to thermomechanically consistent discrete stress and strain fields. 
  Moreover, for the studied examples, the solving strategy GO-ADM yields a generally improved approximation of the globally optimal solution compared to the standard ADM-based direct solver.  \\
\end{abstract}

\begin{highlights}
\item We extend our data-driven solving strategy GO-ADM to the structural analysis of geometrically exact beams formulated using director-based kinematics.
\item We initialize the data for nonlinear systems with results obtained from a conventional finite element analysis of the same structure, performed using a prescribed constitutive model.
\item We propose a penalty formulation to weakly enforce the thermomechanical consistency constraint in the discrete solution.
\item We numerically illustrate via single- and multi-member structures that the proposed solving strategy GO-ADM yields a generally improved approximation of the globally optimal solution.
\end{highlights}


\end{frontmatter}

\section{Introduction}
  
Over the last decade, data-driven computational mechanics (DDCM) has developed into an alternative computational framework that replaces the explicit constitutive relation with the direct use of experimental or synthetic material data. 
In its original form \cite{ortiz_ddcm_2016}, also known as the direct DDCM, the method seeks stress and strain pairs from a given dataset which are closest to those satisfying equilibrium, compatibility, and prescribed boundary or initial conditions \cite{ortiz_ddcm_2016,Kirchdoerfer2018ddcmDyn}. 
This is essentially 
an optimization problem in which the objective is to minimize the distance between the discrete material data and the continuous field variables, known as a discrete-continuous optimization problem. 
The solution strategy originally proposed in \cite{ortiz_ddcm_2016} is an alternating iterative procedure in which admissible mechanical states and nearest material data points are updated successively. 
This procedure is closely related to the alternating direction method (ADM) without initialization \cite{Douglas1956adm,Gabay1976adm} 
(see also \cite{Glowinski2014adm} for a historical overview of ADM). 
ADM is applicable to convex and, in practice, also to certain non-convex problems. 
Its basic form, however, may converge slowly or unreliably because the coupling constraints are enforced only implicitly. 
For convex formulations, this drawback can be alleviated by its extension, the alternating direction method of multipliers (ADMM), which introduces dual variables through an augmented Lagrangian framework and thereby improves stability and ensures convergence under standard assumptions \cite{Fortin1983admm,Boyd2011admm}. 
For non-convex problems, ADM nevertheless remains a practical choice, although no general guarantee of convergence to either local or global optima is available (see e.g. discussions in \cite{Gebhardtddcmsolution2025}).

Besides this direct method, 
other variants of DDCM have also been proposed, including inverse approaches that reconstruct constitutive laws from data \cite{Ibanez2017ddcm,Ibanez2019ddcm} and 
hybrid formulations that combine features of both direct and inverse settings \cite{Gebhardtddcmstatic2020,Kanno2021hybridddcm,Gebhardtddcmcontacts2024}. 
The direct data-driven paradigm has been extended to a wide range of applications, 
such as noisy datasets when combining with the maximum-entropy approaches \cite{Kirchdoerfer2017ddcm} or locally convex reconstruction (LCR) approaches \cite{He2020ddcmnoise}, 
dynamic problems \cite{Kirchdoerfer2018ddcmDyn,Gebhardtddcmdynamic2020}, 
finite strains \cite{Keip_ddcm_nonlinearbar,Conti2020nonlinDDCM,Platzer2021admnonlin}, 
stability problems such as snap-through \cite{Kuang2023ddcmsnapthrough} or buckling \cite{Romero2026ddcm}, 
inelasticity \cite{Eggersmann2019ddcminelastic}, 
fracture \cite{Carrara2020fracture}, 
uncertainty quantification \cite{Zschocke2022ddcmuncertainty,Prume2023ddcmuncertainty}, 
multiscale computations \cite{Korzeniowski2021ddcmmultiscale,Gorgogianni2023ddcmmultiscale,Prume2025ddcm}, 
and coupled with game theory for solid mechanics \cite{Weinberg2023ddcm}. 
It has also been adapted for constitutive identification \cite{Leygue2018datamodel,Stainier2019datamodel,Flaschel2022datamodel} 
and transferred to other fields including magnetism \cite{Gersem2020ddcmmagnetic}, electromechanical coupling \cite{Marenic2022ddcmelectro}, and electrical circuit simulation \cite{Gebhardtelectric2025}.

Various properties of the direct DDCM framework have been studied, and several strategies have been proposed to improve its performance. 
One of the earliest and most important aspects concerns the computation of globally optimal solutions \cite{kanno_data_driven_2019,Galetzka2021ddcm}. 
This is particularly relevant for sparse datasets, three-dimensional systems, and nonlinear constitutive behavior, for which the standard direct solver does not guarantee global optimality. 
To address this, Kanno reformulated the direct DDCM problem as a mixed-integer programming problem that can be solved globally with standard solvers \cite{kanno_data_driven_2019}. 
However, the associated computational cost increases rapidly with the size of the dataset. 
More recently, global optimality for linear systems in certain symmetric cases was established in \cite{Gebhardtddcmsolution2025} by introducing a structure-specific initialization of the stress--strain pairs within the original solver \cite{ortiz_ddcm_2016}. 
Other efforts to improve accuracy through a better approximation of the global optimum include the use of local or adaptive weighting parameters associated with the tangent of the constitutive manifold \cite{Galetzka2021ddcm,Nguyen2022ddcm}, 
algorithms incorporating material tangent information from or into the dataset based on tensor voting \cite{Eggersmann2021ddcmaccuracy,Ciftci2022ddcm}, 
and modified fixed-point iterations designed to escape local minima \cite{Prume2025ddcm}. 
In our previous work \cite{viljar2025,nguyenGoadm2026}, we combined a greedy optimization algorithm \cite{Temlyakov2008,TEMLYAKOV2014greedy} with the standard ADM-based solver \cite{ortiz_ddcm_2016,Keip_ddcm_nonlinearbar} 
for one-dimensional bars with nonlinear strains, introducing the so-called GO-ADM solving strategy. 
GO-ADM generally yields a better approximation of the global optima by reducing the global objective function through an iterative search for alternatives to the initial stress--strain pairs in the dataset, as numerically illustrated in \cite{viljar2025,nguyenGoadm2026}. 
A similar approach is the defective restarting approach, introduced in 
\cite{Rocha2025}, which likewise aims to prevent the ADM-based solver from becoming trapped in local optima. 
Both approaches explore the nearest neighbourhood and consider the second-nearest data point as an alternative, drawing inspiration from dropout regularization in deep neural networks \cite{Rocha2025}. 
However, while GO-ADM systematically performs this search across the stress and strain states of all elements, 
the defective restarting approach 
applies it only to randomly selected parts of the states (see also discussion in \cite{nguyenGoadm2026}). 
Another major aspect of the direct DDCM is the computational cost, which is mainly governed by the search for optimal stress--strain pairs in the dataset. 
To reduce this cost, approximate nearest-neighbor algorithms have been proposed in \cite{Eggersmann2021ddcm}, while tree-based (or $k$-d-tree-based) nearest-neighbor search methods \cite{Bentley1975,Zheng2020}, possibly coupled with neural networks for multi-fidelity data \cite{Bahmani2021}, offer an efficient alternative, especially for high-dimensional datasets. 
In addition, the authors of \cite{Nguyen2022ddcm} showed that adaptive hyperparameters representing the tangent of the constitutive manifold can also lower the computational cost of the distance-minimization procedure. Finally, the mathematical structure of the method and the existence of solutions have been further analyzed in \cite{Gebhardtddcmsolution2025,Conti2020nonlinDDCM,Gebhardtddcmhilbert2025,Conti2018ddcm}.

In this work, we extend the GO-ADM solving strategy introduced in our previous work \cite{viljar2025,nguyenGoadm2026}, to geometrically exact beams formulated using director-based kinematics, based on a greedy optimization algorithm and the alternating direction method. 
For such nonlinear systems, 
we propose to initialize the stress and strain data with discrete fields obtained from a conventional finite element analysis (FEA) of the same structure and loading. 
This corresponds to an approximate nonlinear optimization problem (ANLP) \cite{Gebhardtddcmstatic2020} under the assumption of a linear constitutive manifold. 
Although this initialization is computationally more demanding than a random initialization \cite{ortiz_ddcm_2016}, initialization with the stress-free state \cite{Gebhardtddcmstatic2020}, or a structure-specific initialization \cite{Gebhardtddcmsolution2025}, 
it reduces the risk of divergence when iteratively solving the data-driven problem. 
The resulting discrete stress and strain fields from such an FEA can be employed as a synthetic dataset. 
An alternative is to generate such a dataset with virtual material tests, i.e. the discrete solution of the same structure subjected to different loading scenarios. 
To enforce the thermomechanical consistency \cite{Gebhardtddcmhilbert2025} of the discrete stress and strain fields, we propose a penalty approach to weakly enforce this constraint. In particular, we add a penalty term to the global objective function to enforce a tolerance of the normalized product between the discrete stress and strain fields. 
We perform a numerical parameter study showing that a sufficiently large normalizing factor for the stress--strain scalar product preserves the conditioning of the system, thereby allowing larger penalty factors and a more efficient enforcement of the thermomechanical consistency constraint. 
Numerical examples involving beam structures and frames in two and three dimensions demonstrate that the proposed GO-ADM solver is robust and provides a slightly better approximation of the globally optimal solution than the standard solving strategy for the studied cases. Combined with the proposed penalty approach, it sufficiently enforces thermomechanical consistency for the examples considered.

The outline of the paper is as follows: 
in Section \ref{sec:formulations}, we briefly review the optimization problem of the geometrically exact beam using director-based kinematics.  
Here, we also recall the constraints of hinged and rigid joints for multi-beam structures. 
In Section \ref{sec:solver}, we describe the extension of our GO-ADM solver to the beam formulation and discuss the data initialization approaches for this solver. 
In Section \ref{sec:penalty}, we introduce the penalty approach for weakly enforcing the thermomechanical consistency constraint. 
We also perform a numerical parameter study and discuss how to choose different parameters. 
In Section \ref{sec:results}, we numerically illustrate via single- and multi-beam structures favorable properties of our solving strategy together with the introduced penalty approach. 
In Section \ref{sec:conclusions}, we summarize our results and the main conclusions.

\section{Preliminaries}\label{sec:formulations}

In this section, we 
briefly review the geometrically exact beam formulation in a data-driven continuous setting \cite{Gebhardtddcmstatic2020}, including the coupling constraints for hinged and rigid joints in multi-beam structures. 
We also briefly review the direct data-driven solver \cite{ortiz_ddcm_2016,Keip_ddcm_nonlinearbar} that is based on the alternating direction method (ADM).
We start with a brief
recap of the strain measures of the geometrically exact beams \cite{Betsch2002,Eugster2014}.

\subsection{Data-driven geometrically exact beam elements}

\begin{figure}[htb]
    \centering
    \def\svgwidth{0.9\textwidth}
    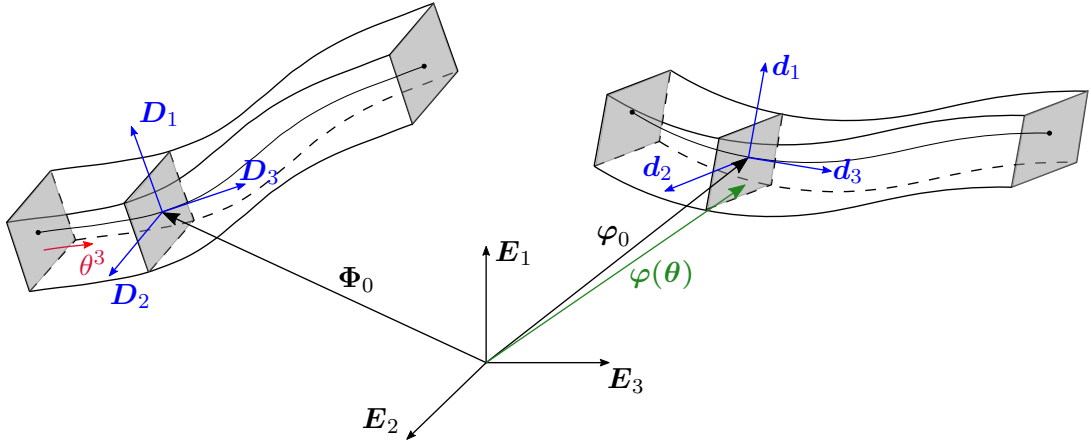

    \caption{Sketch of a three-dimensional geometrically exact beam. Upper- and lowercase letters refer to the reference and current configuration, respectively.}\label{fig:3Dbeam}
\end{figure}

Consider a bounded domain 
$\domain \subset \mathbb{R}$ 
and a physical body that occupies the closure $\bar{\domain}$. 
The position of any point belonging to the beam, illustrated in Figure \ref{fig:3Dbeam}, is:
\begin{equation}
    \curconfig (\vect{\theta}) \,=\, 
        \curconfig_0 \left(\theta^3\right) 
        \,+\, \theta^1 \, \vect{d}_1 \left(\theta^3\right) 
        \,+\, \theta^2 \, \vect{d}_2 \left(\theta^3\right) \; \in \mathbb{R}^3 \,, 
\end{equation}
where 
$\curconfig_0 \in \mathbb{R}^3$ denotes the position vector of the beam axis, 
$\vect{d}_1 \in S^2$, $\vect{d}_2 \in S^2$, and $\vect{d}_3 \in S^2$ the three mutually orthonormal directors belonging to the unit sphere $S^2$. 
Here, $\vect{\theta} = \left(\theta^1,\,\theta^2,\,\theta^3\right)$ is the set of parameters describing the configuration, where $\theta^1$ and $\theta^2$ are the cross-sectional coordinates and $\theta^3$ the coordinate along the beam axis. 
We recall from \cite{Gebhardtddcmstatic2020} the first and second deformation vectors, which are:
\begin{equation}\label{eq:deformVec}
  \gammaV = \begin{bmatrix}
    \vect{d}_1 \cdot \curconfig_0^\prime \\
    \vect{d}_2 \cdot \curconfig_0^\prime \\
    \vect{d}_3 \cdot \curconfig_0^\prime
  \end{bmatrix}\,, \qquad 
  \omegaV = \frac{1}{2} \, \begin{bmatrix}
    \vect{d}_3 \cdot \vect{d}_2^\prime - \vect{d}_2 \cdot \vect{d}_3^\prime \\
    \vect{d}_1 \cdot \vect{d}_3^\prime - \vect{d}_3 \cdot \vect{d}_1^\prime \\
    \vect{d}_2 \cdot \vect{d}_1^\prime - \vect{d}_1 \cdot \vect{d}_2^\prime
  \end{bmatrix}\,,
\end{equation}
respectively, where $(\cdot)^\prime$ denotes the first derivative with respect to the arc length coordinate $\theta^3$. 
The first two components of $\gammaV$ refer to shear responses and the third one to axial response, while the bending responses correspond to the first two components of $\omegaV$ and torsion to its third component. 
The vector of the two strain measures is:
\begin{equation}\label{eq:strainmeasures}
  \eV(\qV) = \begin{bmatrix}
    \gammaV(\qV) - \gammaV_{\text{ref}} \\
    \omegaV(\qV) - \omegaV_{\text{ref}}
  \end{bmatrix} \,,
\end{equation}
where $\qV$ is the vector containing all kinematic fields:
\begin{equation}\label{eq:qField}
  \qV^T = \left[\curconfig_0^T \; \vect{d}_1^T \; \vect{d}_2^T \; \vect{d}_3^T\right]^T\,,
\end{equation}
and the subscript ``ref'' refers to the initial configuration $\refconfig$.

In this work, 
we adapt 
the discrete-continuous nonlinear optimization formulation (DCNLP) for static structural analysis of beam structures \cite{Gebhardtddcmstatic2020}, using our notations. 
For the sake of clarity and notation, we define the following function spaces:
\begin{equation}\label{eq:funcSpaces}
    \mathcal{Q} := H_0^1(\domain;\mathbb{R}^{12}) \,, \quad
    \mathcal{S} := L^2(\domain;\mathbb{R}^{6}) \,, \quad
    \mathcal{V} := L^2(\domain;\mathbb{R}^{12}) \,,
\end{equation}
where \(L^2\) is the space of square-integrable functions, and \(H^1\) the Sobolev space of functions in \(L^2\) with first-order weak derivatives also belonging to \(L^2\). 
We also define 
the following inner product that is weighted by a matrix $\mat{C}$:
\begin{equation}
  \langle\vect{a},\,\vect{b}\rangle_{\mat{C}} := \langle\vect{a},\,\mat{C}\,\vect{b}\rangle_{\mathcal{S}}\,, \quad \text{for } \vect{a},\,\vect{b} \in \mathcal{S} \text{ and } \mat{C} \in \mathbb{R}^{6\times6} \,.
\end{equation}

\noindent
The DCNLP formulation for the data-driven geometrically exact beam \cite[Eq.~14-15]{Gebhardtddcmstatic2020} takes the following form:

\noindent
Find $\xF \in \xSpace$ such that:
\begin{equation}\label{eq:dcnlp}
  \inf_{\xF,\,\ytilde} \; \sup_{\lF \in \lSpace} \;
        \globObjFunc \left(\yF ,\, \ytilde\right)
        \, + \, \map\left( \lF,\, \xF; \, \vect{f} \right)
        \text{ s.t. } 
        \ytilde \in \datafieldset \,,
\end{equation}

\noindent
with
\begin{equation}\label{eq:objFuncNmap}
  \begin{aligned}
        & \globObjFunc \left(\yF ,\, \ytilde\right) := 
        \frac{1}{2} \, \langle\eV-\eTilV,\,\eV-\eTilV\rangle_{\mat{C}} + \frac{1}{2} \, \langle\sV-\sTilV,\,\sV-\sTilV\rangle_{\mat{C}^{-1}} \,, \\
        & \map\left( \lF,\, \xF; \,\vect{f} \right) := \langle \muV, \,\mathcal{B}^T \sV + \mathcal{H}^T \chiV - \vect{f}\rangle_{\mathcal{V}} 
        + \langle \lambV,\, \vect{\epsilon}(\qV) - \eV\rangle_{\mathcal{S}} + \langle \nuV,\, \hV(\qV)\rangle_{\mathcal{V}} \,,
    \end{aligned}
\end{equation}

\noindent
where the subscript "G" denotes the global objective function that is evaluated over the whole domain $\domain$, and
\begin{equation*}
    \begin{aligned}
        & \xF := (\qV,\,\eV,\,\sV) \in \xSpace := \mathcal{Q} \, \times \, \mathcal{S} \, \times \, \mathcal{S} \,, \\
        & \yF := (\eV,\,\sV) \in \ySpace := \mathcal{S} \, \times \, \mathcal{S} \,, \\
        & \lF := (\chiV,\,\muV,\,\lambV,\,\nuV) \in \lSpace := \mathcal{V} \, \times \, \mathcal{Q} \, \times \, \mathcal{S} \, \times \, \mathcal{V} \,.
    \end{aligned}
\end{equation*}

\noindent
We note that our choice of these 
function spaces \eqref{eq:funcSpaces} is based on a linear analysis and might not meet the required regularity of the 
nonlinear strain measure defined in Equation \eqref{eq:strainmeasures}. 
Such an analysis of the regularity requirements is outside the scope of this work and is considered for future work. 
Here, 
$\sV$ is the vector of generalized stress fields corresponding to those in $\eV$ (see also Equation \eqref{eq:strainmeasures}), 
$\datafieldset$ is the space of the corresponding strain and stress fields constructed from a given closed data set of stress--strain pairs $\dataset$ as follows:
\begin{equation}
    \datafieldset := \{ \ytilde := (\tilde{e},\, \tilde{s}) \in L^2 \cross L^2: \left(\tilde{e}(\arclen),\, \tilde{s}(\arclen) \right) \in \dataset \; \forall \, \arclen \in \domain \} \,.
\end{equation}
In general, $\dataset$ consists of available experimental measurements, i.e. discrete data points and the existence of a constitutive manifold is a special case (see also \cite{Gebhardtddcmhilbert2025}).  
$\globObjFunc(\cdot,\cdot)$ denotes the global objective function, 
$\mat{C}$ a constant weighting matrix to ensure unit consistency, 
$\vect{f}$ the generalized external force vector, 
$\vect{h} (\qV)$ the vector of all constraints, 
$\mathcal{B}$ and $\mathcal{H}$ the strain-displacement operator (see also \cite[Eq.~57]{Gebhardtddcmstatic2020}) and the operator regarding the variation of $\vect{h} (\qV)$ with respect to $\qV$, 
(see also \cite[Eq.~66]{Gebhardtddcmstatic2020}), respectively, i.e.:
\begin{equation*}
  \mathcal{B}(\qV) \, \delta \qV = \frac{\partial}{\partial \, \qV} \,\vect{\epsilon}(\qV) \; \delta \qV \,, \qquad
  \mathcal{H}(\qV) \, \delta \qV = \frac{\partial}{\partial \, \qV} \,\vect{h}(\qV) \; \delta \qV \,.
\end{equation*}
For the geometrically exact beam element, $\vect{h} (\qV)$ consists of the orthonormal director constraints \cite[Eq.~65]{Gebhardtddcmstatic2020} that are:
\begin{equation}
    \vect{h} (\qV) = \frac{1}{2} \, \begin{bmatrix}
        \vect{d}_1 \cdot \vect{d}_1 - 1 \\
        \vect{d}_2 \cdot \vect{d}_2 - 1 \\
        \vect{d}_3 \cdot \vect{d}_3 - 1 \\
        2\, \vect{d}_2 \cdot \vect{d}_3 \\
        2\, \vect{d}_1 \cdot \vect{d}_3 \\
        2\, \vect{d}_1 \cdot \vect{d}_2 \\
    \end{bmatrix}\,.
\end{equation}
$\chiV$, $\muV$, $\lambV$, $\nuV$ are the Lagrange multipliers corresponding to the enforcement of the constraints in $\vect{h} (\qV)$ within the equilibrium conditions, compatibility conditions, and the constraints in $\vect{h} (\qV)$ within the optimization problem \eqref{eq:dcnlp}, respectively.

In this work, we do not employ the nullspace approach as in \cite[Eq.~14-15]{Gebhardtddcmstatic2020} since we focus on this DCNLP instead of an approximate nonlinear optimization problem. 
Furthermore, Equation 
\eqref{eq:dcnlp} can be directly adapted for multibeam structures, which we have explicitly formulated in 
\ref{sec:ddcm_multibeam} for the sake of clarity. 
In such cases, the constraint vector $\hV(\qV)$ additionally includes the coupling constraints between the beam members. 
Here, we recall the vector of such constraints for a hinged and rigid joint between the $m$- and $s$-th beam members, that is:
\begin{equation}\label{eq:jointCons}
    \hV_{\text{hinged}} = \begin{bmatrix}
      \curconfig_0^m - \curconfig_0^s 
    \end{bmatrix}\,, \quad \text{and} \quad
    \hV_{\text{rigid}} \left(\qV^m,\,\qV^s\right) = \begin{bmatrix}
      \curconfig_0^m - \curconfig_0^s \\
      \vect{d}_1^m \cdot \vect{d}_1^s - \vect{D}_1^m \cdot \vect{D}_1^s \\
      \vect{d}_1^m \cdot \vect{d}_2^s - \vect{D}_1^m \cdot \vect{D}_2^s \\
      \vect{d}_1^m \cdot \vect{d}_3^s - \vect{D}_1^m \cdot \vect{D}_3^s \\
      \vect{d}_2^m \cdot \vect{d}_1^s - \vect{D}_2^m \cdot \vect{D}_1^s \\
      \vect{d}_2^m \cdot \vect{d}_2^s - \vect{D}_2^m \cdot \vect{D}_2^s \\
      \vect{d}_2^m \cdot \vect{d}_3^s - \vect{D}_2^m \cdot \vect{D}_3^s \\
      \vect{d}_3^m \cdot \vect{d}_1^s - \vect{D}_3^m \cdot \vect{D}_1^s \\
      \vect{d}_3^m \cdot \vect{d}_2^s - \vect{D}_3^m \cdot \vect{D}_2^s \\
      \vect{d}_3^m \cdot \vect{d}_3^s - \vect{D}_3^m \cdot \vect{D}_3^s
    \end{bmatrix} \,, 
\end{equation}
respectively, where $\vect{D}_i$, $i=1,2,3$, are the directors in the reference configuration. 
One can derive 
the first-order necessary optimality conditions (Karush-Kuhn-Tucker (KKT) conditions) following the same procedure in \cite[Eq.~16-17]{Gebhardtddcmstatic2020}, which we adapted for multibeam structures in \ref{sec:ddcm_multibeam}. 
In this work, to tackle the resulting partial differential equations, we employ the finite element method to spatially discretize the kinematic fields, as done in \cite{Gebhardtddcmstatic2020} and our previous work \cite{nguyenGoadm2026}.

\subsection{Data-driven solver based on the alternating direction method}

The well-established data-driven solver was first introduced in \cite{ortiz_ddcm_2016} together with the data-driven computational mechanics paradigm. It 
seeks the stress--strain pairs in a given dataset, $\ytilde \in \dataset$, that are closest to the stress--strain variable fields that satisfy the set of constraints in $\map$. 
This solving strategy is based on the alternating direction method (ADM) \cite{Gebhardtddcmsolution2025,nguyenGoadm2026} since it alternates between two minimization subproblems: 
minimizing the objective function with fixed data points $\ytilde$ to yield optimal variable fields $\xF$, 
and with fixed $\xF$ to yield optimal data $\ytilde$. 
Solvers based on the ADM method are applicable to both convex and certain non-convex problems. 
In general, such solvers do not guarantee a globally optimal solution despite globally optimal solutions of each subproblem \cite{Gebhardtddcmsolution2025,kanno_data_driven_2019}.

\section{Greedy-optimization based alternating direction method}\label{sec:solver}

In this section, we 
extend our solving strategy GO-ADM introduced in \cite{viljar2025,nguyenGoadm2026} to the structural analysis of geometrically exact beams formulated using director-based kinematics (see also the previous section).
We start with a brief review of GO-ADM and describe its application to geometrically exact beams. 
We then discuss a data initialization strategy for nonlinear systems based on a conventional finite element analysis of the same structure using a prescribed constitutive model.

\subsection{Solving strategy}\label{sec:goadm}

\begin{algorithm}[htb]
\textbf{Input}:
	dataset $\dataset$,
	generalized external force vector $\vect{f}$ \\
\textbf{Output}: $\qhat$, $\ytilde$     \Comment{$\qhat$: nodal values of all kinematics fields.}
\begin{algorithmic}[1]
    \State $\qhat^{(0)} = \vect{0}$      \Comment{Initial solution guess.}
\For{$j$ in $1,\ldots,$ number of load steps}
    \State $\vect{f}^{(j)} = \loadFac_j \, \vect{f}$    \Comment{$\loadFac_j$: $j$-th load factor.}
    \State Inititalize $\ytilde^{(j)}$      \Comment{see Algorithm \ref{alg:datainitialize}.}
    \State $\qhat^{(j)}, \, \ytilde^{(j)}$ = GO-ADM-solver $\left(\qhat^{(j-1)}, \, \ytilde^{(j)}, \, \vect{f}^{(j)}, \, \dataset\right)$      \Comment{See Algorithm \ref{alg:goadmsolver}.}
\EndFor
\caption{GO-ADM solving strategy for geometrically nonlinear data-driven problems \cite{nguyenGoadm2026}.}\label{alg:goadmstrategy}
\end{algorithmic}
\end{algorithm}
\begin{algorithm}[htb]
\textbf{Input}:
	solution guess $\qhat_0$, 
    initial selected data $\ytilde_0$,
    dataset $\dataset$, 
	external force vector $\vect{f}$ \\
\textbf{Output}: $\qhat$, $\ytilde$
\begin{algorithmic}[1]
    \State $\yhat \gets \qhat, \, \ytilde$ = ADM-solver $\left(\qhat_0, \, \ytilde_0, \, \vect{f}, \, \dataset\right)$      \Comment{First results.}
    \State dist$^{(0)}$ = dist$_G\left(\yhat,\, \ytilde\right)$
    \State $k=0$     \Comment{Number of ``greedy'' searches.}
    \While{$k \leq k_{\text{max}}$}
        \State $\idset = [i]_{1,\ldots,\mdofs}$ s.t. $\left[\eleObjFunc\left(\yhat_i,\, \ytilde_i\right) \right]_{i \in \idset}$ is descending
        \For{$i$ in $\idset$}
            \State $k \mathrel{+}= 1$, \textcolor{blue1}{$\ytilde^{n}$} = $\ytilde$
            \State $q$, $p$ = $\argmin\left( \, \eleObjFunc\left(\yhat_i,\, \ytilde_j\right) \, \right)$, $j=1,\ldots,\numDataPts$     \Comment{Find 2 nearest data points.}
            \If{\textcolor{blue1}{$\ytilde^{n}_i$} $\neq \, \ytilde_{q}$}
                \State \textcolor{blue1}{$\ytilde^{n}_i$} = $\ytilde_{q}$      \Comment{Get the nearest data point.}
            \Else
                \State \textcolor{blue1}{$\ytilde^{n}_i$} = $\ytilde_{p}$      \Comment{Get the 2nd-nearest data point.}
            \EndIf
            \State $\yhat \gets \qhat$, \textcolor{blue1}{$\ytilde^{n}$} = ADM-solver $\left(\qhat_0, \, \textcolor{blue1}{\ytilde^{n}}, \, \vect{f}, \, \dataset\right)$      \Comment{Recompute with new $\textcolor{blue1}{\ytilde^{n}_i}$ in $\textcolor{blue1}{\ytilde^{n}}$.}
            \State dist$^{(k)}$ = dist$_G\left(\yhat,\, \textcolor{blue1}{\ytilde^{n}}\right)$
            \If{dist$^{(k)}$ $<$ dist$^{(k-1)}$ or dist$^{(k)} \leq \delta$}
                \State $\ytilde_i = \textcolor{blue1}{\ytilde^{n}_i}$
                \If{dist$^{(k)} \leq \delta$}: $k=k_{\text{max}}+1$
                \EndIf
                \State break
            \EndIf
        \EndFor
	\EndWhile
\caption{GO-ADM-solver \cite{nguyenGoadm2026}.}\label{alg:goadmsolver}
\end{algorithmic}
\end{algorithm}

To achieve a better approximation of the globally optimal solution, we have combined a greedy optimization algorithm and standard data-driven solver based on the alternating direction method \cite{viljar2025,nguyenGoadm2026}. 
We refer to the standard solver as the ADM-solver \cite{ortiz_ddcm_2016,Keip_ddcm_nonlinearbar} and to ours as the GO-ADM solver. 
The main idea is to iteratively assign the stress--strain states in each bar element to the second-nearest data point in the neighbourhood, testing whether the global objective function, $\globObjFunc$, decreases \cite{viljar2025,nguyenGoadm2026}. 
In this work, we now extend and apply the latter to three-dimensional geometrically exact beam structures, reviewed in the previous section.

\begin{remark}\label{rmk:dataset}
  We note that the dataset, $\dataset$, of the three-dimensional geometrically exact beam consists of $\numDataPts$ stress--strain states, each of which is a pair of six-components stress and strain vectors. 
  In other words, each stress--strain state in $\dataset$ consists of six stress--strain data pairs corresponding to six components of the stress and strain vectors. 
  In general, such datasets may be obtained from experiments or material tests. While strain or deformation quantities can be measured directly or reconstructed from displacement measurements, stresses are generally not directly measurable and must be inferred, for instance, from measured forces, specimen geometry, and equilibrium relations. The construction of sufficiently accurate paired stress--strain data is therefore nontrivial, particularly for slender structures undergoing large deformations, where geometric nonlinearities, nonuniform deformation, and measurement uncertainties may complicate the reconstruction of the corresponding physical states. 
  In the present work, we focus on the numerical aspects of the proposed solving strategy and therefore employ synthetic datasets generated under the assumptions of homogeneous isotropic materials using virtual material tests, as described in \ref{sec:datageneration}.
\end{remark}

Extending the GO-ADM solver introduced in \cite{viljar2025,nguyenGoadm2026} to 
the considered three-dimensional geometrically exact beam formulation is straightforward. 
For the sake of readability and clarity, 
we include here 
Algorithms \ref{alg:goadmstrategy} and \ref{alg:goadmsolver} to describe the GO-ADM solver and 
discuss the difference in three-dimensional setting. 
The former first shows where 
the GO-ADM solver is employed and the latter then explains this solver step by step in detail. 
We see in Algorithm \ref{alg:goadmstrategy} that at each load step, the GO-ADM solver provides two outputs: 
i) the structural solution, $\qhat$, which now consists of nodal values of all kinematic fields of the beam formulation (see also \eqref{eq:qField}), and 
ii) converged data, $\ytilde$, which now is a pair of 
two six-components vectors: a stress and a strain vector, instead of two scalars (a data point) as in a one-dimensional setting. 
This necessarily means that the dataset, $\dataset$, now consists of pairs of stress and strain vectors, or rather stress--strain states 
(see also \ref{rmk:dataset}). 
One required input of the GO-ADM solver is the initial stress--strain data, 
which we select from the dataset, $\dataset$, using Algorithm \ref{alg:datainitialize}, discussed in the next section.

Algorithm \ref{alg:goadmsolver} describes the GO-ADM solver step by step. 
We see in line 8 the key idea of this solver: 
\begin{equation*}
  q,\, p = \argmin\left( \, \eleObjFunc \left(\yhat_i,\, \ytilde_j\right) \, \right), \; 
  i=1,\,\ldots,\,n_e, \; 
  j=1,\,\ldots,\,\numDataPts \,,
\end{equation*}
that is to systematically search for the two closest data in $\dataset$ and assign $\ytilde$ of each element to the best alternative, testing whether this reduces the global objective function, $\globObjFunc$, \cite{viljar2025,nguyenGoadm2026}. 
We note that to distinguish from $\globObjFunc$, the subscript "E" denotes the local objective function that is evaluated over an element. 
Here, $n_e$ and $\numDataPts$ denote the number of elements and the number of given stress--strain states in $\dataset$, respectively. 
We see in line 14 of Algorithm \ref{alg:goadmsolver} that after 
the second-nearest data is assigned to $\ytilde$ of each element, we recompute the solution:
\begin{equation*}
  \yhat \gets \qhat, \ytilde^{n} = \text{ADM-solver } \left(\qhat_0, \, \ytilde^{n}, \, \vect{f}, \, \dataset\right)\,,
\end{equation*}
using the standard ADM solver and new $\ytilde^n$ as initial data. Here, the superscript $n$ denotes the data $\ytilde$ with new values for the stress--strain states of the considered element.  
For more technical details regarding the ADM solver, we refer to \cite[Alg.~2]{nguyenGoadm2026} and discussions therein.

Regarding the searching procedure for the nearest and second-nearest within both the ADM and GO-ADM solvers, we now consider the generalized stress and strain vector consisting of six components. 
We note that in cases of homogeneous isotropic beams with assumed linear constitutive relation, the axial, shear, bending, and torsional responses are uncoupled, leading to a diagonal generalized constitutive matrix. 
Therefore, one can independently search for the nearest (and second-nearest) data for each pair of corresponding generalized stress and strain components. 
For such cases, one can also generate an independent synthetic dataset for each component (see also discussions in \ref{sec:datageneration}) and perform the search in parallel. 
In this work, unless stated otherwise, 
we generate synthetic datasets using virtual material tests (see also \ref{sec:datageneration}), where the stress and strain components are related to a physical state. 
We note that one can generally obtain constitutive data consisting of stress--strain pairs from material experiments and/or tests, although the acquisition of sufficiently accurate paired stress--strain data can be challenging, particularly for slender structures undergoing large deformations 
(see also Remark \ref{rmk:dataset}). 
Using a synthetic dataset, 
we search for the stress--strain states, i.e. pairs of stress and strain vectors in the dataset, instead of a separate search for each component. 
From now on, we refer to the former as a vector-wise search and the latter a component-wise search.

\begin{remark}
  We note that in the cases of homogeneous isotropic materials, even though the deformations corresponding to each component of the generalized stress and strain vectors are not coupled, these deformations are measured and described in terms of stress--strain states, formulated as tensors or vectors, instead of independent single scalars. 
  This necessarily means that the component-wise searching strategy is physically incorrect. One can consider this as a numerical approach to deal with incomplete data that includes measurements or values of each deformation but not the complete state.
\end{remark}

Regarding the computational cost, 
compared to the data-driven bar formulation studied in \cite{viljar2025,nguyenGoadm2026}, 
the higher dimensionality of the beam formulation generally requires in general more memory and higher computational effort. 
In particular, the evaluation of the objective functions involves vectors instead of scalars and the linearized system of equations has higher dimension for the same mesh. 
When searching for closest data, the same computational effort is expected, given that the vector-wise searching strategy is employed on a data region consisting of the same number of states as data points. 
Using the component-wise searching strategy requires six times more effort per state since six searches are performed instead of one. 
Moreover, the current GO-ADM solver searches for the second nearest neighbour in the entire dataset, which can be restricted to an active neighbourhood to reduce the searching effort. 
As discussed in our previous work \cite{nguyenGoadm2026}, other potential strategies to reduce the computational cost of the GO-ADM solver include, for instance, 
applying this solver to selective load steps, and/or assigning second-nearest data only to a subset of the stress--strain states. 
We plan to explore and investigate these approaches in future work for further improvements of the GO-ADM solver.

Compared with the standard ADM solver, the GO-ADM strategy introduces additional computational effort through the greedy search, since each tested alternative stress--strain state requires an additional solving step using the ADM solver. The number of such repeated solves depends on the stopping criterion and the prescribed maximum number of greedy searches, $k_{\text{max}}$. In our previous work \cite{nguyenGoadm2026}, we have discussed and compared the computational cost of the ADM and GO-ADM solving strategies in more detail. 
In the present beam formulation, the same algorithmic structure is retained, while the higher-dimensional stress--strain states and larger linearized systems generally increase the computational effort of each solver.

\begin{remark}
  We note that both the cardinality and the distribution of the dataset can affect the accuracy and computational cost of the direct data-driven solution. For one-dimensional elasticity, we investigated the influence of nonuniform data distribution on the ADM- and GO-ADM-based solving strategies in our previous work \cite{nguyenGoadm2026}. In the present work, ADM and GO-ADM are intentionally compared using the same datasets in order to isolate the effect of the solving strategy. 
  Whether GO-ADM can achieve a prescribed accuracy with a smaller dataset, or whether it is more or less sensitive than the ADM solver to data sparsity, therefore constitutes a separate question that we leave for future work. Nevertheless, the numerical examples in the next section confirm the qualitative tendency that enriching the dataset, particularly in the relevant stress--strain regions, generally improves the accuracy of the data-driven solution.
\end{remark}

\subsection{Data initialization approaches}\label{sec:datainitialization}

\begin{algorithm}[htb]
\textbf{Input}:
    dataset $\dataset$, 
    load step $j$, 
    number of initializations $n_{I}$ \\
\textbf{Output}: $\ytilde^{(j)}$
\begin{algorithmic}[1]
    \If{$j \leq n_{I}$}
        \If{\textit{random}}
            \State $\idset$ = random selected $\mdofs$-indices $i \in \{1,\ldots,\numDataPts\}$
            \State Select $\ytilde_i^{(j)} \in \dataset$, with $i \in \idset$ for $\mdofs$ elements
        \ElsIf{\textit{stress-free}}
            \State Select $\ytilde_i^{(j)} = (\vect{0},\,\vect{0}) \in \dataset$, $\forall \, i=1,\ldots,\mdofs$
        \ElsIf{\textit{structure-specific}}
            \State $\yhat \gets \qhat$ = Solve Equation \eqref{eq:equilibriumCondMatrixEq}
            \State Select $\ytilde_i^{(j)} \in \dataset$ s.t. $\frac{1}{2} \, \langle\shat-\tilde{\sV},\,\shat-\tilde{\sV}\rangle_{\mat{C}^{-1}}$ = min, $i=1,\ldots,\mdofs$
        \Else           \Comment{Default with the \textit{ANLP} initialization approach}
            \State $\yhat \gets \qhat$ = Solve Equation \eqref{eq:anlpWeakForm}
            \State Select $\ytilde_i^{(j)} \in \dataset$ s.t. $\eleObjFunc(\yhat_i,\ytilde_i^{(j)})$ = min, $i=1,\ldots,\mdofs$
        \EndIf
    \Else
        \State $\ytilde^{(j)} = \ytilde^{(j-1)}$
    \EndIf
    \State Return $\ytilde^{(j)}$
\caption{Data initialization scheme.}\label{alg:datainitialize}
\end{algorithmic}
\end{algorithm}

The GO-ADM solver, based on the alternating direction method, requires initial values of the stress and strain data, $\ytilde$, to solve for the discrete solution, $\qhat$, and update $\ytilde$ until convergence. 
As discussed and reviewed in our previous work \cite{nguyenGoadm2026}, one can initialize $\ytilde$ either by randomly selecting data from $\dataset$ \cite{ortiz_ddcm_2016}, selecting stress-free states \cite{Gebhardtddcmstatic2020}, i.e. zero stresses and strains, or using the structure-specific approach \cite{Gebhardtddcmsolution2025}. 
Despite higher dimensional stress and strain vectors, the application of these three approaches to the geometrically exact beam formulation is straightforward. 
We note that since a 
random initialization might lead to $\ytilde$ that does not correspond to any relevant stress--strain state, it possibly results in 
a higher number of iterations for the ADM solver and/or diverging Newton-Raphson scheme. 
Using the stress-free or structure-specific initialization reduces this risk since they relate to the reference or one specific current configuration, respectively. 
The latter is based on solving a linear system and guarantees global optima 
for geometrically linear structures in certain symmetric cases \cite{Gebhardtddcmsolution2025}. 
For further discussions regarding the three aforementioned initialization approaches, we refer to our previous work \cite{nguyenGoadm2026}. 
For the sake of clarity and completeness, we state here the equations of the equilibrium conditions of the geometrically exact beam that are solved when using the structure-specific initialization:
\begin{equation}\label{eq:equilibriumCondMatrixEq}
  \langle \delta \qV, \,\mathcal{B}^T \sV + \mathcal{H}^T \chiV - \vect{f}\rangle_{\mathcal{V}} = \vect{0} \,,
\end{equation}
where $\delta \qV$ is the variation of $\qV$ (see also Equation \eqref{eq:qField}). 
We note that solving these equations for the stresses, $\sV$, requires a solution of the kinematics fields in $\qV$ since 
the operators $\mathcal{B}(\qV)$ and $\mathcal{H}(\qV)$ depend on these, as well as the Lagrange multipliers $\chiV$. 
Employing the solution of the previous load step, one solves for $\sV$ and then finds the closest stress and corresponding strain states in $\dataset$ as initial values for $\ytilde$.

In this work, 
to further reduce the risk of having a large number of iterations for the ADM solver and diverging Newton-Raphson scheme, particularly for the considered nonlinear beam structure,
we propose a fourth initialization approach 
using the solution of the conventional finite element analysis for the same beam structure based on an assumed constitutive relation. 
In particular, we assume a linear relation and employ the same discretization and loads for this initialization procedure. 
For the sake of clarity and completeness, we state here the weak form of the resulting standard boundary-value problem in a continuous setting:

\noindent
Find $\xF \in \xSpace$ such that:
\begin{equation}\label{eq:anlpWeakForm}
  \begin{aligned}
    \inf_{\xF} \; \sup_{\chiV \in \mathcal{V}} \; &
      \langle \delta \qV, \,\mathcal{B}^T \sV + \mathcal{H}^T \chiV - \vect{f}\rangle_{\mathcal{V}} 
      + \langle \delta \eV,\, \mat{D} \, \eV - \sV\rangle_{\mathcal{S}} 
      + \langle \delta \sV,\, \vect{\epsilon}(\qV) - \eV\rangle_{\mathcal{S}} 
      + \langle \delta \chiV,\, \hV(\qV)\rangle_{\mathcal{V}} \,, \\
      & \forall \left(\delta \xF,\, \chiV \right) \in \xSpace \times \mathcal{V} \,, 
  \end{aligned}
\end{equation}
where $\delta \xF := (\delta \qV,\, \delta \eV,\,\delta \sV)$ denotes the collection of the variation of the kinematics, strain, and stress fields, 
$\delta \chiV$ the variation of the Lagrange multipliers, $\chiV$, 
and $\mat{D}$ the material matrix corresponding to the assumed linear constitutive relation. 
The weak form \eqref{eq:anlpWeakForm} and its discrete form are nonlinear and one can solve this using, for instance, the Newton-Raphson method. 
Here, we employ a three-field mixed finite element formulation in which the kinematic, generalized strain, and generalized stress fields, $\qV,\, \eV,$, and $\sV$, respectively, are treated as independent fields, based on the Hu-Washizu principle. 
The motivation of this choice is to 
make its similarity to the discrete-continuous nonlinear optimization formulation (DCNLP) stated in Equation \eqref{eq:dcnlp} readily apparent. 
Mixed and multi-field formulations for geometrically exact spatial beams have been considered in earlier works, e.g. \cite{Wackerfus2009,Santos2010}, and have also been employed in the data-driven setting with director-based kinematics \cite{Gebhardtddcmstatic2020}. One can find more recent developments of mixed formulations for geometrically exact director-based beams, including their relation to the Hu-Washizu principle, e.g., in \cite{Kinon2025beam,Kinon2026beam}. Since the present formulation is employed only for the data initialization, a detailed derivation of the mixed finite element equations is beyond the scope of this work. 
We note that the boundary-value problem associated with the weak form \eqref{eq:anlpWeakForm} can be considered as a special case of an approximate nonlinear optimization problem (ANLP), introduced in \cite{Gebhardtddcmstatic2020}, when assuming and enforcing a linear constitutive manifold. 
Hence, from now on, we refer to this initialization approach as the ANLP initialization. 
Finding the closest data to the discrete solution of \eqref{eq:anlpWeakForm} in $\dataset$ yields the initial data, $\ytilde$, either only at the first or up to selected load steps. 
Moreover, one can also apply the discrete solution of other kinematics fields in $\qhat$ as initial guess when solving the associated DCNLP.

Regarding the computational cost of the ANLP initialization, it requires a complete finite element analysis of a nonlinear beam formulation. 
One can solve the related ANLP once in advance and only for selected load steps. 
The solution is then stored for the initialization of $\ytilde$ when solving the associated DCNLP. 
We briefly emphasize that the ANLP solution is used only for initialization, without constraining the data-driven solution to satisfy the constitutive relation assumed in the ANLP. 
On the one hand, 
the computational effort and required memory storage are higher than the three aforementioned initialization approaches. 
On the other hand, 
this approach points the data-driven solver to the data neighbourhood of the best solution based on the assumed constitutive relation, for both linear and nonlinear systems. 
Given an accurate assumption, this might guarantee global optima and 
reduce the number of iterations of the data-driven solver, as well as that of the Newton-Raphson scheme when using the obtained discrete solution as an initial solution guess.

Owing to this favorable property of the ANLP initialization approach but being aware of its high computational cost, 
we employ this as our default option only for the first load step and use the converged data $\ytilde$ from the previous load step as initial value for the next one. 
In this work, we perform our numerical studies using this default initialization approach unless stated otherwise. 
We describe how we choose and employ the ANLP and the three aforementioned initialization approaches at each load step 
in Algorithm \ref{alg:datainitialize}. 
Here, the number of initializations $n_{I}$ is the number of load steps for which we choose to initialize $\ytilde$ using one of these four approaches. 
It is an optional input and takes a value of one as a default value.
We note that when using the ANLP initialization and searching for the closest data $\ytilde_i^{(j)} \in \dataset$ for the state in the $i$-th element at the $j$-th load step (see line 12 of Algorithm \ref{alg:datainitialize}), 
we employ the element objective function:
\begin{equation*}
  \eleObjFunc(\yhat_i,\ytilde_i^{(j)}) = \min\,, \quad i=1,\ldots,\mdofs\,,
\end{equation*}
that is the same objective function defined in Equation \eqref{eq:objFuncNmap} evaluated at the $i$-th element.

\section{Thermomechanical consistency constraint}
\label{sec:penalty}

In this section, we introduce an approach to weakly enforce the thermomechanical consistency constraint \cite{Gebhardtddcmhilbert2025} of the solution, using the penalty method. 
This constraint plays an essential role in ensuring a numerical solution that stores non-negative elastic energy in a direct data-driven setting. 
We then numerically illustrate via an exemplary beam structure that the scaling factor of the strain energy in the proposed penalty term significantly affects the condition number of the resulting system matrix. 
A sufficiently large value of this scaling factor leads to a well-conditioned system, allowing larger penalty factors. 
We show that given good conditioning, a sufficiently large penalty factor enforces the thermomechanical consistency constraint efficiently, irrespective of the chosen enforced tolerance. 
We start with the proposed penalty function and discuss its most important properties.

\subsection{A penalty method}

\begin{figure}[htb]
    \centering
    \includegraphics[width=0.47\textwidth]{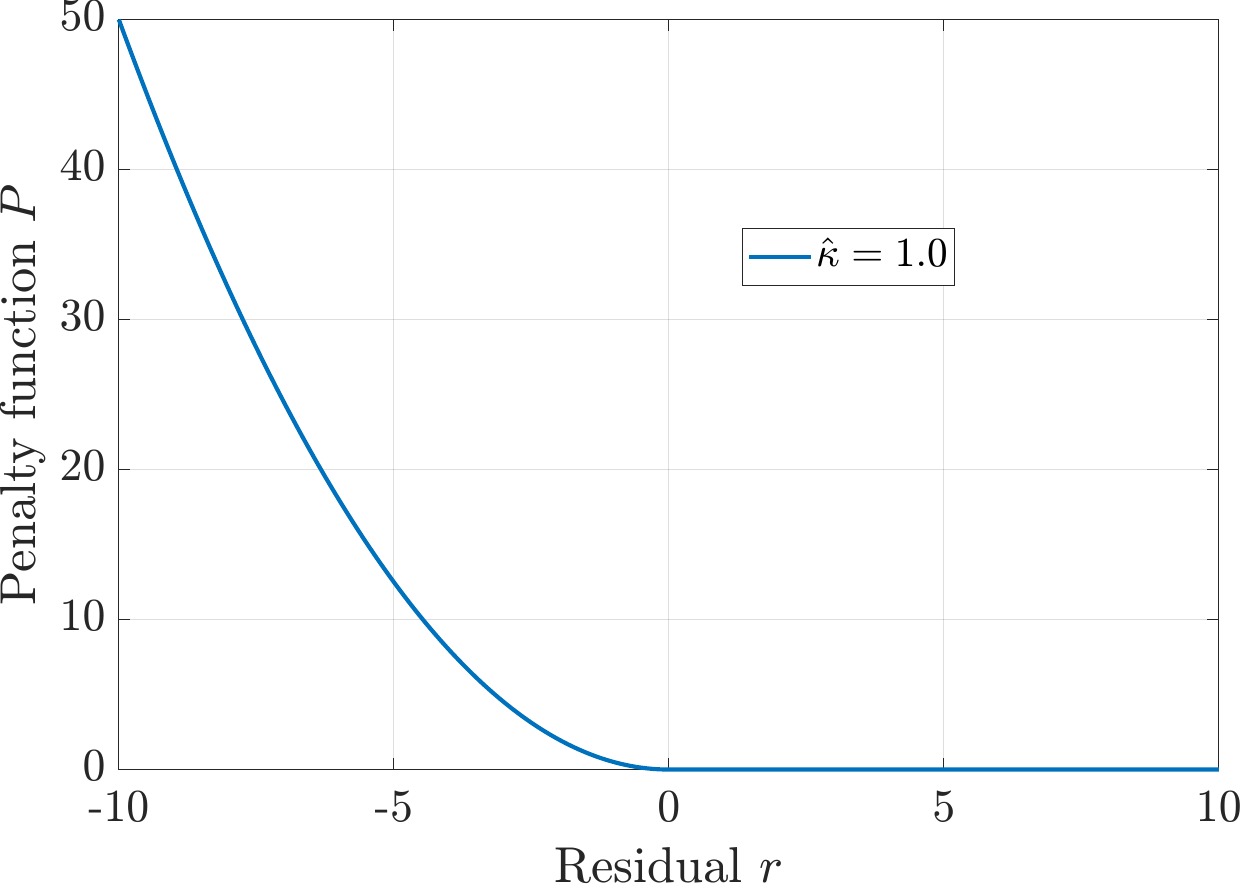}

    \caption{Penalty function \eqref{eq:penFunc}, $\penFunc (\eV,\,\sV)$, with $\penFac=1.0$.} \label{fig:penFunc}
\end{figure}

In this work, we consider an isothermal and quasi-static elastic setting, without dissipative or history-dependent effects. 
In this context, the thermomechanical consistency condition considered in \cite{Gebhardtddcmhilbert2025} reduces to requiring a non-negative generalized strain energy, i.e. $\frac{1}{2} \eV \cdot \sV \geq 0$. 
Physically, this condition excludes generalized stress--strain states associated with a negative stored elastic energy in the present setting. 
We note that in a data-driven setting, 
since the discrete stress and strain fields are obtained through the constrained distance-minimization problem instead of an explicitly prescribed constitutive law, 
they may violate this condition, even if the material dataset itself is thermomechanically consistent. 
To weakly enforce the thermomechanical consistency condition described above, 
we add the following penalty term to the global objective function, $\globObjFunc$, defined in Equation \eqref{eq:objFuncNmap}:
\begin{equation}\label{eq:penFunc}    
        \penFunc (\eV,\,\sV) = \frac{1}{2} \, \hat{\penFac} \, \left[\min(r,\,0)\right]^2 \,, \quad \text{with }         
        r = \frac{\eV \cdot \sV}{\penSca} + \penEps \,, \; \penSca > 0\,, \, \penEps \geq 0 \,,
\end{equation}
where 
$\hat{\penFac}$, $\penSca$, and $\penEps$ are the penalty factor, scaling factor of the product $(\eV \cdot \sV)$ that is twice the strain energy, and the enforced 
lower bound of this normalized product, respectively. 
The key idea is to weakly enforce a non-negative residual, $r \geq 0$, that is equivalent to a lower bound of the strain energy: 
\begin{equation}
    r = \frac{\eV \cdot \sV}{\penSca} + \penEps \geq 0 
    \; \rightarrow \; \frac{1}{2} \, \eV \cdot \sV \geq - \frac{1}{2}\, \penSca \, \penEps \,.
\end{equation}
We note that the thermomechanical consistency constraint strictly requires $\penEps=0$, i.e. non-negative strain energy. 
Choosing a positive but small value of $\penEps$ necessarily means a relaxed tolerance of this constraint. 
Figure \ref{fig:penFunc} shows the penalty function $\penFunc(\eV ,\, \sV)$ for different values of the residual $r$ with a penalty factor of $\hat{\penFac}=1.0$. 
We see that $\penFunc(\eV ,\, \sV)$ is zero for non-negative residual $r$, i.e. when the thermomechanical consistency constraint is satisfied, as expected.

For completeness, we state here the modified global objective function:
\begin{equation}\label{eq:objFuncModi}
    \globObjFuncP \left(\yF ,\, \ytilde\right) := 
    \frac{1}{2} \, \langle\eV-\eTilV,\,\eV-\eTilV\rangle_{\mat{C}} + 
    \frac{1}{2} \, \langle\sV-\sTilV,\,\sV-\sTilV\rangle_{\mat{C}^{-1}} + \frac{1}{2} \, \hat{\penFac} \, \left[\min(r,\,0)\right]^2
    \,. \\
\end{equation}
We note that $\globObjFuncP$ is no longer a distance function as $\globObjFunc$, since the added penalty term does not measure any distance. 
While the first two terms are inner products and hence involve integration over the physical domain, 
the penalty term is evaluated at each material point. 
This necessarily means that the thermomechanical consistency constraint is weakly enforced pointwise at the same points. 
Furthermore, since we normalize the product $(\eV \cdot \sV)$ in the residual $r$ by the scaling factor $\penSca$, $r$ is dimensionless. 
To preserve the unit consistency 
between all terms of the modified objective function \eqref{eq:objFuncModi}, we choose the penalty factor $\hat{\penFac}$ as follows: 
\begin{equation}
    \hat{\penFac} = h \, \penFac \, \max(\eTilV \cdot \sTilV)\,,
\end{equation} 
where $h$ is the element size of the discretization and $\penFac$ is a dimensionless penalty factor. 
From now on, we simply refer to $\penFac$ as the penalty factor. 
We note that the last factor, $\max(\eTilV \cdot \sTilV)$, where $\eTilV$ and $\sTilV$ are the stress and strain vectors in the dataset in $\dataset$, 
is essentially twice the strain energy obtained from a given dataset. 
On the one hand, 
a large penalty factor effectively enforces the thermomechanical consistency constraint. 
On the other hand, it might lead to an ill-conditioned system. 
Note that while 
in general, one can adaptively choose a scaling factor $\penSca$ for each material point, we simply choose a constant scaling factor for all computations in this work.

\begin{remark}
    We note that in general, 
    one cannot ensure that 
    the stress and strain data 
    resulting from measurements and experiments     
    are thermomechanically consistent. However, 
    the given dataset should not be modified or constrained to any predefined properties. 
    In this work, we employ synthetic data from virtual material tests based on a linear constitutive relation (see also \ref{sec:datageneration}), which are thermomechanically consistent. 
\end{remark}

\subsection{Numerical parameter study}\label{sec:penaltyPara}

\begin{figure}[htb]
\centering
    \subfloat[Snapshots]{
    \def\svgwidth{0.33\textwidth}
\begingroup%
  \makeatletter%
  \providecommand\color[2][]{%
    \errmessage{(Inkscape) Color is used for the text in Inkscape, but the package 'color.sty' is not loaded}%
    \renewcommand\color[2][]{}%
  }%
  \providecommand\transparent[1]{%
    \errmessage{(Inkscape) Transparency is used (non-zero) for the text in Inkscape, but the package 'transparent.sty' is not loaded}%
    \renewcommand\transparent[1]{}%
  }%
  \providecommand\rotatebox[2]{#2}%
  \newcommand*\fsize{\dimexpr\f@size pt\relax}%
  \newcommand*\lineheight[1]{\fontsize{\fsize}{#1\fsize}\selectfont}%
  \ifx\svgwidth\undefined%
    \setlength{\unitlength}{337.5bp}%
    \ifx\svgscale\undefined%
      \relax%
    \else%
      \setlength{\unitlength}{\unitlength * \real{\svgscale}}%
    \fi%
  \else%
    \setlength{\unitlength}{\svgwidth}%
  \fi%
  \global\let\svgwidth\undefined%
  \global\let\svgscale\undefined%
  \makeatother%
  \begin{picture}(1,1.11111111)%
    \lineheight{1}%
    \setlength\tabcolsep{0pt}%
    \put(0,0){\includegraphics[width=\unitlength,page=1]{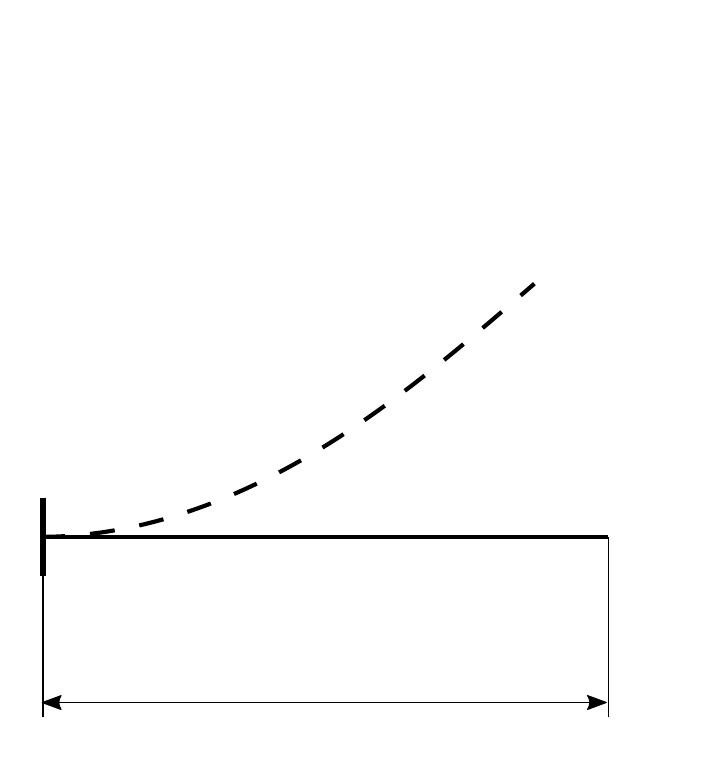}}%
    \put(0.82513231,0.64922529){\color[rgb]{0,0,0}\makebox(0,0)[lt]{\lineheight{1.25}\smash{\begin{tabular}[t]{l}$F$\end{tabular}}}}%
    \put(0,0){\includegraphics[width=\unitlength,page=2]{figs/cantileverSimoSnaps-sketch.pdf}}%
    \put(0.88427965,0.23602587){\color[rgb]{0,0,0}\makebox(0,0)[lt]{\lineheight{1.25}\smash{\begin{tabular}[t]{l}$F$\end{tabular}}}}%
    \put(0.32190362,0.136706){\color[rgb]{0,0,0}\makebox(0,0)[lt]{\lineheight{1.25}\smash{\begin{tabular}[t]{l}$L_0 = 100$ m\end{tabular}}}}%
    \put(0.02596232,1.00889543){\color[rgb]{0,0,0}\makebox(0,0)[lt]{\lineheight{1.25}\smash{\begin{tabular}[t]{l}$EI = 3.5\cdot10^7$ Nm$^2$\end{tabular}}}}%
    \put(0.02674497,0.88577474){\color[rgb]{0,0,0}\makebox(0,0)[lt]{\lineheight{1.25}\smash{\begin{tabular}[t]{l}$GA = 1.61538\cdot10^8$ N\end{tabular}}}}%
    \put(0.02804242,0.76478343){\color[rgb]{0,0,0}\makebox(0,0)[lt]{\lineheight{1.25}\smash{\begin{tabular}[t]{l}$F_{\text{max}} = 130$ kN\end{tabular}}}}%
  \end{picture}%
\endgroup%
} \hspace{0.3cm}
    \subfloat[Minimal value of $(\vect{e}_h \cdot \vect{s}_h)$]{\includegraphics[width=0.5\textwidth]{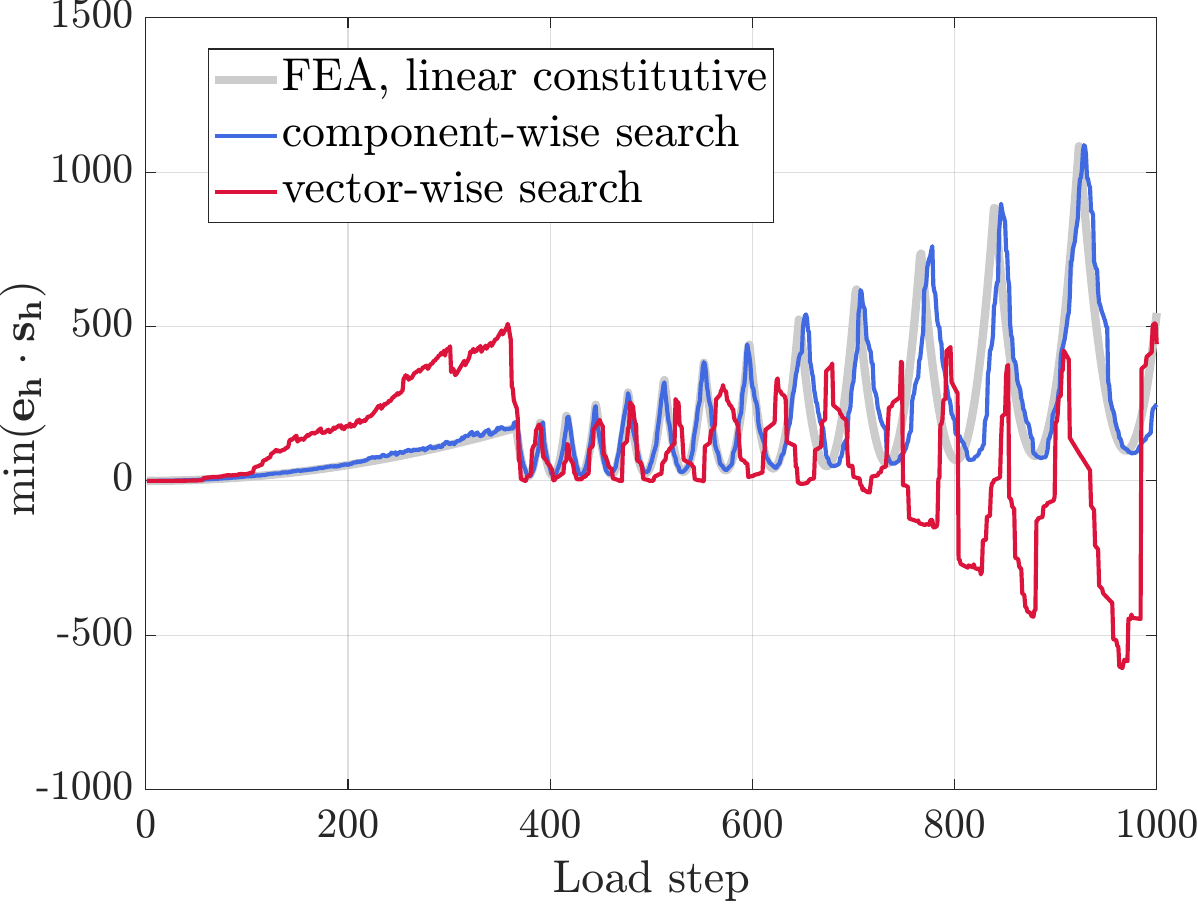}}

    \caption{Sketch of a cantilever beam subjected to a follower end force \cite{Simo1986} and the minimal values of the product $(\vect{e}_h \cdot \vect{s}_h)$ over all elements, computed with \textbf{our GO-ADM solving strategies} and $\penFac=0$, using a synthetic dataset of 5000 stress--strain states, generated from virtual material tests.}\label{fig:simobeamOriMinE}
\end{figure}

\begin{figure}[htb]
\centering
    \subfloat[$\penEps=0.0$]{\includegraphics[width=0.49\textwidth]{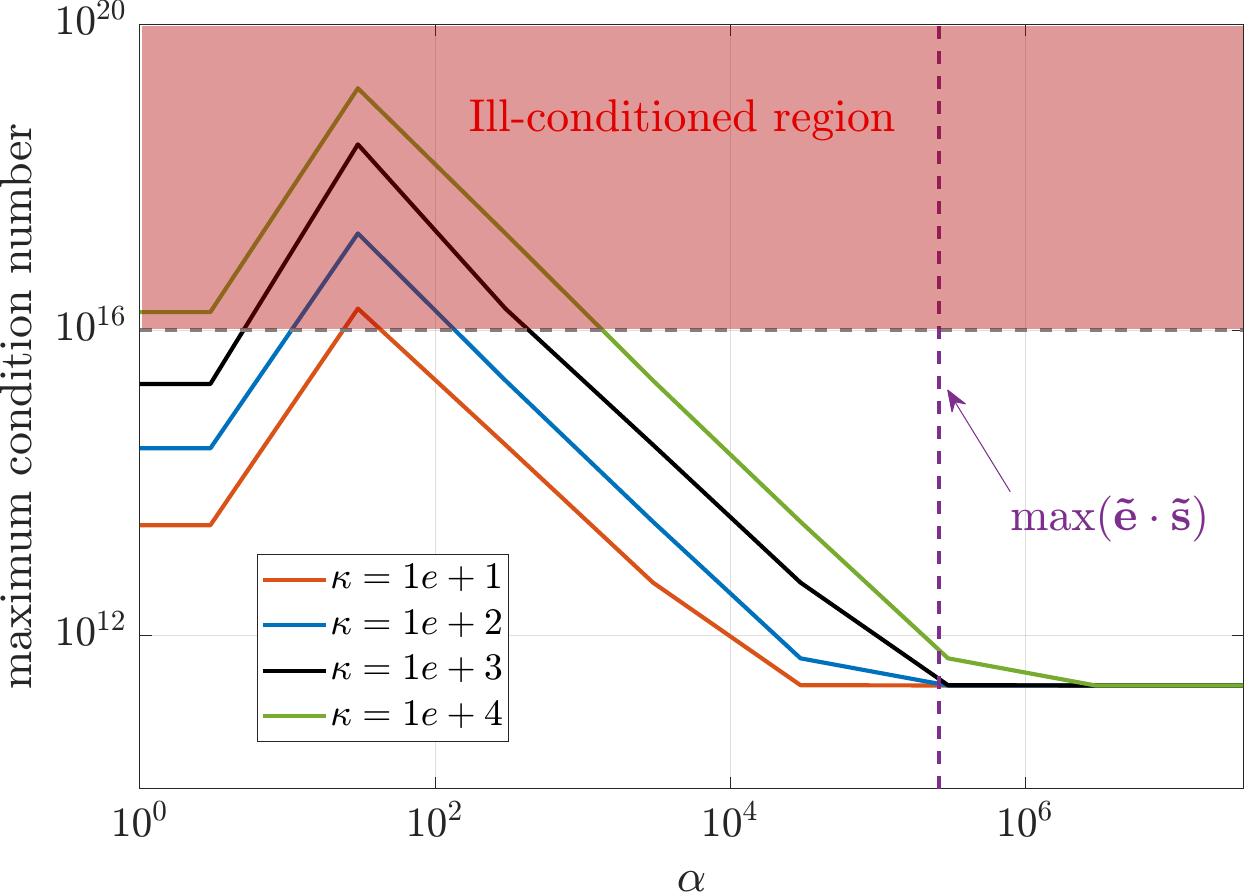}} \hspace{0.2cm}
    \subfloat[$\penEps=0.1$]{\includegraphics[width=0.49\textwidth]{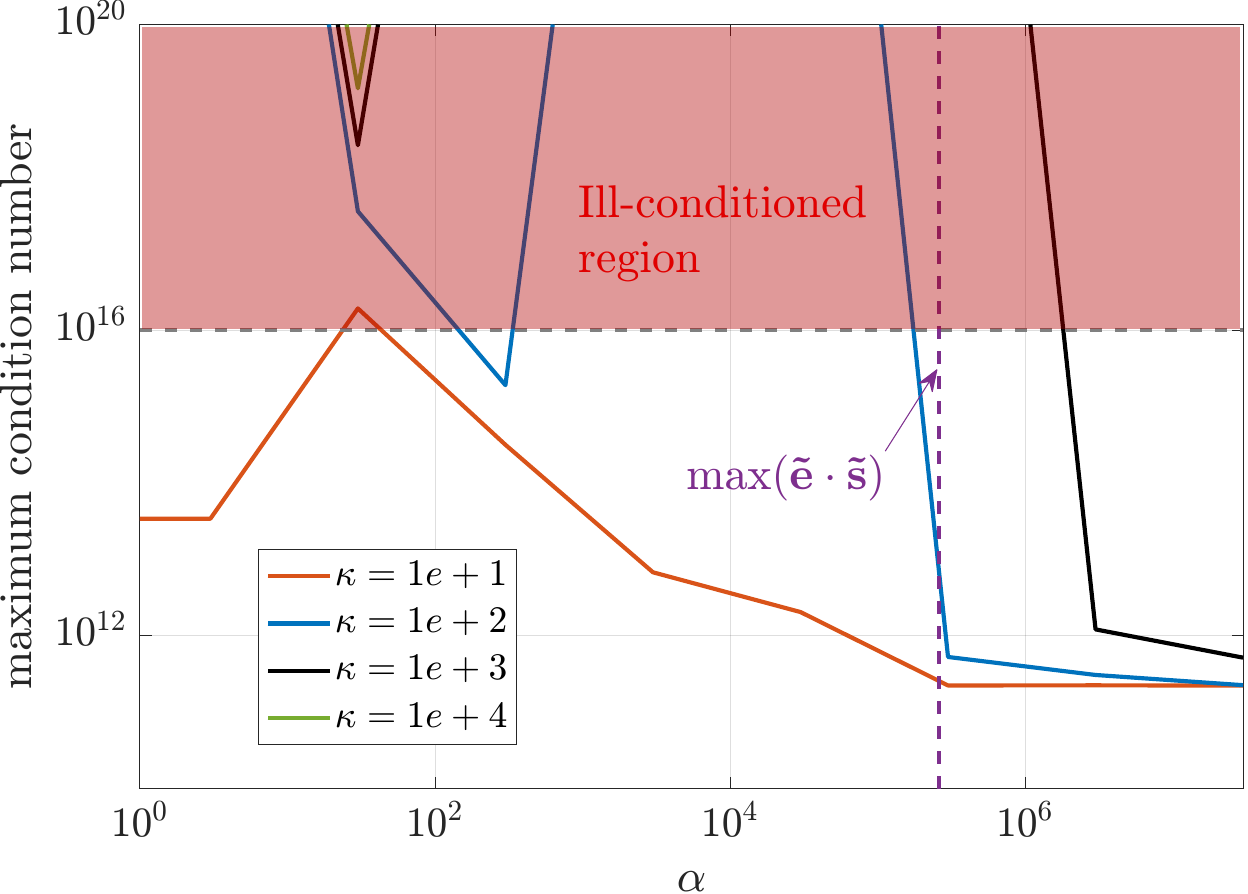}}

    \caption{Maximum condition number of the resulting system matrix over all iterations and load steps, computed with different penalty factors $\penFac$ and scaling factor $\penSca$ for the cantilever beam in Figure \ref{fig:simobeamOriMinE}a.}\label{fig:simobeamCondA}
\end{figure}

\begin{figure}[htb]
\centering
    \subfloat[$\penSca=\max(\eTilV \cdot \sTilV)$]{\includegraphics[width=0.49\textwidth]{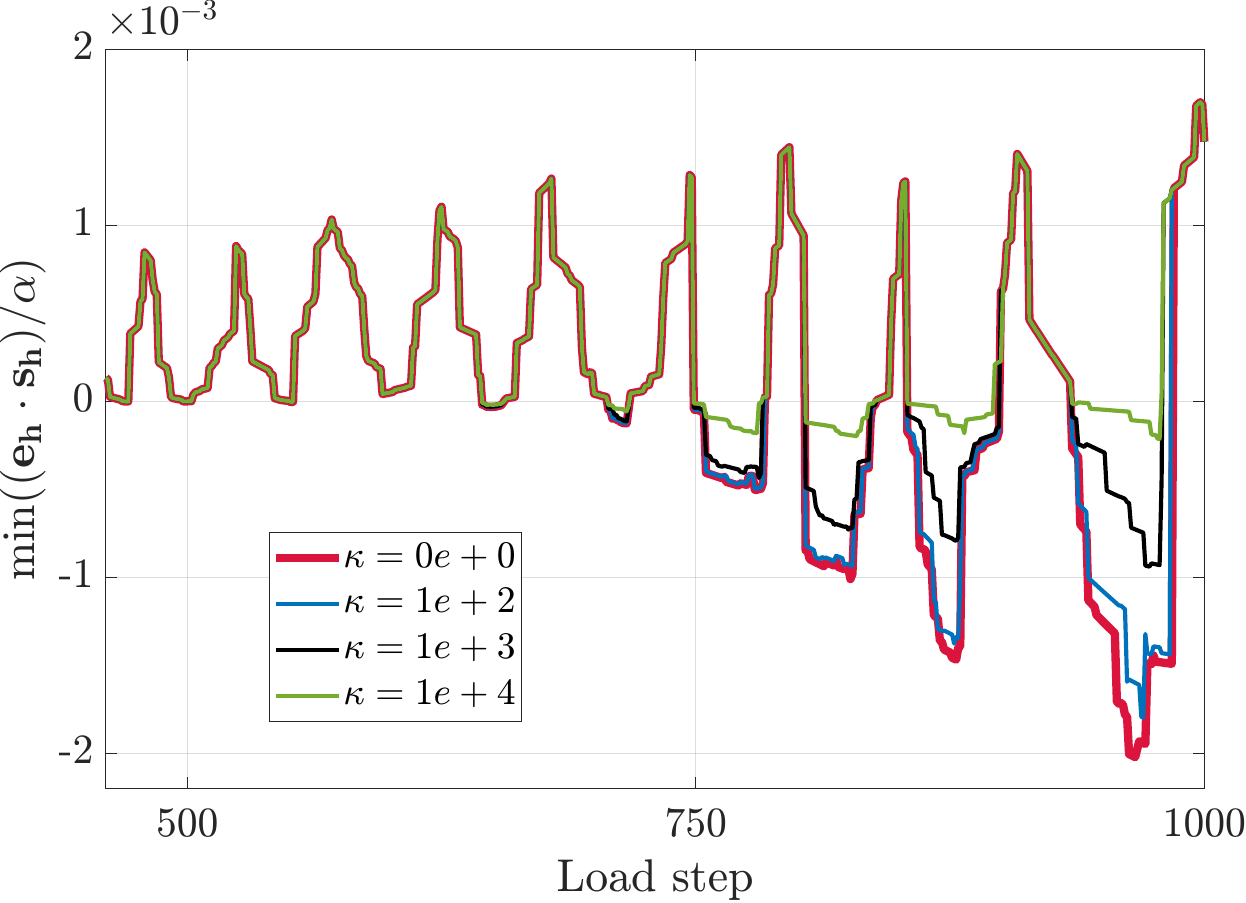}}\hspace{0.2cm}
    \subfloat[$\penSca=100\,\max(\eTilV \cdot \sTilV)$]{\includegraphics[width=0.49\textwidth]{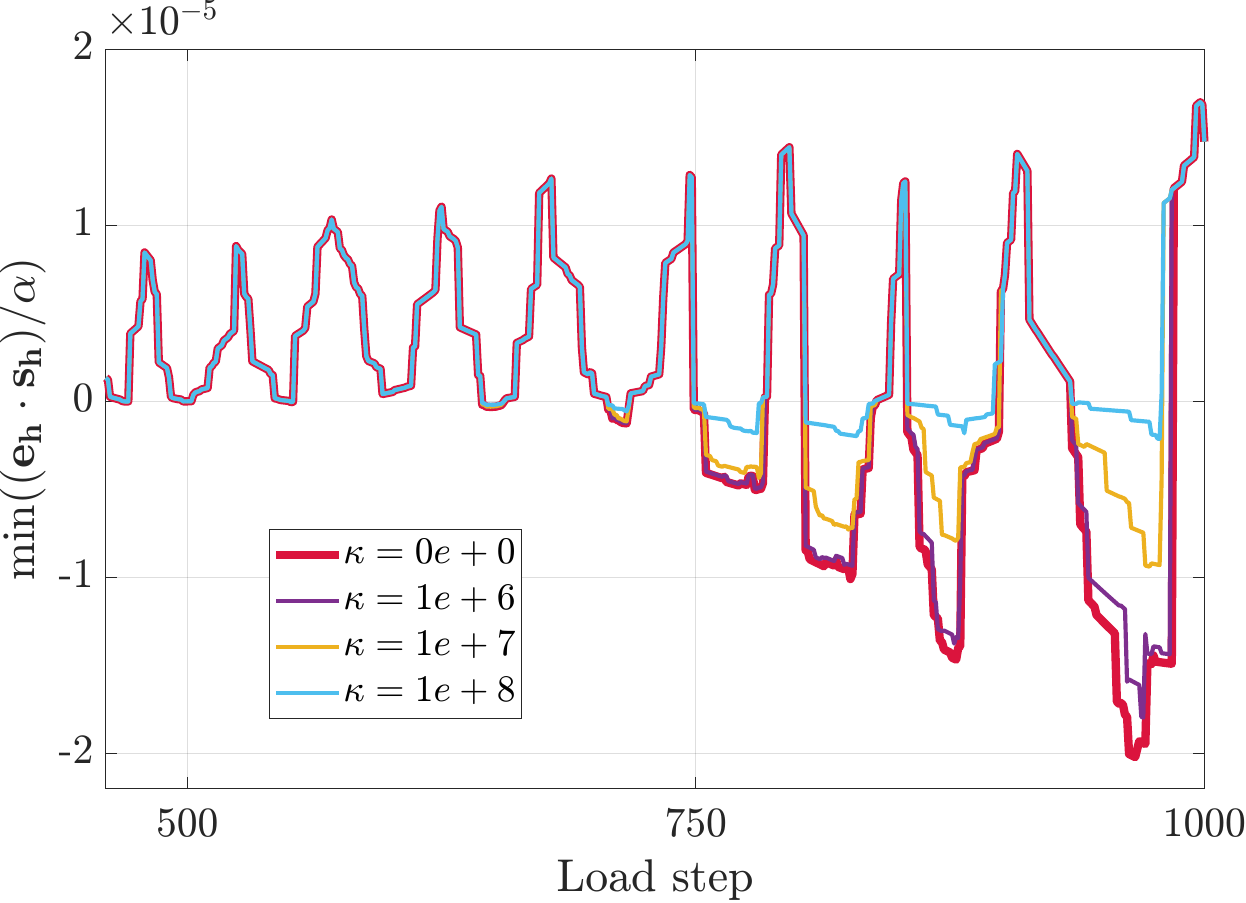}}

    \caption{Minimal values of the product $(\vect{e}_h \cdot \vect{s}_h)/\penSca$ over all elements of the cantilever beam in Figure \ref{fig:simobeamOriMinE}a, computed with $\penEps=0$ and different penalty factors $\penFac$.}\label{fig:simobeamMinEPara}
\end{figure}


We now numerically illustrate that the scaling factor $\penSca$ of the penalty function \eqref{eq:penFunc} plays a crucial role in ensuring a well-conditioned system matrix and, consequently, accuracy and the numerical stability. 
When $\penSca$ is chosen to be sufficiently large, the resulting system becomes well-conditioned, which allows larger penalty factors, $\penFac$, without compromising numerical stability. 
Given good conditioning, we show that selecting a sufficiently large value of $\penFac$ provides an effective enforcement of the thermomechanical consistency constraint, and that this effectiveness is generally independent of the chosen tolerance $\penEps$. 
To this end, we consider a cantilever beam subjected to a follower end force, illustrated in Figure \ref{fig:simobeamOriMinE}a, 
which is a well-established example of geometrically exact beams with linear constitutive relation, studied, for instance, in \cite{Simo1986}. 
We perform our study with a mesh of 32 elements and 1000 load steps, using 
our GO-ADM solver, introduced in Section \ref{sec:solver}, and a dataset of 5000 stress--strain states that is a subset of a synthetic dataset generated with virtual material tests (see also \ref{sec:datageneration}).  
In this section, we focus on the thermomechanical consistency of the discrete solution and the effect of different penalty parameters on the enforcement of this constraint.

\begin{remark}
    We note that the choice of using 1000 load steps is consistent with \cite{Simo1986} and ensures that the converged solution from the preceding step provides a sufficiently close initial guess for the next step and facilitates the local quadratic convergence of the Newton-Raphson scheme. 
    Fewer load steps generally require more Newton iterations per step and may lead to non-convergence, so they do not necessarily reduce the overall computational cost. Provided that the same equilibrium branch is reached and convergence is achieved at each step, the load-step size primarily affects convergence and robustness rather than spatial discretization accuracy.
\end{remark}

Figure \ref{fig:simobeamOriMinE}b illustrates the minimal values of the product $(\eV_h \cdot \sV_h)$, that is twice the discrete strain energy, across all elements at each load step, obtained with $\penFac=0$, i.e. without the enforcement of the thermomechanical consistency constraint. 
It shows the result obtained with a component- and vector-wise search in blue and red, respectively. 
As a reference solution, 
we include the result obtained from the standard finite element analysis (FEA) of the same beam structure, in grey. 
We see that using the component-wise search leads to a thermomechanically consistent solution (blue curve) at all load steps which is a good approximation of the reference values. 
Using the vector-wise search, however, leads to inconsistency at larger load steps (red curve), as well as a large difference when comparing to the reference values. 
This is because, within the employed dataset, certain individual stress or strain components may lie in a small neighborhood of their corresponding reference values, while the complete stress--strain vectors do not necessarily lie close to the full reference states. 
The component-wise searching strategy finds these components, which are not necessarily associated with any stress--strain vector included in the given dataset but with the best possible vector closest to the reference one. 
The vector-wise search only finds the closest vectors among those included in the dataset and will approach those found by the component-wise search when the dataset is enriched with more data, i.e. with increasing number of data points. 
We also recall that the component-wise search is only valid for cases of uncoupled stress--strain components, 
i.e. isotropic materials with assumed linear constitutive relation. 
Moreover, neither the component- nor vector-wise search strategy guarantees the convergence of the ADM solver to the same data. 
Our numerical experiments for the studied beam showed that the latter generally requires a higher number of iterations for this data search and is more likely to diverge than the former. 
We note that using synthetic dataset generated directly with linear constitutive relation leads to the same observation when using either one of these two searching strategies. 
In the following, we focus on the results obtained with the vector-wise search strategy when weakly enforcing the thermomechanical consistency constraint using the introduced penalty function.

\begin{remark}
    If the dataset includes the reference solution, i.e. the reference stress--strain states, both the component- and vector-wise searching strategies find these states. For the considered beam example and also those in our numerical studies in the next section, the resulting discrete solution in such cases is thermomechanically consistent.
\end{remark}

Figure \ref{fig:simobeamCondA} shows the maximum condition number of the resulting system matrix over all iterations and load steps as a function of the scaling factor $\penSca$, obtained with different values of the penalty factor $\penFac$. 
Focusing on the results for a tolerance $\penEps=0$ (see Figure \ref{fig:simobeamCondA}a), we observe that increasing $\penFac$ increases the condition number as expected. 
We see that increasing the scaling factor $\penSca$ to a sufficiently large value 
reduces the condition number, leading to a well-conditioned system. 
Moreover, choosing $\penSca$ that is larger than the maximal value of the product $(\eTilV \cdot \sTilV)$ (see the purple line in Figure \ref{fig:simobeamCondA}a), i.e. twice the strain energy obtained from the stress and strain data, 
reduces the effect of $\penFac$ on the condition number. 
This necessarily means that it allows larger values of $\penFac$ without compromising the conditioning of the resulting system. 
We note that our numerical experiments show the same results for small values of the tolerance $0 \leq \penEps \leq 10^{-3}$. 
For a larger tolerance, we illustrate its effect on the condition number for one exemplary value $\penEps = 0.1$ in Figure \ref{fig:simobeamCondA}b. 
We observe that when choosing such a significant tolerance, a large penalty factor $\penFac$ requires a significantly larger scaling factor $\penSca$ to ensure a well-conditioned system. 
Based on these observations, we focus on the results obtained with 
the tolerance $\penEps=0$ and scaling factors $\penSca \geq \max(\eTilV \cdot \sTilV)$ in the following.

Figure \ref{fig:simobeamMinEPara} illustrates the minimal values of the product $(\eV_h \cdot \sV_h)$ across all elements at each load step for the studied beam structure, 
obtained with different penalty factors $\penFac$ and a tolerance $\penEps=0$. 
We focus on later load steps where the discrete solution is thermomechanically inconsistent. 
For comparison purposes, 
we include here again the result from Figure \ref{fig:simobeamOriMinE}b that is obtained with $\penFac=0$ (pink curve), i.e. without the penalty term. 
We observe that increasing $\penFac$ enforces the thermomechanical consistency constraint more efficiently, as expected, in all cases. 
We see in Figure \ref{fig:simobeamMinEPara}b that 
using a larger scaling factor $\penSca$ requires a larger penalty factor $\penFac$ to sufficiently enforce this constraint since it leads to smaller residual $r$ (see also Equation \eqref{eq:penFunc}). 
For this case, our numerical experiments show that smaller values of $\penFac \leq 10^5$ do not enforce this constraint at any load step and therefore are not included here. 
Based on this parameter study, in this work, 
we choose a zero tolerance $\penEps=0$, scaling factor $\penSca = \max(\eTilV \cdot \sTilV)$, and penalty factor $\penFac$ such that $\hat{\penFac}$ is at least three or four orders of magnitude larger than that of $\penSca$ for our numerical studies in the next section.

\section{Numerical studies}\label{sec:results}

In this section, we numerically demonstrate the favorable numerical behavior of our solving strategy GO-ADM and the proposed penalty function discussed in Section \ref{sec:penalty}. 
To this end, we consider two- and three-dimensional single- and multi-beam structures, which are well studied in the literature. 
We show that for the studied examples, the GO-ADM solver remains robust, generally reduces the global objective function, leading to a better approximation of the global optima, and using the introduced penalty function weakly enforces the thermomechanical consistency constraint. 
For our computations, 
we employ synthetic (sub-)datasets generated with virtual material tests as described in \ref{sec:datageneration}. 
We also numerically illustrate via one example of a three-dimensional frame that 
our solving strategy GO-ADM is also robust in the case of noisy data. 
All computations reported in this work were performed in MATLAB R2024a on a Dell Latitude 7440 laptop running Ubuntu 22.04.5 LTS, equipped with a 13th Gen Intel Core i5-1345U processor (10 cores, 12 threads) and 14 GiB of usable system memory.

\subsection{Single-beam structures}


\begin{figure}[htb]
    \centering
    \begin{tikzpicture}

    \node[anchor=south west, inner sep=0] (img1)
    at (0,0)
    {\subfloat[Snapshots with GO-ADM, $F$ in kN]{
        \def\svgwidth{0.33\textwidth}
        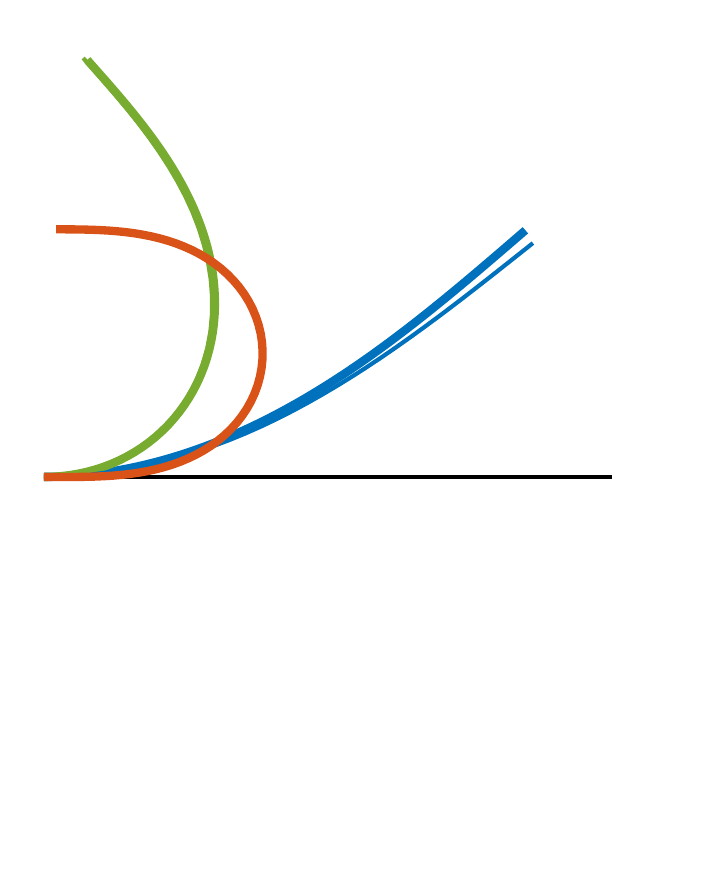}};
    
    \begin{scope}[x={(img1.south east)},y={(img1.north west)}]


    \node[
        draw,
        rounded corners=2pt,
        fill=none,
        text opacity=1,
        inner sep=2pt,
        anchor=north west
    ]
    at (0.56,0.99)
    {
        \begin{tikzpicture}

            \draw[line width=2.5pt]
                (0,0) -- (0.1,0);

            \node[anchor=west]
                at (0.09,0)
                {\footnotesize : with $\dataset_1$};

            \draw[line width=0.8pt]
                (0,-0.05) -- (0.1,-0.05);

            \node[anchor=west]
                at (0.09,-0.05)
                {\footnotesize : with $\dataset_2$};

        \end{tikzpicture}
    };

    \end{scope}


    \node[anchor=south west, inner sep=0] (img2)
    at (0.37\textwidth,0)
    {\subfloat[$\globObjFuncP$ with $\dataset_2$]{\includegraphics[width=0.6\textwidth]{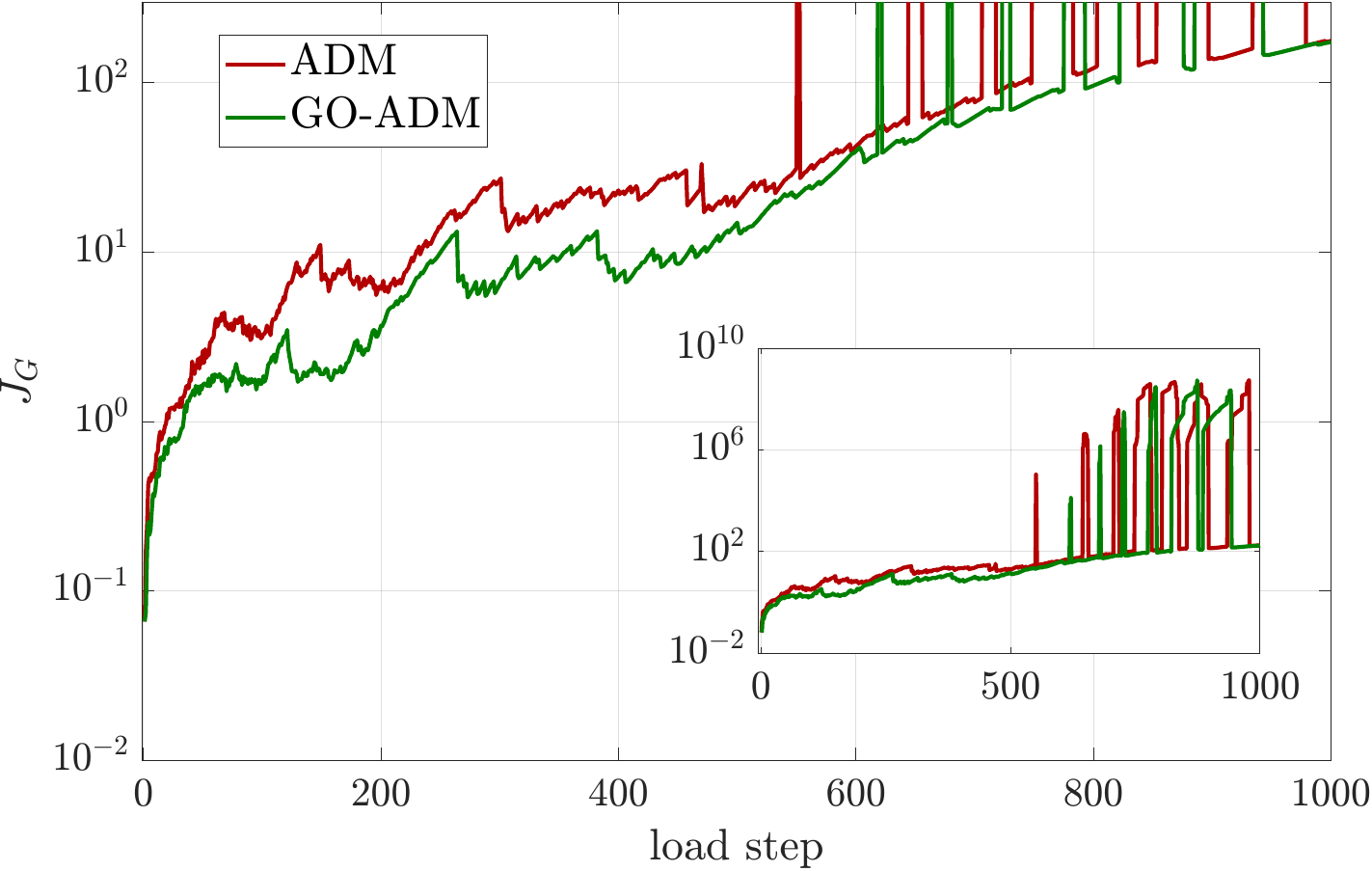}}};
    \end{tikzpicture}

    \caption{Snapshots and the value of the global objective function, $\globObjFuncP$, of the cantilever beam in Figure \ref{fig:simobeamOriMinE}a, computed with \textbf{the ADM and our GO-ADM solving strategies}, using synthetic datasets with ($\dataset_1$) and without ($\dataset_2$) the solution from a standard FEA.}\label{fig:simobeamSnapNcost}
\end{figure}

\begin{figure}[htb]
    \centering
    \begin{tikzpicture}
        \node[anchor=south west, inner sep=0] (img) at (0,0)
        {\includegraphics[width=0.9\textwidth]
        {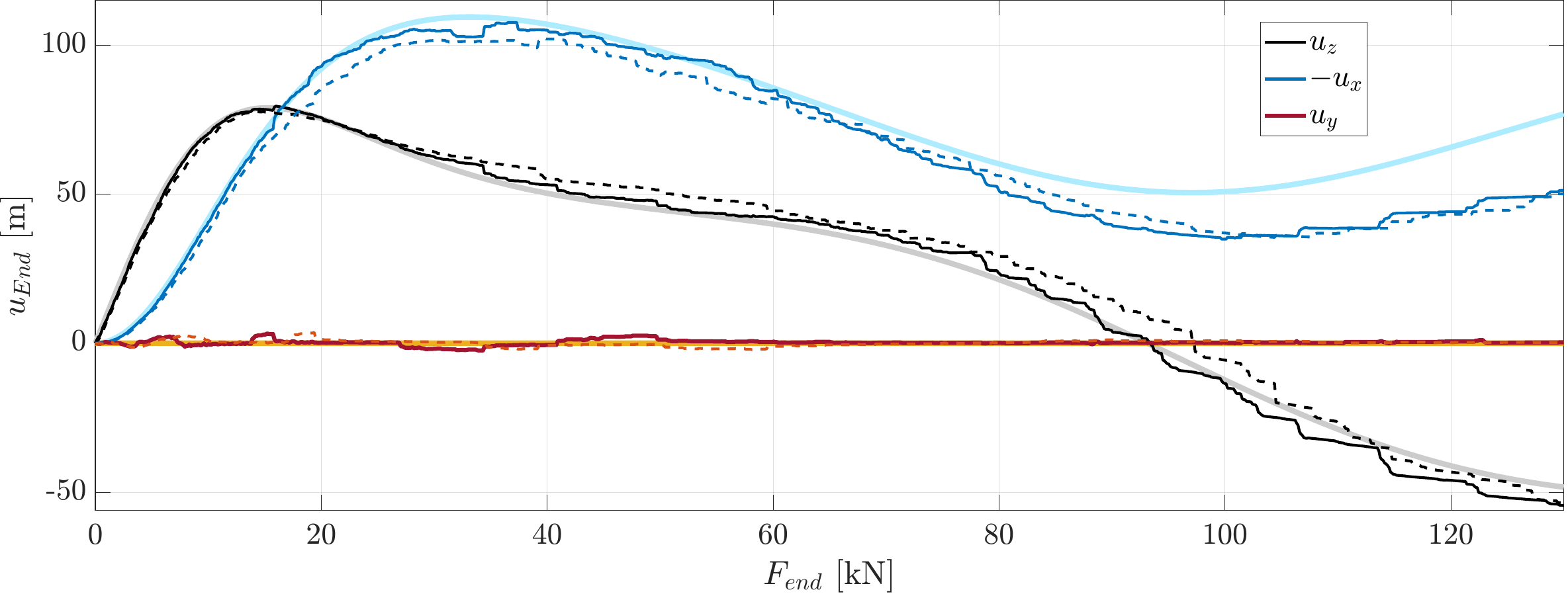}};


        \draw[line width=2.5pt]
            (1.5,1.5) -- (2,1.5);

        \node[anchor=west]
            at (1.95,1.5)
            {\footnotesize{: GO-ADM, $\dataset_1$}};

        \node[anchor=west]
            at (1.15,1.1)
            {\footnotesize{\textit{(thick solid curves)}}};

        \draw[line width=0.8pt]
            (5,1.5) -- (5.5,1.5);

        \node[anchor=west]
            at (5.45,1.5)
            {\footnotesize{: GO-ADM, $\dataset_2$}};

        \node[anchor=west]
            at (4.65,1.1)
            {\footnotesize{\textit{(thin solid curves)}}};

        \draw[line width=0.8pt,dashed]
            (8.5,1.5) -- (9,1.5);

        \node[anchor=west]
            at (8.95,1.5)
            {\footnotesize{: ADM, $\dataset_2$}};

        \node[anchor=west]
            at (8.15,1.1)
            {\footnotesize{\textit{(thin dashed curves)}}};
    \end{tikzpicture}

    \caption{Load-deflection curve of the tip displacement of the cantilever beam in Figure \ref{fig:simobeamOriMinE}a, computed with \textbf{the ADM and our GO-ADM solving strategies}, using synthetic datasets with ($\dataset_1$) and without ($\dataset_2$) the solution from a standard FEA.}\label{fig:simobeamUFcurve}
\end{figure}

We first consider a single beam structure that is a cantilever beam subjected to a follower end force, studied in \cite{Simo1986}, for which we performed the numerical parameter study in the previous section. 
We now focus on the performance of our GO-ADM solver in terms of accuracy and reducing the global objective function, $\globObjFuncP$. 
We perform our analysis on the same mesh of 32 elements with 1000 load steps, using the vector-wise searching strategy and the same synthetic dataset of 5000 stress--strain states as in the previous section. 
Based on the results of this parameter study, we choose a penalty factor $\penFac=10^4$, a scaling factor $\penSca = \max(\eTilV \cdot \sTilV)$, and a tolerance $\penEps=0$. 
For comparison purposes, we consider a second dataset that is the original synthetic one enriched with the discrete solution from the standard finite element analysis (FEA) based on a linear constitutive relation. 
We refer to the enriched dataset as $\dataset_1$ and the original synthetic one as $\dataset_2$. 
When using $\dataset_1$, we obtain the discrete solution from the standard FEA, irrespective of the employed data-driven solver or the searching strategy, as discussed in the previous section, which can be considered as a reference solution for this example. 
We emphasize that we employ $\dataset_1$ primarily as a verification dataset to assess whether the data-driven solver can recover a known discrete solution. It is not intended to represent the practical data-driven setting. 
In contrast, $\dataset_2$ provides the more relevant assessment of the solver when the target solution is not contained in the material dataset since it 
does not contain the FEA solution of the considered loading problem and is generated from virtual material tests under different loading scenarios.

Figure \ref{fig:simobeamSnapNcost}a illustrates the snapshots of the considered beam obtained with our GO-ADM solver using the datasets $\dataset_1$ (thick curves) and $\dataset_2$ (thin curves). 
We observe a very good agreement between the former and those in \cite[Figure 4a]{Simo1986}, while the latter differs more significantly with increasing load.  
Figure \ref{fig:simobeamSnapNcost}b shows the value of the modified global objective function, $\globObjFuncP$, at each load step, obtained with the ADM (red curve) and GO-ADM (green curve) solver, using the dataset $\dataset_2$. For visual purposes, we focus on values up to $10^2$ and show the complete value range in the inset figure. 
We see that using our GO-ADM solver generally reduces the values of $\globObjFuncP$, as expected. 
We also see that at some load steps, approximately after the 500$^\text{th}$ step, 
we obtain jumps in $\globObjFuncP$. The reason behind this observation is that the employed dataset $\dataset_2$ does not include the reference stress--strain state, i.e. that from the standard FEA, or any state in its neighbourhood. 
Therefore, the underlying linear search could not converge to a state that is closest to the discrete solution, leading to significantly large values of $\globObjFuncP$. 
This is also consistent with different snapshots at larger loads observed in Figure \ref{fig:simobeamSnapNcost}a. 
We note that for this example, the contribution of the penalty term to $\globObjFuncP$ is insignificant, owing to our choice of the penalty parameters.

In Figure \ref{fig:simobeamUFcurve}, we plot the load-deflection curve of three components of the tip displacement, $-u_x$ (blue curves), $u_y$ (orange curves), and $u_z$ (black curves), using the dataset $\dataset_2$ with either the ADM (dashed curves) or the GO-ADM solver (thin solid curves). 
We also include the results obtained with the GO-ADM solver using the dataset $\dataset_1$ (thick solid curves) as the reference solution which is the same as that from a standard FEA and shows a very good agreement with \cite[Figure 4b]{Simo1986}. 
We observe that for end forces approximately $F_{\text{end}} \lesssim 65$ kN, that is equivalent to the 500$^\text{th}$ load step, using our GO-ADM solver leads to load-deflection curves (thin solid curves) that are closer to the reference solution (thick solid curves), i.e. a better approximation of the latter, compared to those obtained with the ADM solver (thin dashed curves). 
For larger forces, using either one of these two solvers leads to approximately the same results which differ more significantly from the reference solution. 
These observations are consistent with those from the global objective function above. 
We note that our numerical experiments show that increasing the number of stress--strain states in $\dataset_2$ improves the accuracy when using either the ADM or the GO-ADM solver. 
We conclude that for this example, the accuracy significantly depends on the employed dataset. 
Using our GO-ADM solver generally leads to a better approximation of the reference solution, i.e. the global optima, and reduces the global objective function.


\begin{figure}[htb]
    \centering
    \begin{tikzpicture}
        \node[anchor=south west, inner sep=0] (img) at (0,0)
        {\includegraphics[width=0.6\textwidth]{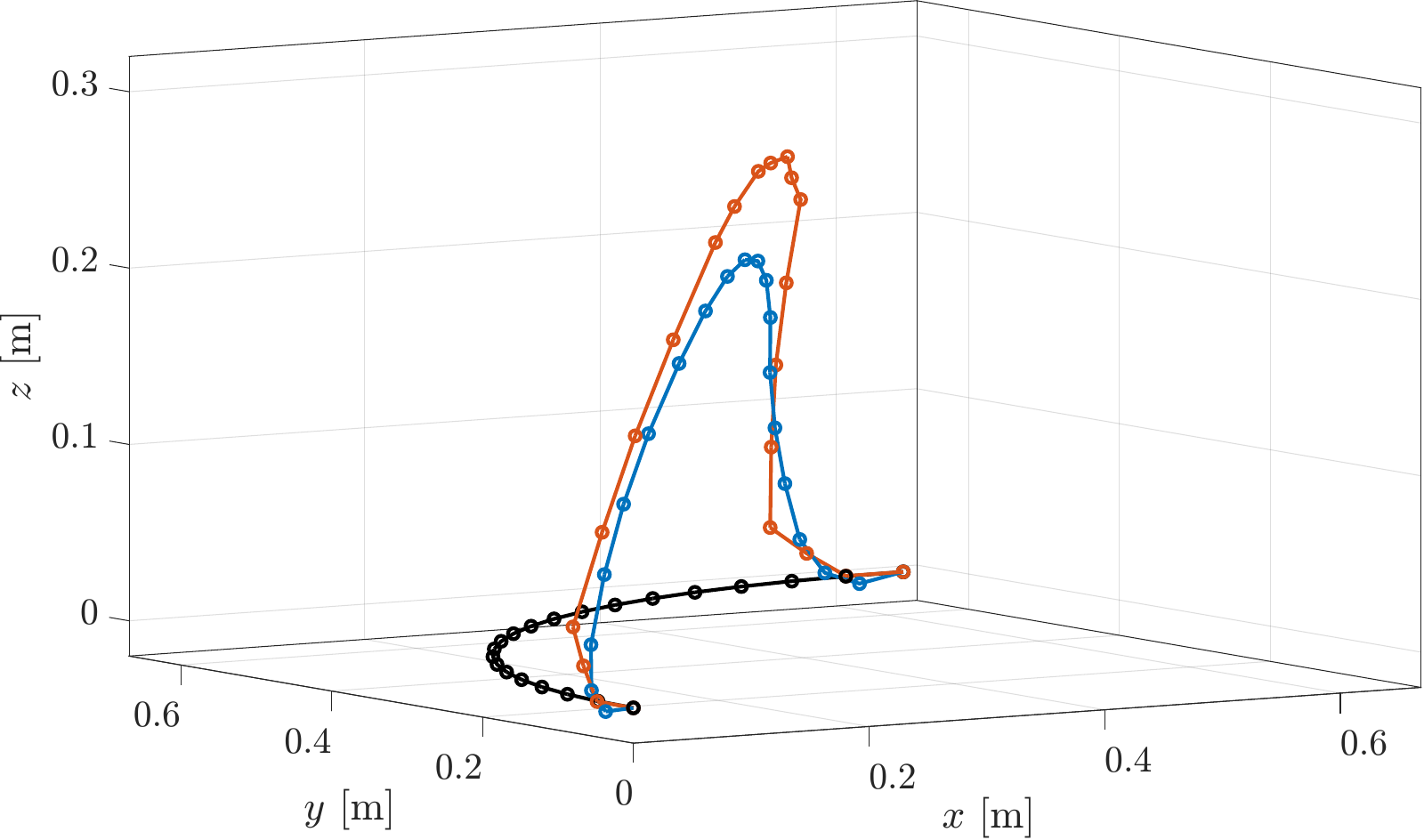}};


        \draw[line width=1pt]
            (1.3,5) -- (1.8,5);

        \node[anchor=west]
            at (1.75,5)
            {\footnotesize{: ref. configuration}};

        \draw[color=blue2, line width=1pt]
            (1.3,4.5) -- (1.8,4.5);

        \node[anchor=west]
            at (1.75,4.5)
            {\footnotesize{: \textcolor{blue2}{with $\dataset_1$}}};

        \draw[color=burntorange, line width=1pt]
            (1.3,4) -- (1.8,4);

        \node[anchor=west]
            at (1.75,4)
            {\footnotesize{: \textcolor{burntorange}{with $\dataset_2$}}};
    \end{tikzpicture}

    \caption{The reference (black) and deformed configuration of the clamped-clamped curved beam studied in \cite[Fig.~6]{Gebhardtddcmstatic2020}, computed with \textbf{our GO-ADM solving strategy} using synthetic dataset with (\textcolor{blue2}{$\dataset_1$, blue}) and without (\textcolor{burntorange}{$\dataset_2$, orange}) the solution from a standard FEA.}\label{fig:curvedbeam3Ddeformed}
\end{figure}
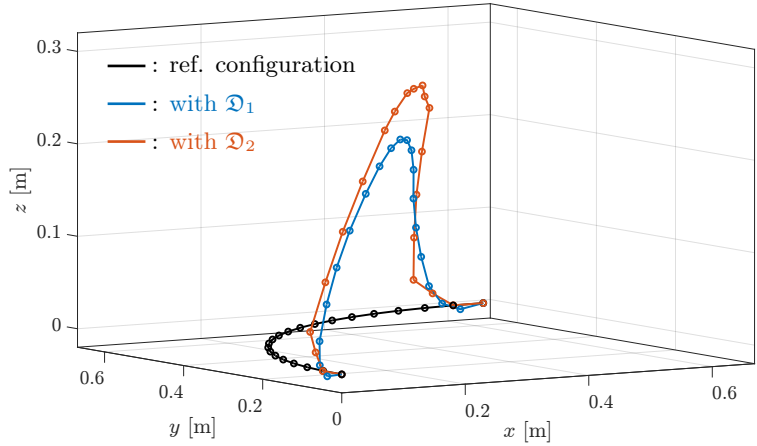

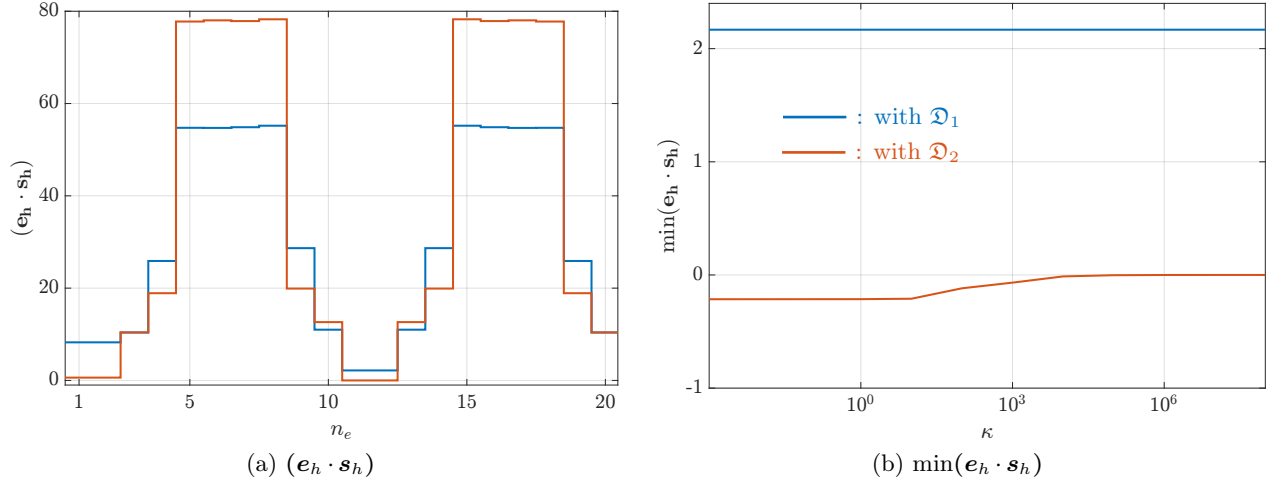
\begin{figure}[htb]
    \centering
    \begin{tikzpicture}

    \node[anchor=south west, inner sep=0] (img1)
    at (0,0)
    {\subfloat[$(\eV_h \cdot \sV_h)$]{
            \includegraphics[width=0.49\textwidth]{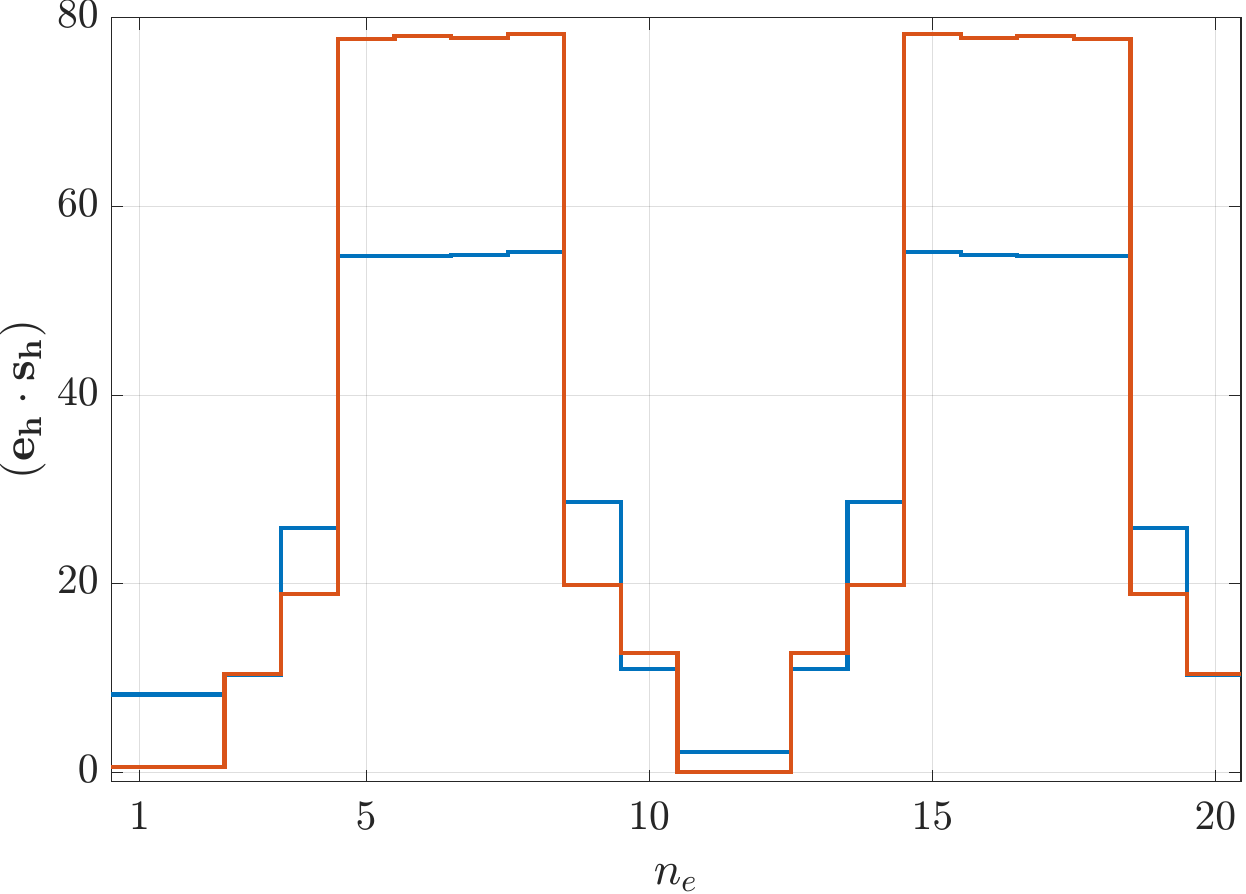}}};


    \node[anchor=south west, inner sep=0] (img2)
    at (0.52\textwidth,0)
    {\subfloat[$\min(\eV_h \cdot \sV_h)$]{
        \includegraphics[width=0.49\textwidth]{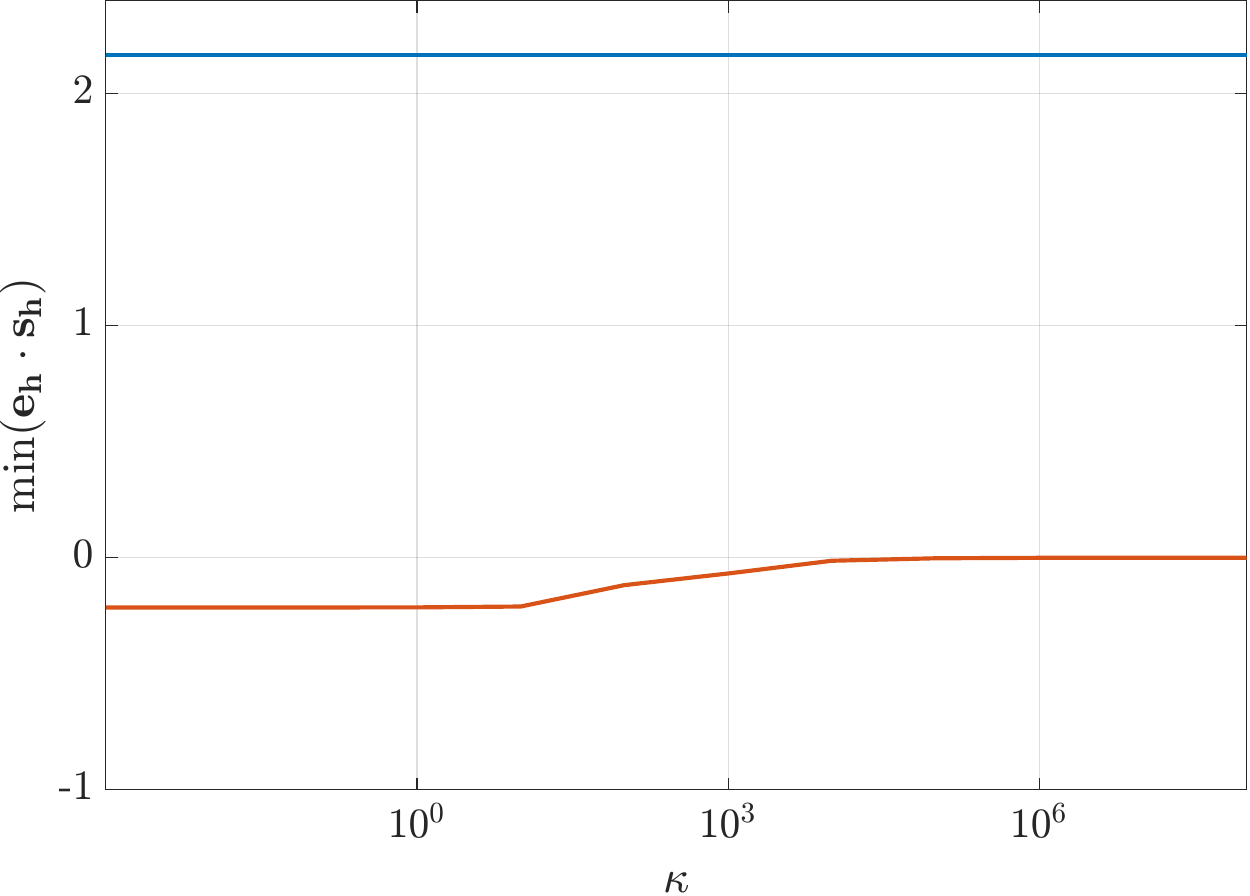}}};

    \begin{scope}[x={(img2.south east)},y={(img2.north west)}]

        \draw[blue2, line width=1.0pt]
            (0.25,0.72) -- (0.3,0.72);

        \node[anchor=west, text=blue2]
            at (0.3,0.72)
            {\footnotesize{: \textcolor{blue2}{with $\dataset_1$}}};

        \draw[burntorange, line width=1.0pt]
            (0.285,0.65) -- (0.335,0.65);

        \node[anchor=west, text=burntorange]
            at (0.335,0.65)
            {\footnotesize{: \textcolor{burntorange}{with $\dataset_2$}}};

    \end{scope}
    \end{tikzpicture}

    \caption{The product $(\eV_h \cdot \sV_h)$ and its minimal value over all elements of the beam illustrated in Figure \ref{fig:curvedbeam3Ddeformed}, computed with \textbf{our GO-ADM solving strategy} using synthetic dataset with (\textcolor{blue2}{$\dataset_1$, blue}) and without (\textcolor{burntorange}{$\dataset_2$, orange}) the solution from a standard FEA.}\label{fig:curvedbeam3Denergy}
\end{figure}

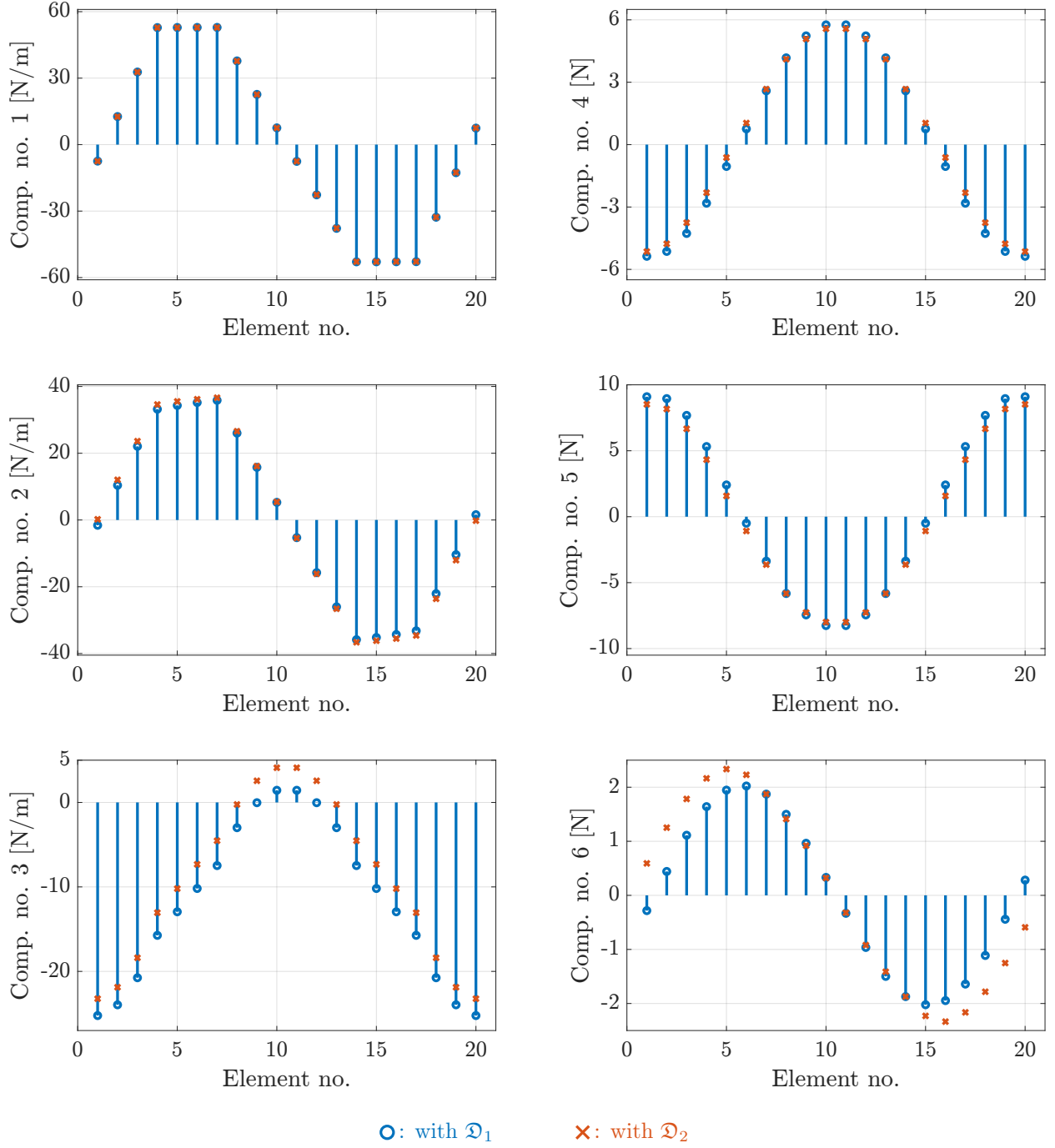
\begin{figure}[!htbp]
    \centering
    \begin{tikzpicture}
        \node[anchor=south west,inner sep=0] (img)
        at (0,0){\includegraphics[width=1\textwidth]{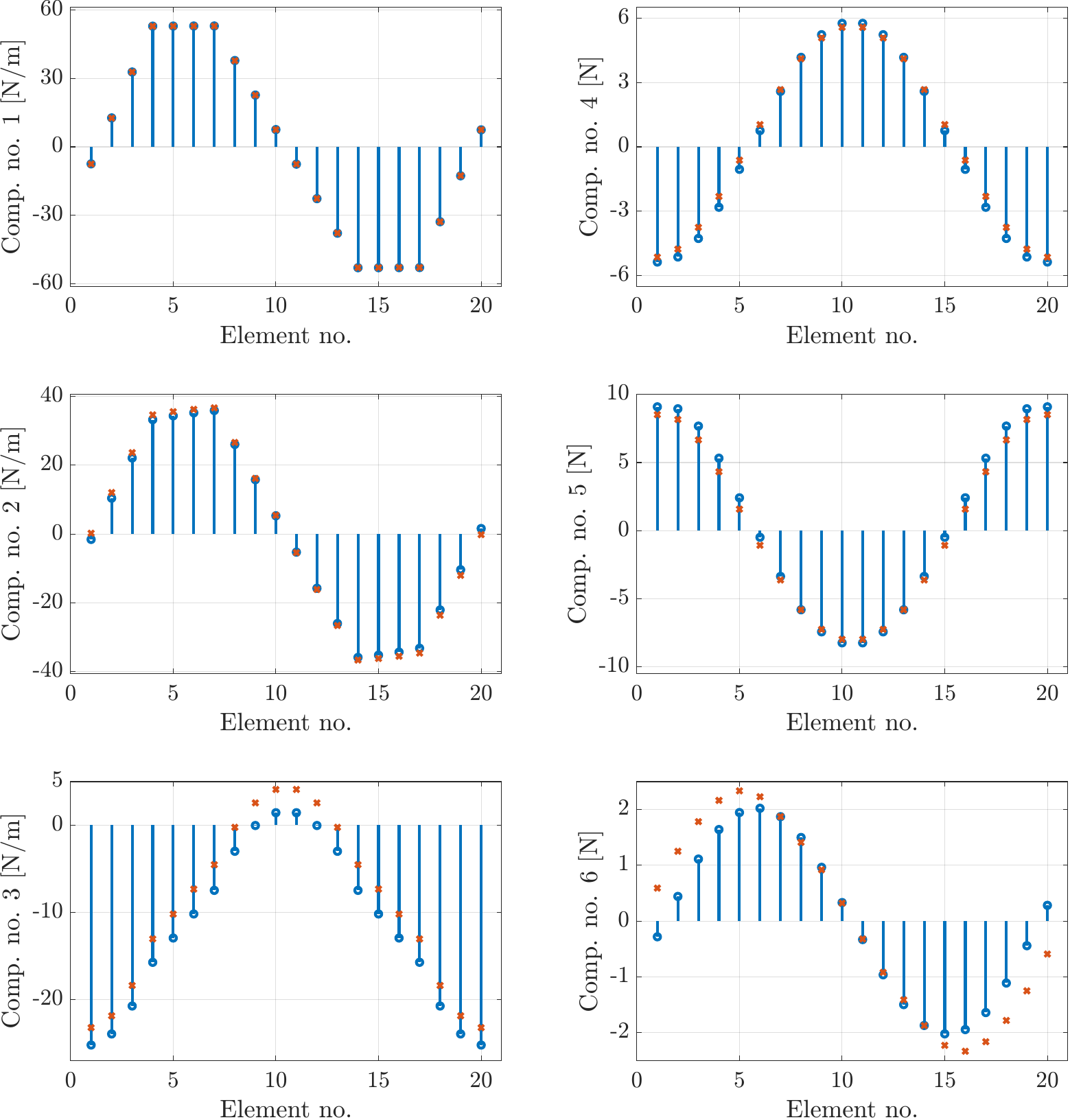}};

        \draw[blue2,line width=1.2pt]
        ($(img.south west)+(6cm,-0.7cm)$)
        circle (0.10cm);

        \node[anchor=west,text=blue2]
        at ($(img.south west)+(6.05cm,-0.7cm)$)
        {\small{: with $\dataset_1$}};

        \draw[burntorange,line width=1.2pt]
        ($(img.south west)+(9.0cm,-0.8cm)$)
        --
        ($(img.south west)+(9.2cm,-0.6cm)$);

        \draw[burntorange,line width=1.2pt]
        ($(img.south west)+(9.0cm,-0.6cm)$)
        --
        ($(img.south west)+(9.2cm,-0.8cm)$);

        \node[anchor=west,text=burntorange]
        at ($(img.south west)+(9.15cm,-0.7cm)$)
        {\small{: with $\dataset_2$}};
    \end{tikzpicture}

    \caption{Distribution of the stress resultant components of the beam illustrated in Figure \ref{fig:curvedbeam3Ddeformed}, computed with \textbf{our GO-ADM solving strategy} using synthetic dataset with (\textcolor{blue2}{$\dataset_1$, blue}) and without (\textcolor{burntorange}{$\dataset_2$, orange}) the solution from a standard FEA. See also \cite[Fig.~8]{Gebhardtddcmstatic2020}.}\label{fig:curvedbeam3DstressComp}
\end{figure}

\begin{table}[htb]
    \centering
    \renewcommand{\arraystretch}{1.2} 
    \setlength{\tabcolsep}{4pt} 
{\footnotesize
    \begin{tabularx}{0.8\textwidth}{c | c | Y| Y| Y| Y}
        & comp. & $\curconfig_0$  & $\vect{d}_1$  & $\vect{d}_2$  & $\vect{d}_3$ \\
        \hline
        \multirow[c]{3}{*}{\cite[Tab.~1]{Gebhardtddcmstatic2020}} 
        & $x$ & 0.30620367 &  0.00761487 &  0.70706900 & 0.70711000 \\
        & $y$ & 0.33041647 & -0.00761485 & -0.70706899 & 0.70711000 \\
        & $z$ & 0.21515428 &  0.99994201 & -0.01076909 & 0.00000000 \\
        \hline
        \multirow[c]{3}{*}{$\dataset_1$} 
        & $x$ & 0.30623724 &  0.00761681 &  0.70706576 & 0.70710678 \\
        & $y$ & 0.33038254 & -0.00761681 & -0.70706576 & 0.70710678 \\
        & $z$ & 0.21521159 &  0.99994198 & -0.01077180 & 0.00000000 \\
        \hline
        \multirow[c]{3}{*}{$\dataset_2$} 
        & $x$ & 0.31588638 &  0.00617972 &  0.70707978 & 0.70710678 \\
        & $y$ & 0.32073339 & -0.00617972 & -0.70707978 & 0.70710678 \\
        & $z$ & 0.27381084 &  0.99996181 & -0.00873944 & 0.00000000 \\
    \end{tabularx}}

    \caption{Nodal variables of the 11$^\text{th}$ node at the deformed configuration of the beam illustrated in Figure \ref{fig:curvedbeam3Ddeformed}, computed with our GO-ADM solving strategy using synthetic dataset with ($\dataset_1$) and without ($\dataset_2$) the solution from a standard FEA.}
    \label{tab:curvedbeam3Dnode11}
\end{table}

\begin{table}[htb]
    \centering
    \renewcommand{\arraystretch}{1.2} 
    \setlength{\tabcolsep}{4pt} 
    
{\footnotesize
    \begin{tabularx}{0.8\textwidth}{c | Y | Y | Y || Y | Y | Y}
        \multirow[c]{2}{*}{comp.} & \multicolumn{3}{c||}{$\eV_h$} & \multicolumn{3}{c}{$\sV_h$} \\
        \cline{2-7}
         & \cite[Tab.~2]{Gebhardtddcmstatic2020} & $\dataset_1$ & $\dataset_2$ & \cite[Tab.~2]{Gebhardtddcmstatic2020} & $\dataset_1$ & $\dataset_2$ \\
        \hline
        1 &  0.50365604 &  0.50365795 &  0.34872924 & 37.77420295 & 37.77434648 & 37.73081550 \\
        2 &  0.34693365 &  0.34693384 &  0.20442204 & 26.02002347 & 26.02003836 & 26.44688047 \\
        3 & -0.02990630 & -0.02990580 & -0.07866797 & -2.99062982 & -2.99057979 & -0.82040143 \\
        4 &  0.04172209 &  0.04172298 &  0.10333812 &  4.17220925 &  4.17229835 &  3.62596082 \\
        5 & -0.05823387 & -0.05823538 & -0.16115903 & -5.82338675 & -5.82353799 & -5.61910141 \\
        6 &  0.00749371 &  0.00749367 & -0.00344413 &  1.49874166 &  1.49873402 &  1.51055449 \\
    \end{tabularx}}

    \caption{Stress--strain state of the 8$^\text{th}$ element at the deformed configuration of the beam illustrated in Figure \ref{fig:curvedbeam3Ddeformed}, computed with \textbf{our GO-ADM solving strategy} using synthetic dataset with ($\dataset_1$) and without ($\dataset_2$) the solution from a standard FEA.}
    \label{tab:curvedbeam3Dele8}
\end{table}

\begin{table}[htbp]
  \centering  
  {\footnotesize
  \begin{tabular}{l|l|c|c}
     & & ADM & GO-ADM \\
    \hline
    \multirow[c]{2}{*}{$\globObjFuncP$} & $\dataset_1$ & 8.4619987e-33 & 8.4619987e-33 \\
    & $\dataset_2$ & 3.6158636e-02 & 3.4966070e-02 \\
  \end{tabular}}
  \caption{Value of the modified global objective function obtained with \textbf{the ADM and GO-ADM solving strategies} for the beam illustrated in Figure \ref{fig:curvedbeam3Ddeformed}, using synthetic dataset with ($\dataset_1$) and without ($\dataset_2$) the solution from a standard FEA.}
  \label{tab:distGcurvedbeam3D}
\end{table}


The second example studied in this section is a clamped-clamped curved beam that is a quarter of a circular arc, studied in \cite{Gebhardtddcmstatic2020}, subjected to symmetric three-dimensional forces (see also \cite[Figure 6]{Gebhardtddcmstatic2020}). 
We consider the same beam geometry, forces, number of load steps, and discretization of 20 finite elements. 
We adapt the same material properties for generating synthetic datasets, as described in \ref{sec:datageneration}, as well as for the standard FEA of the ANLP initialization approach. 
We again consider two datasets of 4000 stress--strain states each: $\dataset_1$ including the FEA solution that is considered as a reference solution, and $\dataset_2$ without this solution that results directly from virtual material tests (see also \ref{sec:datageneration}). 
In Figure \ref{fig:curvedbeam3Ddeformed}, we plot the final deformed configuration obtained with these two datasets, using our GO-ADM solver. 
We observe that using $\dataset_2$ leads to larger deformations than $\dataset_1$, which is due to those associated with the closest stress--strain states included in $\dataset_2$. 
For comparison purposes, in Table \ref{tab:curvedbeam3Dnode11}, we also compare the nodal variables of the 11$^\text{th}$ node obtained with these two datasets with those given in \cite[Table 1]{Gebhardtddcmstatic2020}. 
We see a very good agreement between the results obtained with $\dataset_1$ and the given reference, explaining again our consideration of the results obtained with this dataset as the reference solution. 
Using the dataset $\dataset_2$ leads to different axial position vector, $\curconfig_0$, and director $\vect{d}_1$, while the obtained directors $\vect{d}_2$ and $\vect{d}_3$ show a good agreement with the reference results. 
These observations are consistent with a higher maximal value of the product $(\eV_h \cdot \sV_h)$ 
obtained with $\dataset_2$, illustrated in Figure \ref{fig:curvedbeam3Denergy}a, and with differing stress--strain states at the 8$^\text{th}$ element, given in Table \ref{tab:curvedbeam3Dele8}. 
We see this difference again in each stress component in each element in Figure \ref{fig:curvedbeam3DstressComp}, particularly in the third and sixth components, which correspond to axial and torsional deformations, respectively. 
Nevertheless, 
we observe a good agreement when using either one of these two datasets in stress components 1, 2, 4, and 5, which are related to shear and bending deformations. 
The observed differences are due to stress--strain states included in $\dataset_1$ and $\dataset_2$ which correspond to different loading settings and forces. 
For this example, 
one can improve the accuracy by increasing the number of stress--strain states in $\dataset_2$, i.e. enriching $\dataset_2$, with material tests of dominant axial and torsional deformations, and/or with more rotations (see also \ref{sec:datageneration}).

We note that for the sake of clarity in the figures regarding the clamped-clamped curved beam, we discard the results obtained with the ADM solver using the dataset $\dataset_2$ since they are approximately the same as those obtained with our GO-ADM solver. 
This is reflected in the obtained value of the modified global objective function, $\globObjFuncP$, given in Table \ref{tab:distGcurvedbeam3D}. 
We observe that using our GO-ADM solver only slightly reduces $\globObjFuncP$. 
We also see that when using $\dataset_1$, we obtain the reference solution resulting from the standard FEA included in this dataset when using either the ADM or GO-ADM solver, leading to approximately zero global objective function. 
We recall that the FEA solution is thermomechanically consistent and hence no enforcement of this constraint is required, i.e. the penalty term remains zero during the computation. 
When using $\dataset_2$, however, this is not the case, as illustrated via the minimal value of the product $(\eV_h \cdot \sV_h)$ in Figure \ref{fig:curvedbeam3Denergy}b. 
We see that increasing the penalty factor, $\penFac$, increases the minimal value of this product, as expected. 
For the results obtained with $\dataset_2$, discussed above, we choose a penalty factor of $\penFac=10^4$, a scaling factor of $\penSca = \max(\eTilV \cdot \sTilV)=6 \cdot 10^2$, and a tolerance of $\penEps=0$ (see also discussion in Section \ref{sec:penaltyPara}). 
We note that this choice of parameters leads to insignificant contribution of the penalty term to the objective function, $\globObjFuncP$, shown in Table \ref{tab:distGcurvedbeam3D}. 
We draw the same conclusions as for the previous exemplary cantilever beam.

\subsection{Multi-beam structures}


\begin{figure}[htb]
    \centering
        \def\svgwidth{0.63\textwidth}
        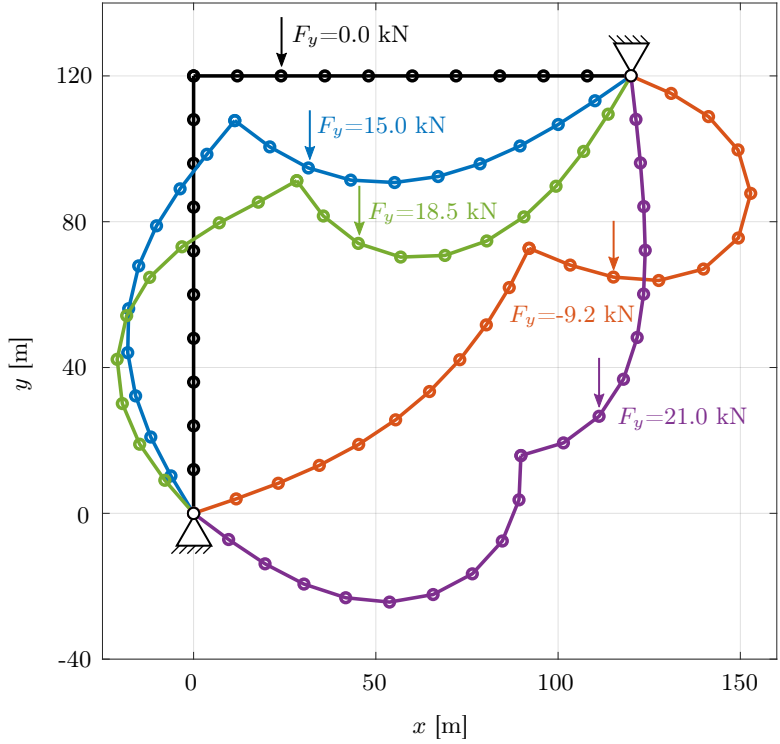

    \caption{The reference configuration (black) and four snapshots of a hinged right-angle frame, studied in \cite{Betsch2002,Simo1986}, computed with \textbf{our GO-ADM solving strategy}, using \textbf{synthetic datasets including the FEA solution} and \textbf{ANLP initialization approach at each load step}.}\label{fig:hingedFrameSnap}
\end{figure}

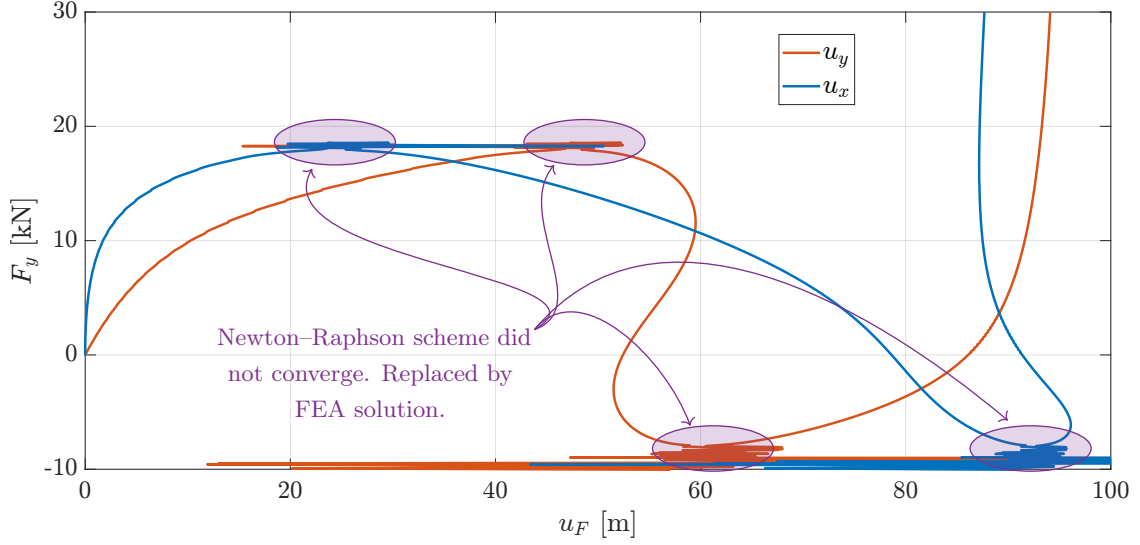
\begin{figure}[htb]
    \centering
    \begin{tikzpicture}
    \node[anchor=south west, inner sep=0] (img) at (0,0)
    {\includegraphics[width=0.9\textwidth]
    {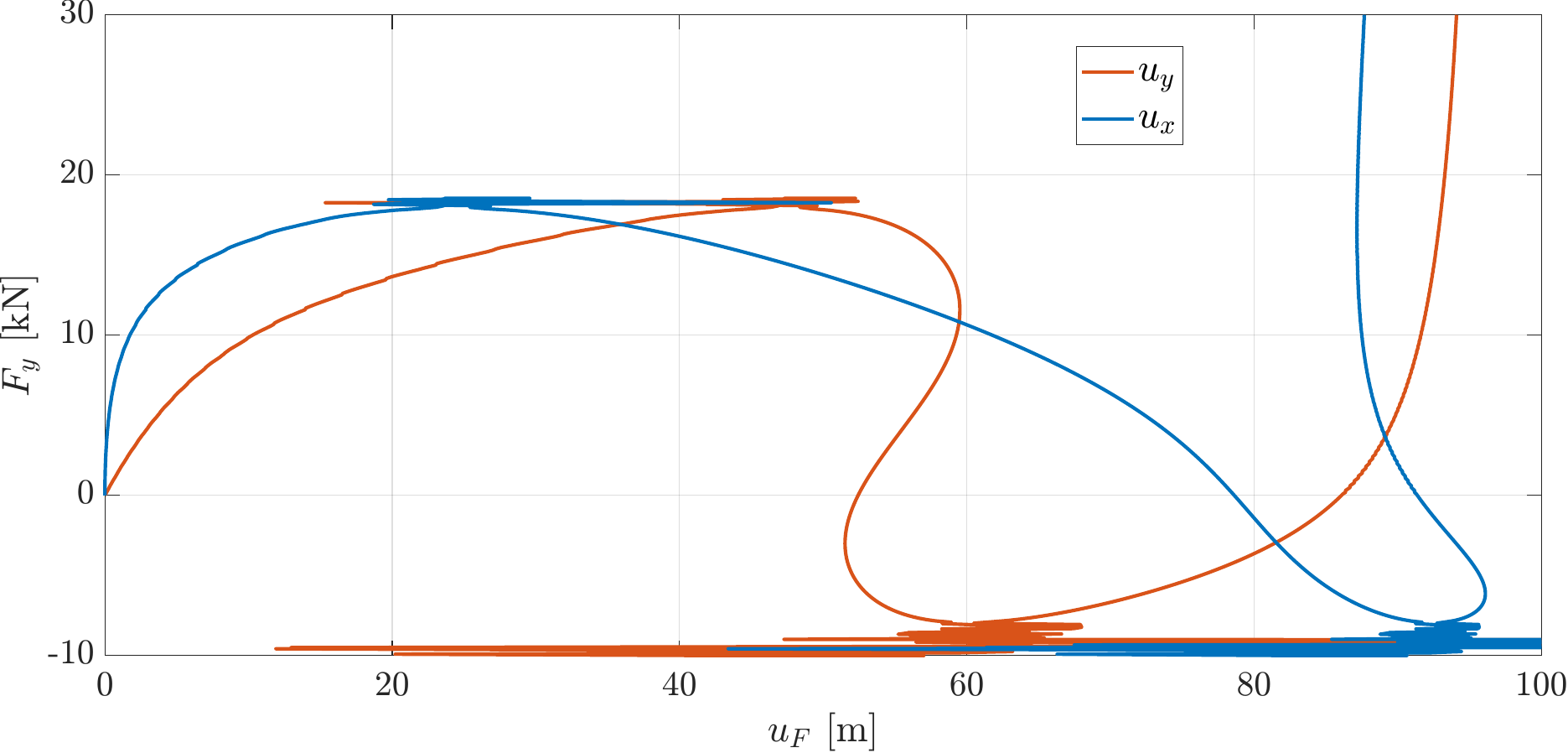}};

    \draw[
        purple1,
        line width=0.5pt,
        fill=purple1,
        fill opacity=0.2]
        (4.3,5.25) ellipse [
        x radius=0.8,
        y radius=0.3];

    \draw[
        purple1,
        line width=0.5pt,
        fill=purple1,
        fill opacity=0.2]
        (7.6,5.25) ellipse [
        x radius=0.8,
        y radius=0.3];

    \draw[
        purple1,
        line width=0.5pt,
        fill=purple1,
        fill opacity=0.2]
        (9.3,1.2) ellipse [
        x radius=0.8,
        y radius=0.3];

    \draw[
        purple1,
        line width=0.5pt,
        fill=purple1,
        fill opacity=0.2]
        (13.5,1.2) ellipse [
        x radius=0.8,
        y radius=0.3];


    \node[
        anchor=west,
        text=purple1,
        align=center](txt)
        at (2.5,2.2)
        {\footnotesize{
        Newton--Raphson scheme did}\\
        \footnotesize{not converge. Replaced by}\\
        \footnotesize{FEA solution.}
        };


    \draw[->,
        purple1,
        line width=0.5pt
        ]($(txt.north east)+(-0.1,-0.15)$)
        to[out=30,in=-120]
        (4.0,4.9);

    \draw[->,
        purple1,
        line width=0.5pt
        ]($(txt.north east)+(-0.1,-0.15)$)
        to[out=20,in=-130]
        (7.2,4.95);
    
    \draw[->,
        purple1,
        line width=0.5pt
        ]($(txt.north east)+(-0.1,-0.15)$)
        to[out=50,in=-250]
        (9,1.5);
    
    \draw[->,
        purple1,
        line width=0.5pt] ($(txt.north east)+(-0.1,-0.15)$)
        to[out=50,in=-220]
        (13.2,1.57);
    \end{tikzpicture}

    \caption{Load-deflection curve of the node under the load of the frame illustrated in Figure \ref{fig:hingedFrameSnap}, computed with \textbf{our GO-ADM solving strategy}, using \textbf{synthetic datasets including the FEA solution} and \textbf{ANLP initialization approach at each load step}.}\label{fig:hingedFrameUF}
\end{figure}

\begin{figure}[htb]
    \centering    
    \includegraphics[width=0.55\textwidth]{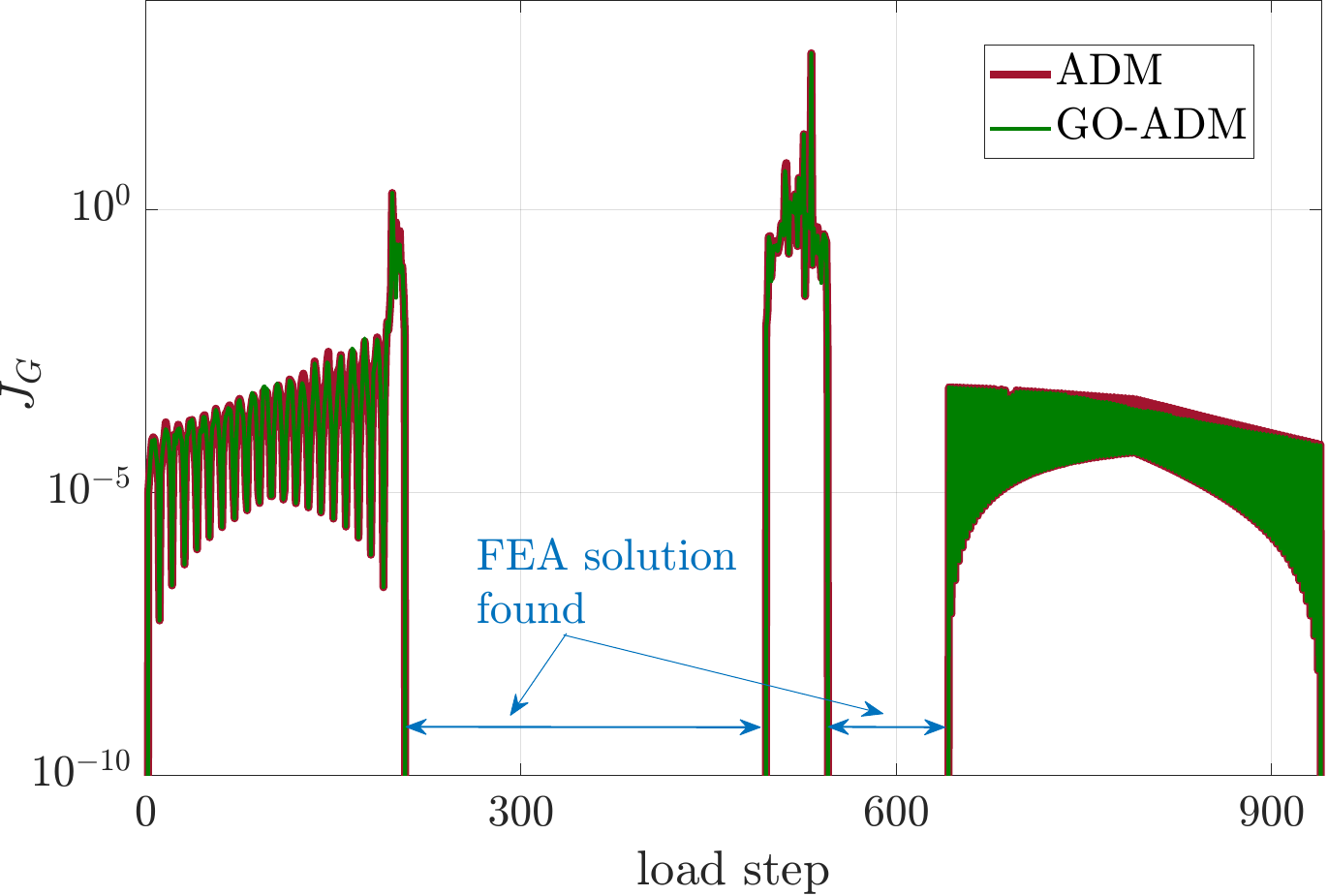}

    \caption{Value of the modified global objective function, $\globObjFuncP$, of the hinged right-angle frame, illustrated in Figure \ref{fig:hingedFrameSnap}, computed with \textbf{the ADM and our GO-ADM solving strategies}, using \textbf{synthetic datasets including the FEA solution} and \textbf{ANLP initialization approach at each load step}.}\label{fig:hingedFrameCostFunc}
\end{figure}


We now numerically illustrate that our GO-ADM solver together with the introduced penalty term remains robust when applying to multi-beam structures. 
To this end, we consider in the following two exemplary frames with planar and three-dimensional deformations in the following. 
The first example is  
a hinged right-angle frame consisting of one vertical and one horizontal leg of 120 m each, illustrated in black in Figure \ref{fig:hingedFrameSnap}, which is established in the literature, for instance, in \cite{Betsch2002,Simo1986}. 
The frame is subjected to a vertical force, $F_y$, at $(x,y) = (24,120)$, whose values follow the trajectory, e.g., in \cite[Fig. 7]{Simo1986}. 
We note that 
this example showcases the buckling and also the complex post-buckling behavior of the considered frame, and in order to investigate such behavior, the authors of \cite{Betsch2002,Simo1986} computed this using quadratic finite elements and the arc-length method. 
In this section of this work, we aim to investigate the robustness of our GO-ADM solver in such stability-losing scenario, even with the simplest discretization and the most standard method. 
Therefore, we discretize the frame with linear finite elements and employ the standard Newton-Raphson method.  
We model the considered frame with two beam structures, one for each leg, which are coupled with a rigid joint (see also Equation \eqref{eq:jointCons}), and discretize each of these with 10 elements. 
To investigate whether our GO-ADM solver can recover a known discrete solution of this buckling example, we employ here a synthetic (thermomechanically-consistent) verification dataset constructed from a finite element analysis (FEA) of the same loading problem. 
We performed the FEA with the aforementioned discretization of 10 linear elements for each leg and the standard load-controlled Newton-Raphson method, assuming a linear constitutive relation with the material properties given in \cite{Betsch2002,Simo1986}. 
In contrast to the computations in \cite{Betsch2002,Simo1986}, which employ quadratic finite elements and an arc-length method, this deliberately simple numerical setting does not reliably trace a single equilibrium branch through the critical buckling and post-buckling regions around $F_y = 18.5$ kN and $F_y = -10$ kN, respectively. 
Consequently, the resulting dataset does not contain an accurate converged reference state on the desired equilibrium branch at every load level in these regions. Between and beside these critical regions, however, the FEA follows a single equilibrium branch and provides reference states for the data-driven search.

In Figures \ref{fig:hingedFrameSnap} and \ref{fig:hingedFrameUF}, we illustrate four snapshots of the studied frame and the corresponding load-deflection curves of the node under the load, respectively. 
We obtained these results using our GO-ADM solver and the ANLP initialization approach at all load steps. 
We observe a very good agreement in the snapshots and the load-deflection curves with the results shown in \cite{Betsch2002,Simo1986}, except the jumps in the latter at the two critical load levels of $F_y = 18.5$ kN and $F_y = -10$ kN (see regions highlighted in purple in Figure \ref{fig:hingedFrameUF}). 
Our numerical experiments show that for this buckling example, the initialization approach and whether it is employed once at the first load step or up to a certain load step play a crucial role in the performance of 
both the ADM and GO-ADM solvers. 
In particular, using the random initialization, whether once or at every load step, does not guarantee convergence of either the Newton-Raphson scheme or the data-driven solver, already in the pre-buckling range. 
We note that this occurs despite the fact that the randomly selected data in this case are from the FEA solution at one of the load steps, which, however, can be very different from the solution associated with the current step due to complex behavior of the studied frame. 
Initializing the data once at the first load step with 
either one of the other three initialization approaches: the stress-free, structure-specific, or the ANLP initialization, 
leads to very good results in the pre-buckling range but jumps in the load-deflection curves in the buckling range (see also the purple-highlighted regions at $F_y=18.5$ kN in Figure \ref{fig:hingedFrameUF}). 
However, the solver could not converge to a discrete solution matching the reference response in the post-buckling range. In the critical buckling and post-buckling regions, this difficulty has two related causes: the employed dataset does not contain an accurate reference state on the desired equilibrium branch at every load level, and the solution from the preceding load step may belong to a different equilibrium branch, as discussed above, and therefore provide an inadequate initial guess. Consequently, convergence of both the Newton-Raphson scheme and the data search of the ADM solver is not guaranteed.

Initializing the data at every load step with either the stress-free, structure-specific, or the ANLP approach, on the one hand, leads to convergence in the post-buckling range before reaching the load level $F_y=-10$ kN. 
On the other hand, it leads to jumps in the load-deflection curve when reaching this load level (see also the purple-highlighted regions at $F_y=-10$ kN in Figure \ref{fig:hingedFrameUF}), and divergence in the solution at subsequent load levels. 
This is because the solution from the preceding load step can similarly provide an inadequate initial guess for continuing along the desired equilibrium branch, leading to divergence. 
Hence, in the present computation, when the Newton-Raphson scheme does not converge at this load level $F_y=-10$ kN, we replace the discrete solution with that from the standard FEA. 
This also includes the solution in the buckling range and is highlighted in purple in Figure \ref{fig:hingedFrameUF}. 
We note that the replaced solution from FEA might not be a converged solution on the right equilibrium branch but is a sufficient solution guess since the FEA converged in the following ranges. 
A more appropriate treatment of these critical regions would require a continuation strategy capable of tracing the desired equilibrium branch, such as the arc-length method employed in \cite{Betsch2002}, for both the FEA and data-driven computations. 
Another interesting approach is to employ the reference load-deflection curve as another source of data of the DCNLP, as introduced in \cite{Romero2026ddcm}, i.e. enhancing the current DCNLP to minimize both the distance regarding the material data and the load-deflection data. 
We plan to combine the arc-length method with our GO-ADM solver and/or the latter approach in future work.

\begin{remark}
    For the studied hinged right-angle frame, we additionally performed numerical experiments for this frame in which the reference FEA solution of the same loading scenario was not included in the material dataset. Instead of the ANLP initialization, we employed the structure-specific initialization only at the first load step and subsequently used the converged data from the preceding load step. Thus, the nonlinear reference FEA solution of the considered loading problem was not used for initialization or solution correction and was retained only for validation. In these experiments, we observed that the convergence of the Newton-Raphson and ADM-solver was not guaranteed, particularly in the critical buckling and post-buckling regions. This reveals a limitation of the present formulation, which does not explicitly incorporate information for distinguishing and following different equilibrium branches, and motivates the extensions based on continuation and additional load-deflection data.
\end{remark}

Figure \ref{fig:hingedFrameCostFunc} illustrates the value of the modified global objective function, $\globObjFuncP$, at each load step, when using the ADM (red) and our GO-ADM (green) solver. 
We observe that for the studied frame, these two solvers approximately lead to the same value of $\globObjFuncP$, which is expected when using a dataset including the FEA solution. 
The GO-ADM solver slightly reduces the objective function only at some load steps. 
We see that even in the pre-buckling range, these two solvers did not find the exact FEA solution, leading to $\globObjFuncP$ that is larger than the machine accuracy ($10^{-12}$). 
We also observe that the data-driven solver finds the right FEA solution corresponding to the considered load step only at some parts of the post-buckling range, where $\globObjFuncP << 10^{-12}$ (see indicated ranges with blue arrows in Figure \ref{fig:hingedFrameCostFunc}). 
In the buckling range around load step 200, as well as at the load level $F_y=-10$ kN around load step 500 
$\globObjFuncP$ shows jumps to higher values since we did not compute this with the replaced solution from FEA but with the diverged solution and data at the corresponding load step. 
The replaced solution is crucial as a solution guess for the subsequent load step. 
Moreover, we see that the data-driven solver could find the right FEA solution after overcoming the load level $F_y=-10$, leading to $\globObjFuncP$ close to the machine accuracy. 
However, it jumps to larger values from load steps 640 onward despite the exact FEA solution included in the employed dataset. 
The reason for these jumps, and also of the higher values in the pre-buckling range, is likely due to the complex behavior of the studied frame, which includes stress and strain states that are close to each other but are associated with different deformations. 
Employing the load-deflection curve as an additional dataset \cite{Romero2026ddcm} might improve this issue, which we plan to study in future work.

Regarding the thermomechanical consistency of the discrete solution for the studied frame, our numerical experiments show that it also depends on the initialization approach, despite the dataset consisting of the FEA solution that is thermomechanically consistent. 
While initializing the data randomly or with the stress-free state, whether once at the first load step or at each load step, the obtained solution is thermomechanically inconsistent at early load steps but not entirely within the pre-buckling range. 
Initializing the data with the structure-specific or the ANLP approach, whether once at the first load step or at each load step, 
leads to a consistent solution in the pre-buckling range but not in the buckling and post-bucking ranges. 
This is due to the fact that the data-driven solver does not always find the true FEA solution at all load steps for the studied frame. 
For this example, we choose a penalty factor of $\penFac=10^4$, a scaling factor of $\penSca = \max(\eTilV \cdot \sTilV)=2 \cdot 10^5$, and a tolerance of $\penEps=0$ (see also discussion in Section \ref{sec:penaltyPara}), to weakly enforce the thermomechanical consistency constraint. 
This choice again leads to insignificant contribution of the penalty term to the objective function, $\globObjFuncP$, shown in Figure \ref{fig:hingedFrameCostFunc}. 
We conclude that for such stability-losing structures as the studied frame, the initialization approach plays a crucial role in ensuring convergence and a thermomechanically consistent solution. 
For the former, the frequency of the initialization also plays an essential role. 
While the ANLP initialization approach requires the highest computational effort, employing it at each load step provides the highest chance of convergence, compared to the three aforementioned approaches. 



\begin{figure}[htb]
    \centering
    
        \def\svgwidth{0.38\textwidth}
        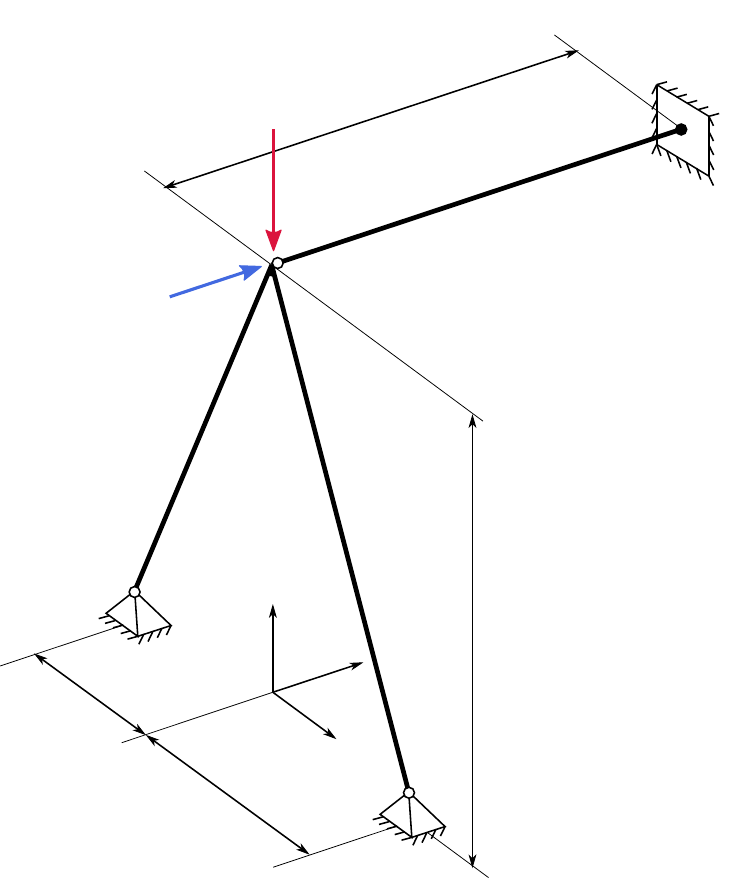 \hspace{0.2cm}
    \includegraphics[width=0.5\textwidth]{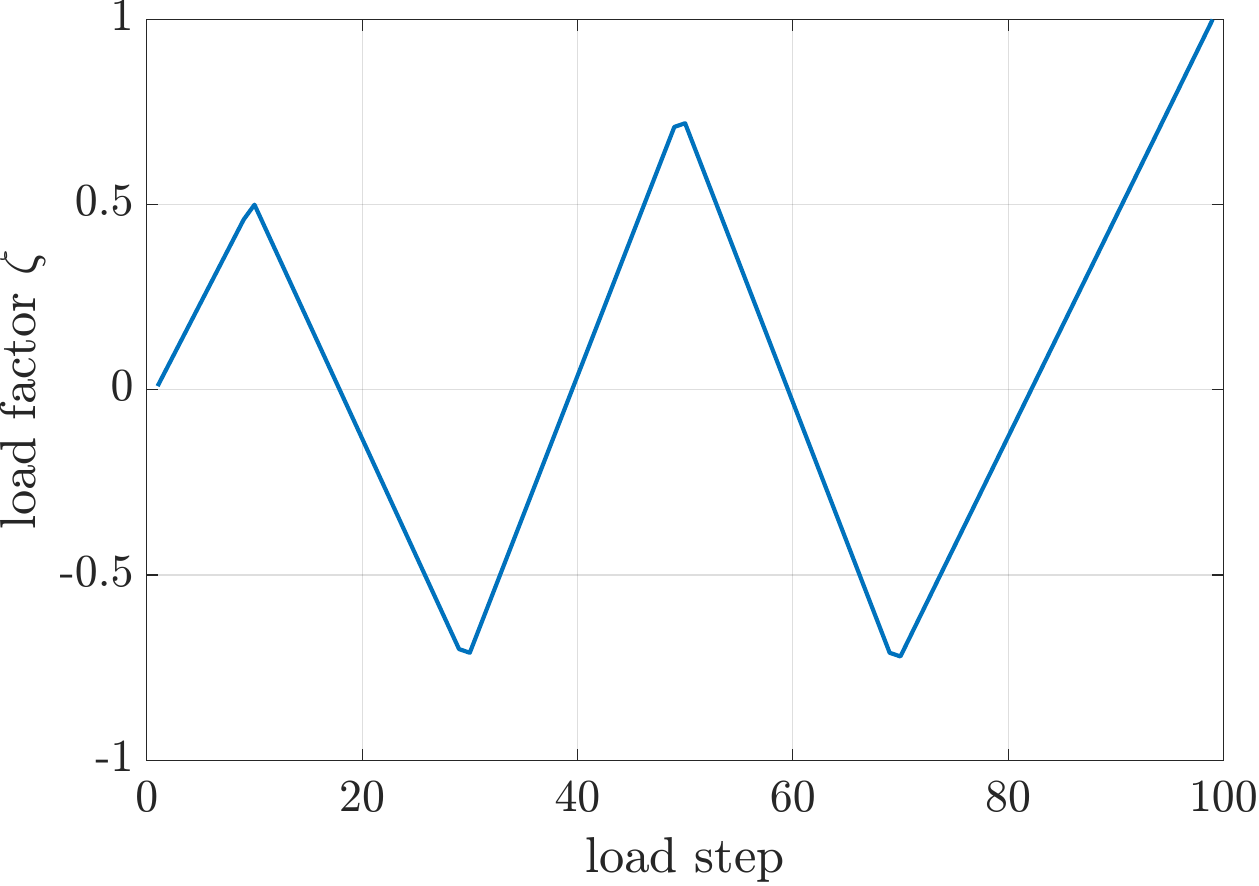}

    \caption{Sketch and the employed load factors at each step for a three-dimensional frame, modified based on the three pinned-bar structure studied in \cite[Sec. 9.7.1]{cardona2001}.}\label{fig:frame3dSketch}
\end{figure}

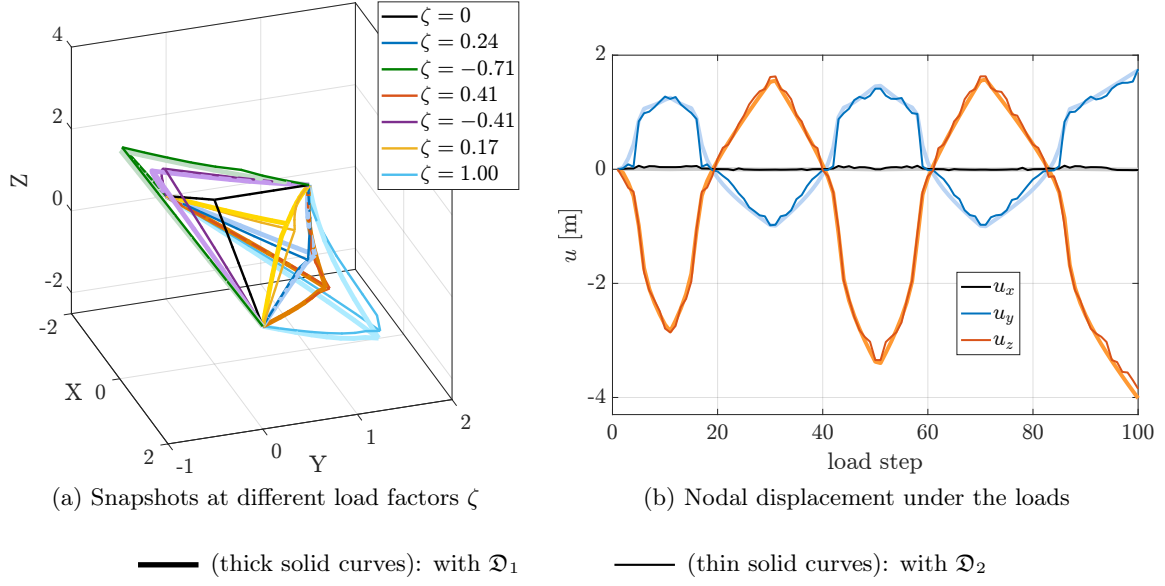
\begin{figure}[htb]
    \centering
    \begin{tikzpicture}


        \node[anchor=south west,
            inner sep=0] (img1) at (0,0)
        {\subfloat[Snapshots at different load factors $\loadFac$]{\includegraphics[width=0.41\textwidth]{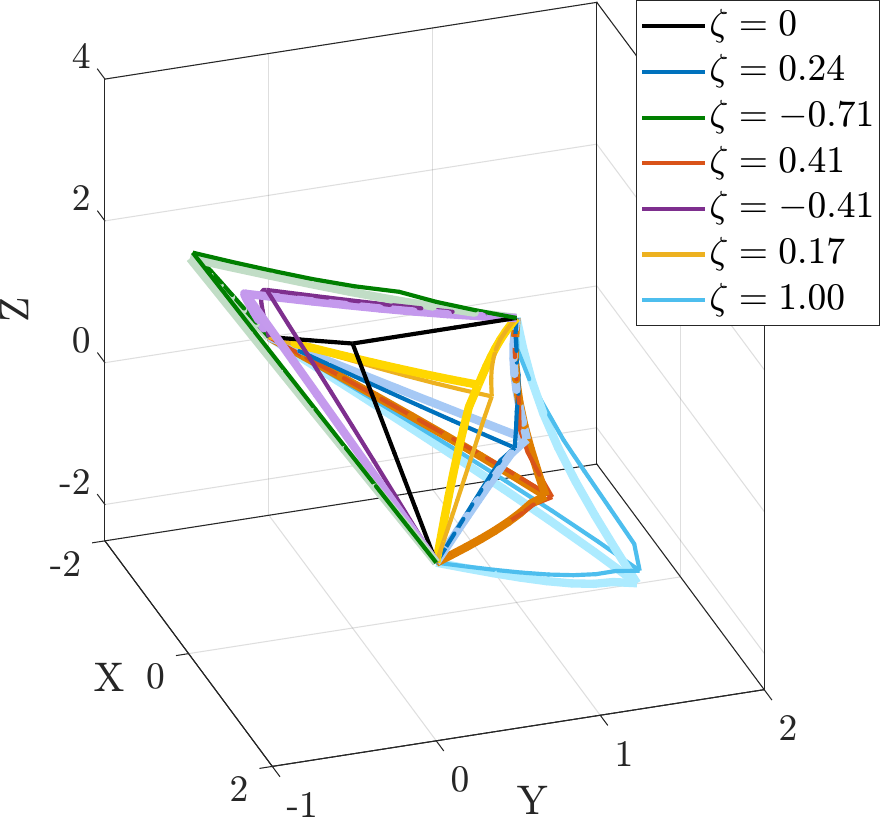}}};


        \node[anchor=south west,
            inner sep=0] (img2) at (0.44\textwidth,0)
        {\subfloat[Nodal displacement under the loads]{\includegraphics[width=0.48\textwidth]{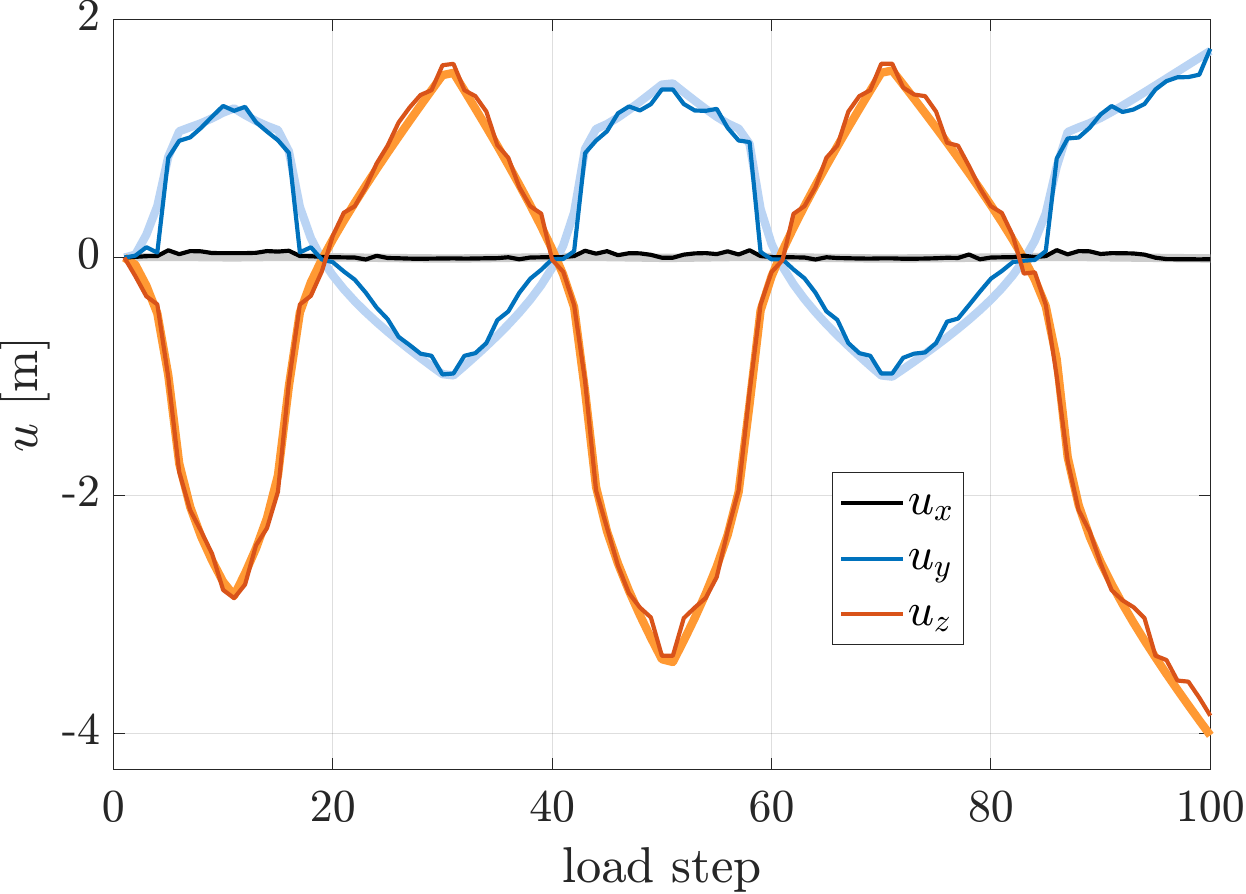}}};


        \node[
            anchor=north,
            inner sep=0pt] at ($(img1.south)!0.5!(img2.south)+(0,-0.5cm)$)
        {
            \begin{tikzpicture}

                \draw[line width=2pt]
                (0,0) -- (0.8,0);

                \node[anchor=west] at (0.95,0)
                {\footnotesize {(thick solid curves): with $\dataset_1$}};

                \draw[line width=0.8pt]
                (6.3,0) -- (7.1,0);

                \node[anchor=west] at (7.25,0)
                {\footnotesize (thin solid curves): with $\dataset_2$};

            \end{tikzpicture}
        };
    \end{tikzpicture}
    
\caption{Snapshots and nodal displacements at point A of the frame illustrated in Figure \ref{fig:frame3dSketch}, computed with \textbf{our GO-ADM solving strategy}, using synthetic datasets with ($\dataset_1$) and without ($\dataset_2$) the solution from a standard FEA.}\label{fig:frame3dSnapsNuCurves}
\end{figure}

\begin{figure}[!htb]
    \centering
    \begin{tikzpicture}

    \node[anchor=south west, inner sep=0] (img1)
    at (0,0)
    {\subfloat[Minimum across all beam members]{\includegraphics[width=0.48\textwidth]{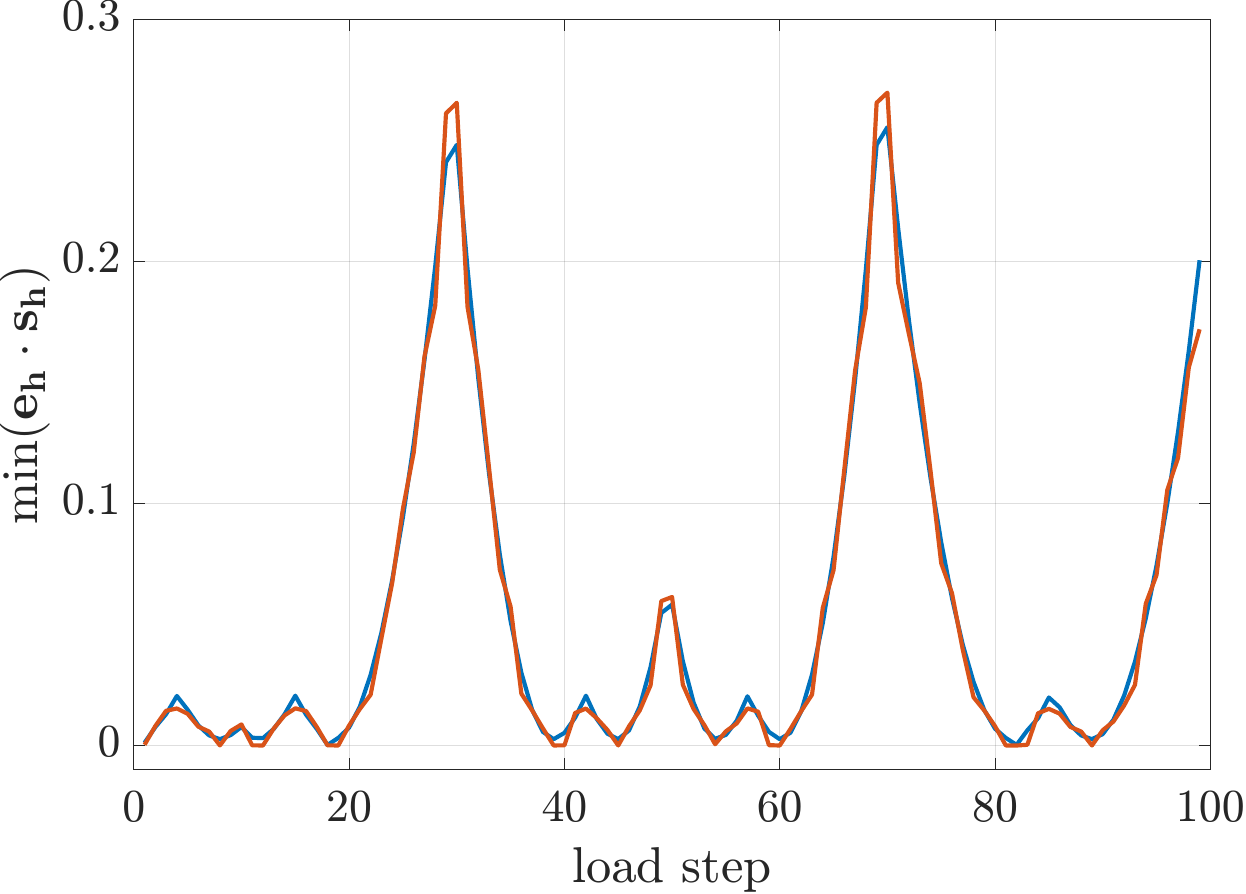}}};


    \node[anchor=south west, inner sep=0] (img2)
    at (0.5\textwidth,0)
    {\subfloat[Minimum across all beam members and load steps]{\includegraphics[width=0.48\textwidth]{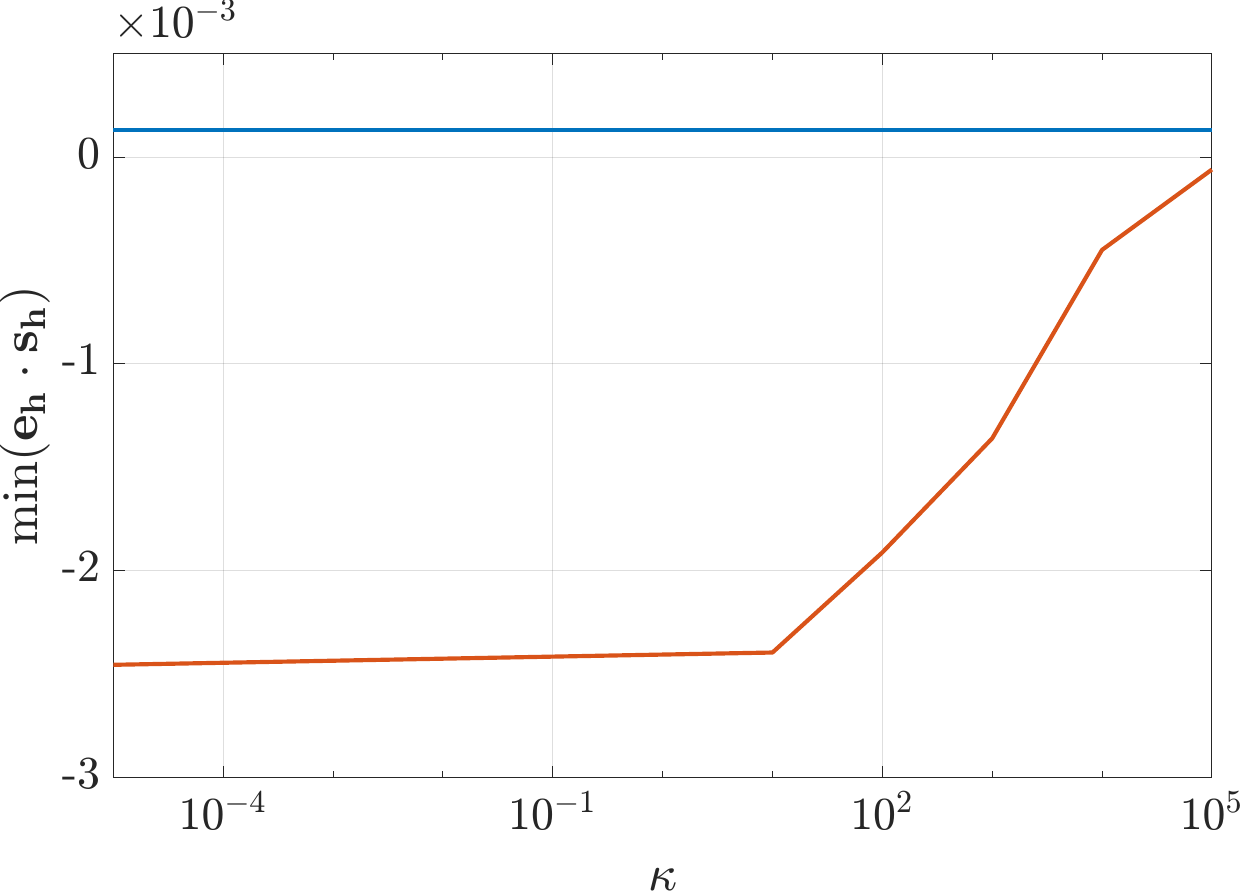}}};

    \begin{scope}[x={(img2.south east)},y={(img2.north west)}]

        \draw[blue2, line width=1.0pt]
            (0.25,0.72) -- (0.3,0.72);

        \node[anchor=west, text=blue2]
            at (0.3,0.72)
            {\footnotesize{: \textcolor{blue2}{with $\dataset_1$}}};

        \draw[burntorange, line width=1.0pt]
            (0.285,0.65) -- (0.335,0.65);

        \node[anchor=west, text=burntorange]
            at (0.335,0.65)
            {\footnotesize{: \textcolor{burntorange}{with $\dataset_2$}}};

    \end{scope}
    \end{tikzpicture}

\caption{The minimal values of the product $(\eV_h \cdot \sV_h)$ across all beam members (a) and the whole computation (b) of the frame illustrated in Figure \ref{fig:frame3dSketch}, computed with \textbf{our GO-ADM solving strategy} using synthetic dataset with (\textcolor{blue2}{$\dataset_1$, blue}) and without (\textcolor{burntorange}{$\dataset_2$, orange}) the solution from a standard FEA.}\label{fig:frame3dminE}

\vspace{1cm}

    \centering    
    \includegraphics[width=0.55\textwidth]{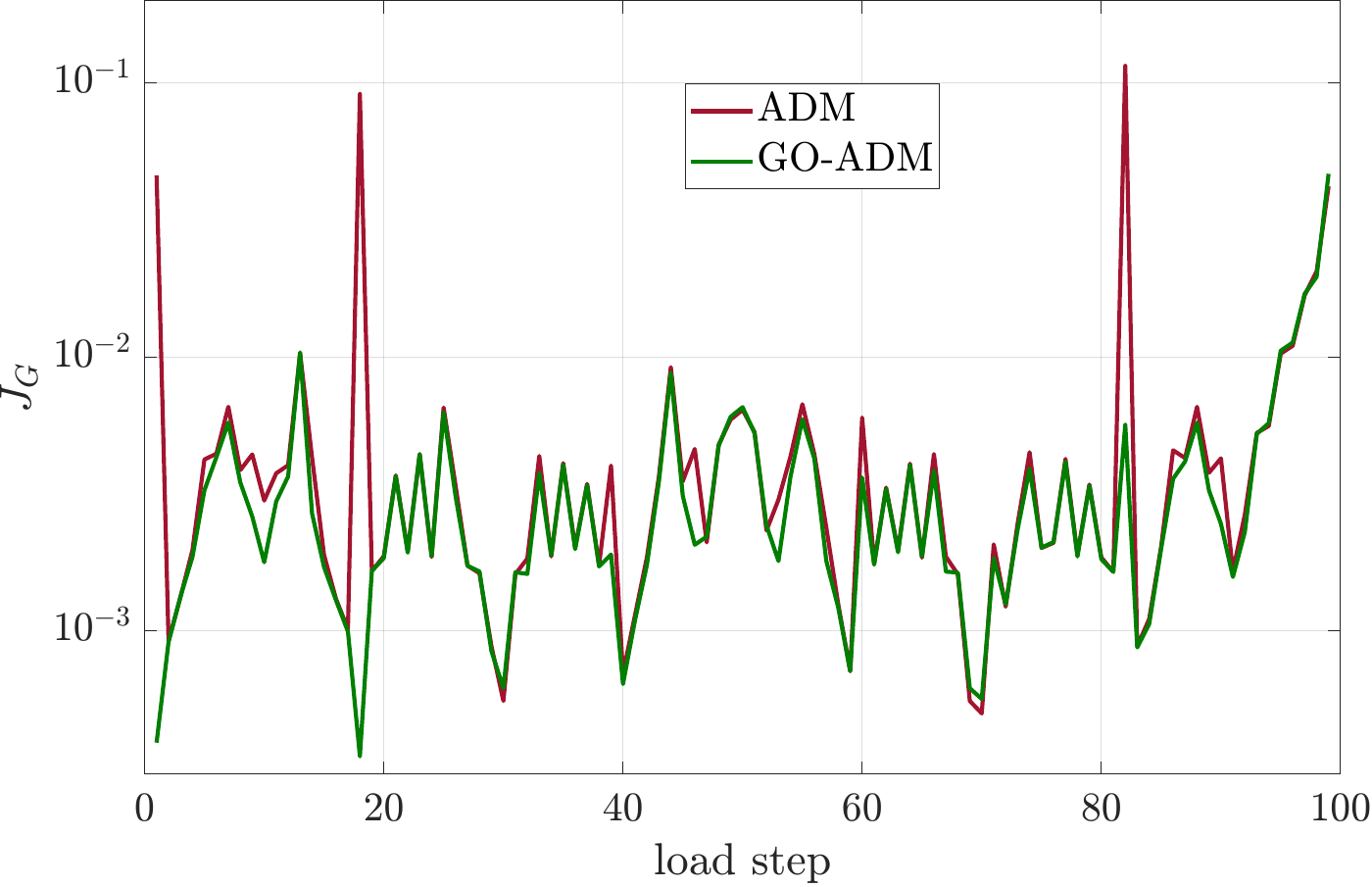}

    \caption{Value of the modified global objective function, $\globObjFuncP$, of the frame illustrated in Figure \ref{fig:frame3dSketch}, computed with \textbf{the ADM and our GO-ADM solving strategies}, using the synthetic dataset $\dataset_2$ \textbf{without the solution from a standard FEA}.}\label{fig:frame3dCostFunc}
\end{figure}


The second example is a three-dimensional (3D) frame consisting of an L-shaped frame in the $xz-$plane that is simply supported by a cantilever beam aligned with the $y$-axis, illustrated in Figure \ref{fig:frame3dSketch}. 
This is a modified version of the three pinned-bar structure studied in \cite[Sec. 9.7.1]{cardona2001}, in which we replace the two-member truss in the $xz-$plane by an L-shaped frame. 
In this section, since we aim at illustrating that the GO-ADM solver together with the introduced penalty function is robust when applied to three-dimensional frame structures,  
we also modify the geometry and the loading, including the load factor, $\loadFac$, of the considered 3D frame, as illustrated in Figure \ref{fig:frame3dSketch}, to avoid buckling and/or snap-through behavior. 
We model the frame with three beam members, numbered in Figure \ref{fig:frame3dSketch}, and 
adapt the stiffness given in \cite[Sec. 9.7.1]{cardona2001} that is unit axial, shear, bending, and torsional stiffness for beam 1 and 3, and 40\% of these values for beam 2. 
We discretize each beam member with 8 linear elements, which lead to sufficiently accurate results based on our convergence study. 
For comparison purposes, we consider here two datasets: dataset $\dataset_1$ which consists of the discrete solution from a standard FEA using the same discretization based on a linear constitutive relation with the same material properties and has 800 stress--strain states for each beam member; 
and dataset $\dataset_2$ which consists of the discrete solution of the material tests, described in \ref{sec:datageneration}, and has 5000 stress--strain states.

Based on our numerical experiments, we choose the ANLP initialization approach at each load step for this example. 
Initializing either randomly or with the stress-free state leads to non-convergence of the Newton-Raphson scheme at the first load steps, irrespective of the employed dataset or whether initializing once at the first load step or at every load step. 
Initializing with either the structure-specific or the ANLP approach once at the first load step leads to the same numerical instability for the studied three-dimensional frame. 
Initializing with these two approaches at every load step leads to a converged solution when using the dataset $\dataset_1$, while only the latter leads to convergence when using the dataset $\dataset_2$. 
We note that our modified frame does not show buckling or snap-through behavior with the chosen loads, but geometrically large deformations. 
These observations indicate that for the studied multi-beam structures, the ANLP initialization approach performs better, particularly when employed at every load step, compared to the three aforementioned initialization approaches.

Figure \ref{fig:frame3dSnapsNuCurves}a illustrates the snapshots at different load levels of the studied three-dimensional (3D) frame, obtained with our GO-ADM solver and the dataset $\dataset_1$ (thick curves) and $\dataset_2$ (thin curves). 
We also include here the reference configuration in black for comparison purposes. 
We observe a good match between the results obtained with these two datasets, except at the last two snapshots (yellow and light blue), which show different displacements at the node under the load, denoted as point A at $(x,y,z) = (0.0,0.0,1.5)$ m. 
We see this reflected in the displacement at point A at each step in Figure \ref{fig:frame3dSnapsNuCurves}b, where the results obtained with $\dataset_1$ (thick curves) and $\dataset_2$ (thin curves) match well at early load steps and differ more from each other at larger load levels, for instance, from load step 80. 
This is due to different stress--strain states included in $\dataset_1$ and $\dataset_2$, as discussed in the previous section of our numerical studies for the single beam structures. 
One can improve the accuracy obtained with $\dataset_2$ by enriching this dataset with more material tests and/or more rotations (see also \ref{sec:datageneration}).

Figure \ref{fig:frame3dminE}a shows the minimal value of the product $(\eV_h \cdot \sV_h)$ across all beam members at each load step when using $\dataset_1$ (blue curve) and $\dataset_2$ (orange curve). 
We observe a good agreement between these results, except for load steps 30, 70, and the last step, where they differ from each other. 
This is consistent with the observed results in Figure \ref{fig:frame3dSnapsNuCurves}, discussed above. 
Focussing on the results obtained with $\dataset_2$, we see that the product $(\eV_h \cdot \sV_h)$ is approximately non-negative at all load steps. 
Figure \ref{fig:frame3dminE}b also shows the minimal value of this product over all load steps, obtained with different values of the penalty factor $\penFac$. 
We include here the constant positive value obtained with $\dataset_1$ (blue curve) for comparison purposes. 
We see that as expected, increasing $\penFac$ increases the minimal value of $(\eV_h \cdot \sV_h)$, weakly enforcing the thermomechanical consistency constraint more effectively. 
Based on these results, 
we choose a penalty factor of $\penFac=10^4$ for all beam members, a scaling factor of $\penSca = \max(\eTilV \cdot \sTilV)$ that is $1$ for beam 1 and 3 and $10$ for beam 2, and a tolerance of $\penEps=0$ (see also discussion in Section \ref{sec:penaltyPara}), for all our computations of the studied 3D frame. 
Figure \ref{fig:frame3dCostFunc} illustrates the value of the modified global objective function, $\globObjFuncP$, obtained with the ADM (red) and GO-ADM (green) solver, using the dataset $\dataset_2$. 
We observe that as expected, the latter generally reduces $\globObjFuncP$, compared to the former. 
For this case, the contribution of the penalty term to $\globObjFuncP$ remains insignificant with our choice of parameters.  
We note that when using $\dataset_1$, both of these two solvers find the FEA solution at all load steps, leading to $\globObjFuncP$ of the machine accuracy. 
We conclude that for the studied 3D frame, the GO-ADM solver remains robust and generally reduces the value of the global objective function, achieving a better approximation of the global optima. 
The introduced penalty term weakly enforces the thermomechanical consistency constraint and is effective with a sufficiently large penalty factor.

\subsection{Application to noisy data}


\begin{figure}[htb]
    \centering    
    \includegraphics[width=0.8\textwidth]{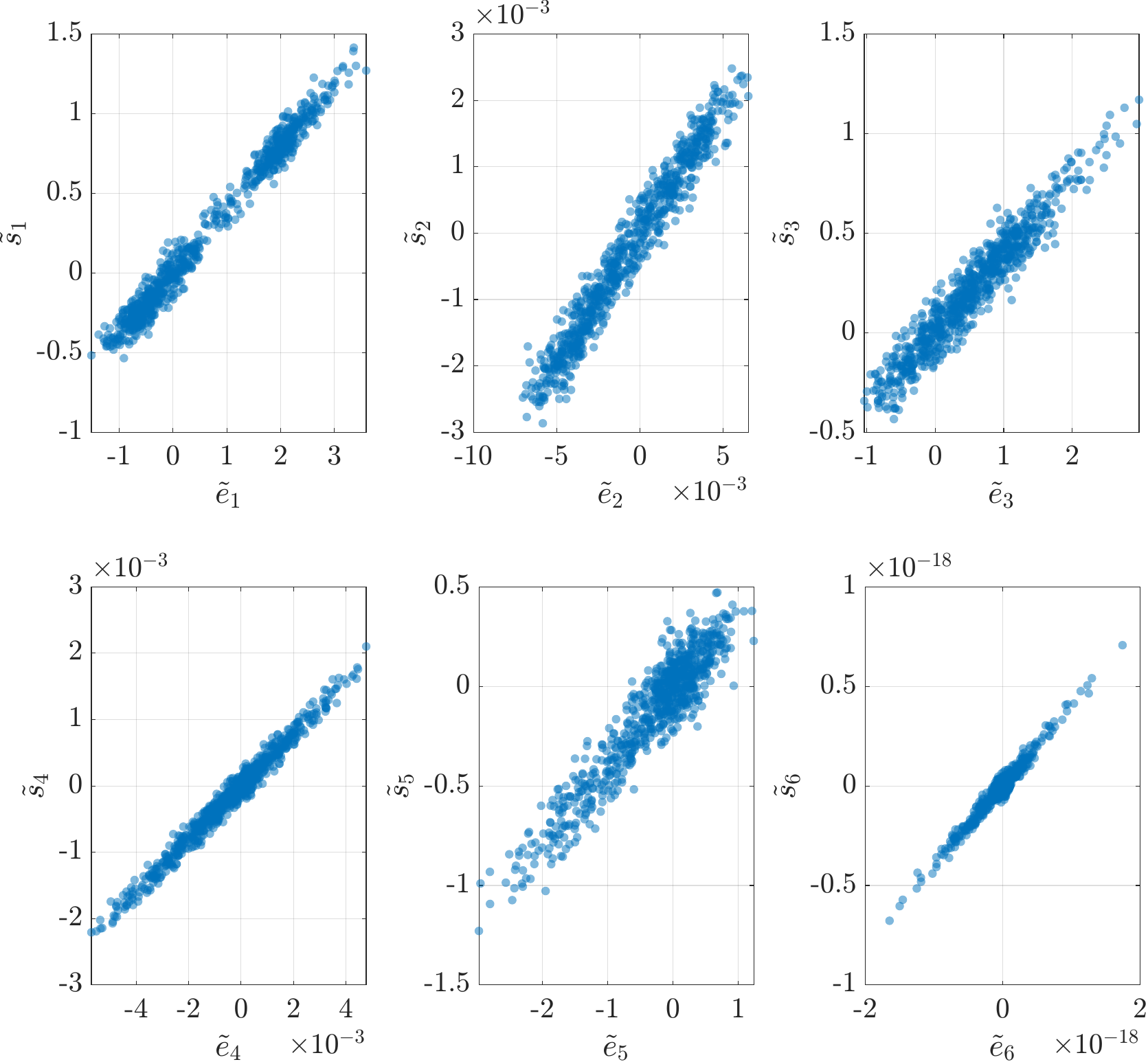}

    \caption{The noisy dataset, $\dataset_{1}$ with noise, of each stress and strain component of beam 2 of the three-dimensional frame illustrated in Figure \ref{fig:frame3dSketch}. \textbf{The dataset of each component consists of 800 data points}.}\label{fig:frame3dnoisyD}
\end{figure}

\begin{figure}[!htb]
    \centering
    \begin{tikzpicture}

    \node[anchor=south west, inner sep=0] (img1)
    at (0,0)
    {\subfloat[Nodal displacement under the loads]{
            \includegraphics[width=0.49\textwidth]{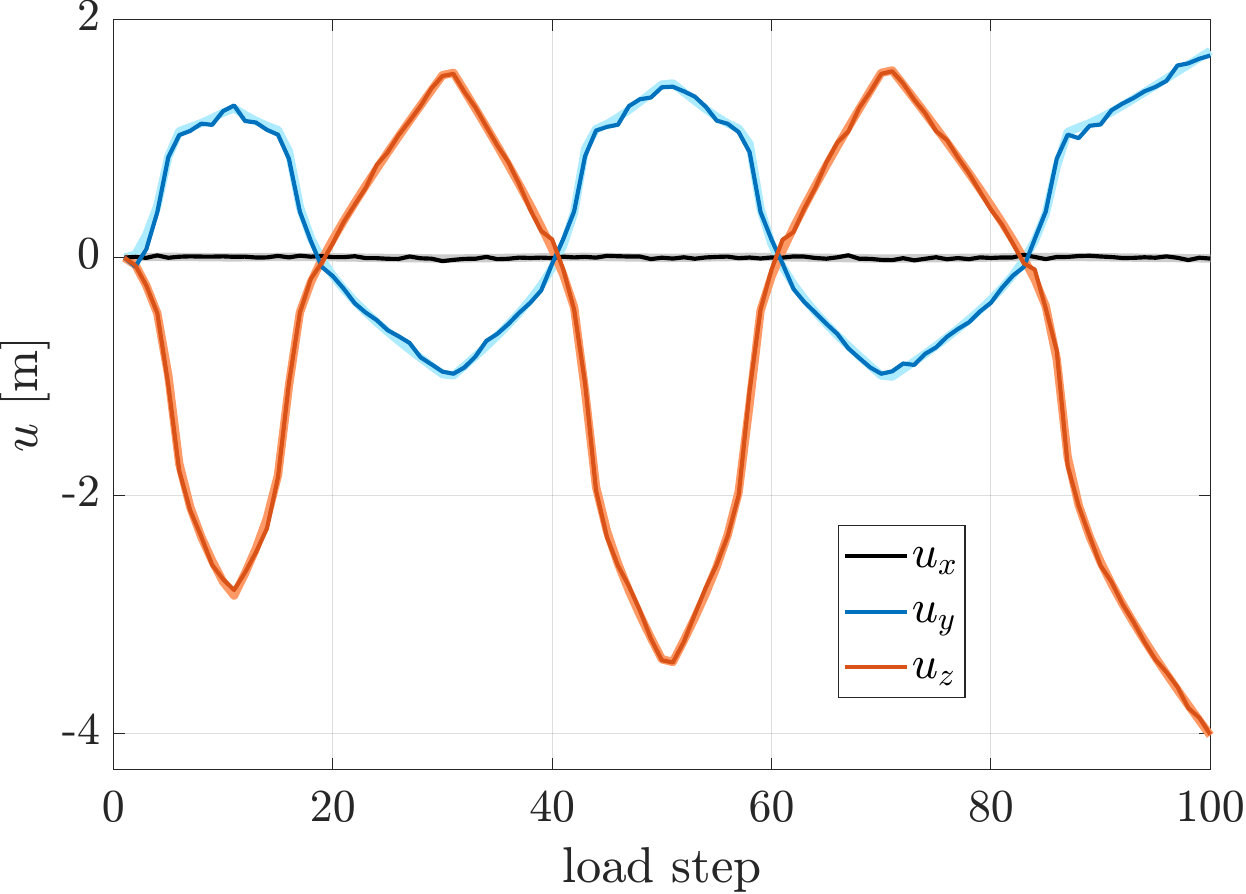}}};

    \begin{scope}[x={(img1.south east)}, y={(img1.north west)}]


    \draw[line width=2.5pt] (0.24,0.45) -- (0.29,0.45);

    \node[anchor=west] at (0.28,0.45)
    {\footnotesize{:w/o noise}};

    \node[anchor=west] at (0.23,0.39)
    {\footnotesize{\textit{(thick curves)}}};


    \draw[line width=0.8pt] (0.24,0.32) -- (0.29,0.32);

    \node[anchor=west] at (0.28,0.32)
    {\footnotesize{:w noise}};

    \node[anchor=west] at (0.23,0.26)
    {\footnotesize{\textit{(thin curves)}}};

    \end{scope}


    \node[anchor=south west, inner sep=0] (img2)
    at (0.52\textwidth,0)
    {\subfloat[$\min(\eV_h \cdot \sV_h)$]{
        \includegraphics[width=0.49\textwidth]{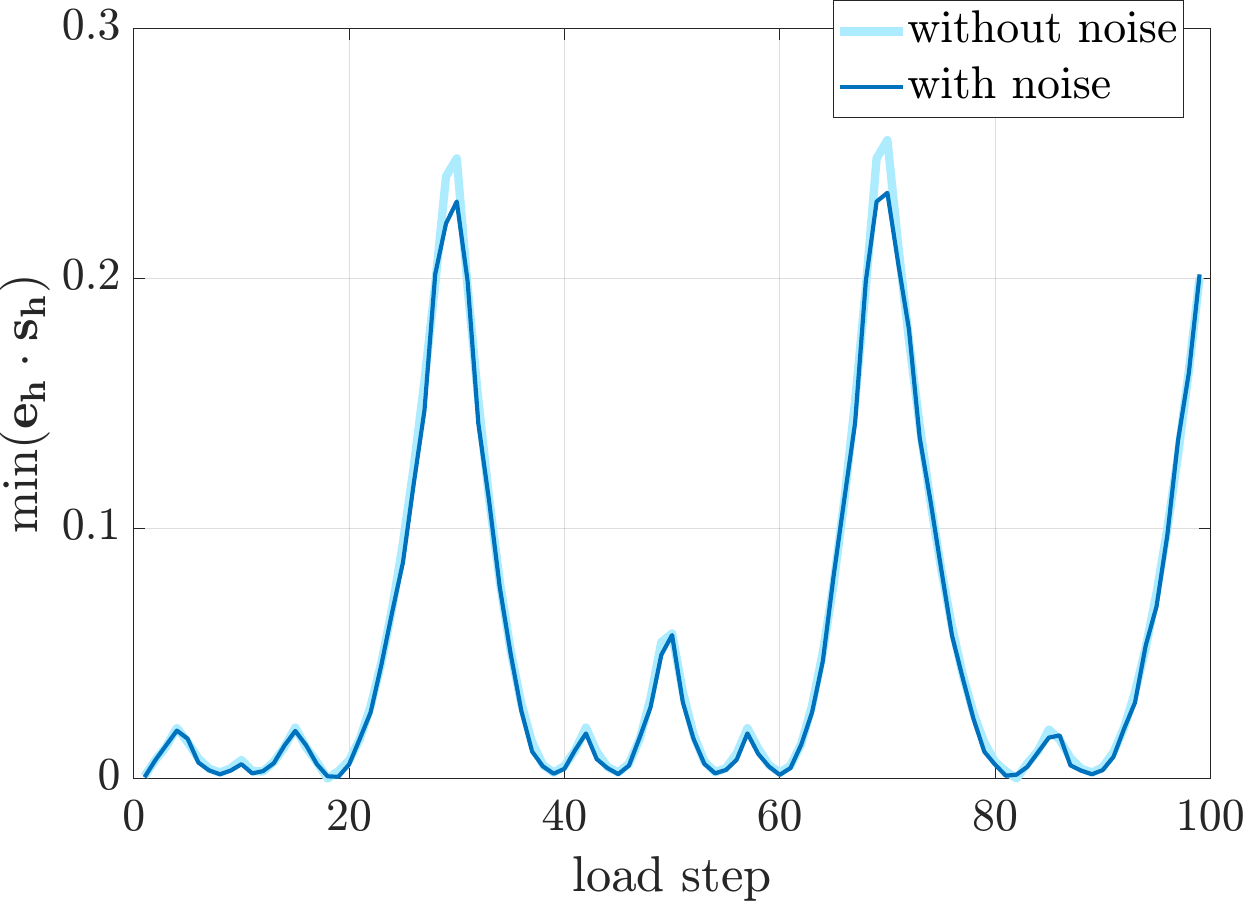}}};

    \end{tikzpicture}

    \caption{Nodal displacements at point A and the minimal value of the product $(\eV_h \cdot \sV_h)$ for the frame illustrated in Figure \ref{fig:frame3dSketch}, computed with \textbf{our GO-ADM solving strategy} using synthetic dataset $\dataset_1$ \textbf{including the FEA solution and noise}, illustrated in Figure \ref{fig:frame3dnoisyD}.}\label{fig:frame3dnoiseUCurvesNminE}

\vspace{1cm}

    \centering    
    \includegraphics[width=0.55\textwidth]{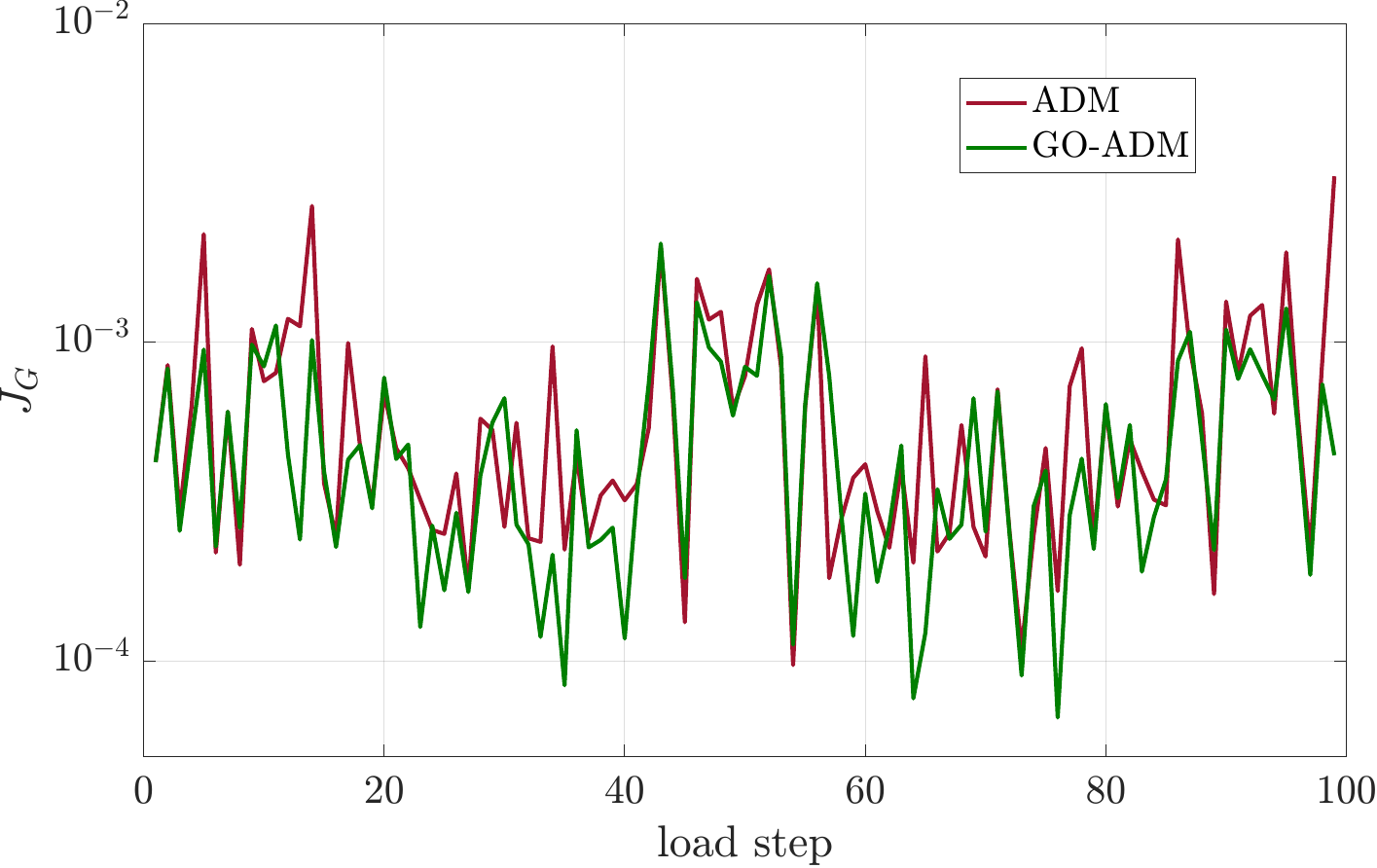}

    \caption{Value of the modified global objective function, $\globObjFuncP$, of the frame illustrated in Figure \ref{fig:frame3dSketch}, computed with \textbf{the ADM and our GO-ADM solving strategies}, using synthetic dataset $\dataset_1$ \textbf{including the FEA solution and noise}, illustrated in Figure \ref{fig:frame3dnoisyD}.}\label{fig:frame3dnoiseCostFunc}
\end{figure}


We now numerically illustrate that the GO-ADM solver together with the introduced penalty term is also robust in the case of noisy data. 
To this end, we choose again the three-dimensional (3D) frame, studied in the previous section and illustrated in Figure \ref{fig:frame3dSketch}, and 
add noise to the dataset $\dataset_1$. 
In particular, we add normally distributed noise to each stress and strain component included in $\dataset_1$, using a noise standard deviation, referred to as the noise level, that is a fraction of the standard deviation of the original stress and strain data. 
For example, a chosen noise level of 0.1 for the data of the first stress component, $\tilde{s}_1$, means adding noise with a standard deviation that is 10\% of the standard deviation of $\tilde{s}_1$ to each data value of $\tilde{s}_1$. 
The noise levels applied to the stress and strain data of the corresponding components, $\tilde{s}_1$, $\tilde{s}_2$, $\tilde{s}_3$, $\tilde{s}_4$, $\tilde{s}_5$, and $\tilde{s}_6$, 
are 10\%, 15\%, 20\%, 10\%, 25\%, and 15\%, respectively. 
In Figure \ref{fig:frame3dnoisyD}, we plot the resulting noisy dataset of each component. 
We compute the studied 3D frame using the same parameters and discretization as in the previous section.

Figure \ref{fig:frame3dnoiseUCurvesNminE}a illustrates the nodal displacements of point A at each load step, obtained with our GO-ADM solving strategy and using the noisy dataset illustrated in Figure \ref{fig:frame3dnoisyD} (thin curves). 
For comparison purposes, we include here the results obtained with the original dataset $\dataset_1$ without noise (thick curves). 
We observe that despite the noisy dataset, we obtain approximately the same displacements. 
Figure \ref{fig:frame3dnoiseUCurvesNminE}b shows the corresponding minimal value of the product $(\eV_h \cdot \sV_h)$ across all beam members at each load step. 
We see approximately the same results when using the original and the noisy dataset, except at load steps 30 and 70, where the noisy dataset leads to slightly smaller minimal values of $(\eV_h \cdot \sV_h)$. 
In Figure \ref{fig:frame3dnoiseCostFunc}, we also plot the value of the modified global objective function, $\globObjFuncP$, at each load step, obtained with the ADM (red) and GO-ADM (green) solver when using the noisy dataset illustrated in Figure \ref{fig:frame3dnoisyD}. 
We observe that as expected, the latter generally reduces $\globObjFuncP$, achieving a better approximation of the global optima.

\section{Summary and conclusions}\label{sec:conclusions}

In this work, we extended our GO-ADM solving strategy introduced in \cite{viljar2025,nguyenGoadm2026}, which combines a greedy optimization algorithm with the alternating direction method, for geometrically exact beams formulated using director-based kinematics. 
To initialize the stress and strain data, 
we proposed the ANLP initialization approach which employs the discrete solution of a conventional finite element analysis (FEA) of the same structure and loading, based on an assumed constitutive model, to identify the closest data points as initial values. 
Since the underlying boundary-value problem is equivalent to an approximate nonlinear optimization problem (ANLP) with a linear constitutive manifold \cite{Gebhardtddcmstatic2020}, we have referred to this approach as an ANLP initialization. 
Choosing either this or one of three existing approaches: a random, a stress-free, or a structure-specific initialization, one can initialize the data once at the first load step or repeatedly at selected load steps. 
Although ANLP initialization is computationally the most expensive among these approaches, it significantly reduces the risk of non-convergence, especially for stability-losing structures, such as structures undergoing buckling.
Our numerical studies showed that, for such structures as well as for multi-beam systems undergoing geometrically large deformations, reinitialization at every load step is generally required. 
Moreover, we considered different synthetic datasets: one constructed from the discrete stress and strain fields of the same FEA used for ANLP initialization, 
one obtained from multiple loading scenarios (virtual material tests), and one combining these two datasets.
We also discussed two searching strategies: 
a component-wise search which seeks pairs of each stress and strain component independently, 
and a vector-wise search which seeks pairs of the complete stress and strain vectors. 
On the one hand, the former requires more search operations and is only admissible for uncoupled generalized constitutive responses, for instance, in cases of isotropic materials with an assumed linear constitutive relation. 
Since the actual physical state relates to a stress and strain tensor or vector, this component-wise search is essentially a numerical treatment for situations in which data for only some of the components are available. 
On the other hand, our numerical results indicated that this searching strategy can yield thermomechanically consistent discrete stress and strain fields in cases where the vector-wise search does not.

To enforce the thermomechanical consistency of the discrete stress and strain fields, $\eV_h$ and $\sV_h$, 
we introduced a penalty approach that weakly enforces a prescribed tolerance on the normalized product $(\eV_h \cdot \sV_h)$. 
Our numerical parameter study showed that normalizing this product by $\max(\eTilV \cdot \sTilV)$ keeps the resulting system well conditioned and allows sufficiently large penalty factors for effective constraint enforcement. 
It further showed that arbitrarily small tolerances, including zero tolerance as required by strict thermomechanical consistency, 
can be imposed without deteriorating the conditioning of the system. 
By contrast, significantly larger tolerances require a correspondingly larger normalization factor to preserve good conditioning. 
As expected for a penalty method, increasing the penalty factor improves the enforcement of the thermomechanical consistency constraint, and a dimensionless penalty factor of order $10^3$ or $10^4$ proved sufficient in all computations considered in this work.
Through numerical examples involving single- and multi-beam structures in both two and three dimensions, we illustrated that the GO-ADM strategy is generally robust and consistently reduces the value of the global objective function, thereby providing a better approximation of the globally optimal solution than the standard ADM-based direct solver. These favorable properties were also observed in the presence of noisy data for a three-dimensional frame.

Several directions for future research emerge from the present work. 
A first one is the extension of the current discrete--continuous nonlinear optimization formulation (DCNLP) to a multi-data setting able to incorporate different data types. 
A particularly relevant example is the load--deflection curve, studied in \cite{Romero2026ddcm}, which is essential in the presence of multiple solutions or equilibrium branches/paths. 
Combining heterogeneous data sources requires a dimensionless objective function and a carefully designed weighting matrix, especially for the closest-point search, i.e. the feedback operator. 
A further extension concerns rate-dependent, history-dependent, and inelastic material behavior. The present formulation considers rate-independent material states described by stress--strain pairs. For more general material behavior, the data-driven state space would need to be augmented by appropriate rate, history, and/or internal variables, together with corresponding thermodynamic admissibility or dissipation conditions, as considered, for instance, in data-driven inelasticity \cite{Eggersmann2019ddcminelastic}. The GO-ADM search strategy could in principle be applied to such an augmented state space, while its detailed formulation and validation are left for future work. 
Another direction concerns the combination of GO-ADM with other iterative schemes, such as the arc-length method. 
This is especially promising for stability-losing structures and may reduce the need for costly reinitialization at every load step. 
Finally, an important topic is the acceleration of the underlying direct data-driven solver and/or the greedy optimization procedure. 
One can employ, for instance, 
approximate nearest-neighbor algorithms \cite{Eggersmann2021ddcm}, or 
adaptive hyperparameters for the distance-minimizing method \cite{Nguyen2022ddcm}, 
or pathfinding algorithms, such as Dijkstra's algorithm and its extended version, the A$^*$ algorithm, for the greedy search. 
Such improvements would substantially broaden the applicability of GO-ADM, particularly to more complex structural models such as shells and solids.

\section*{Acknowledgments}

T.-H. Nguyen, B.A. Roccia, and C.G.\ Gebhardt gratefully acknowledge the financial support from the European Research Council through the ERC Consolidator Grant “DATA-DRIVEN OFFSHORE” (Project ID 101083157). 
AI tools were utilized to enhance the language and readability of the manuscript.

\appendix

\section{Multi-beam structures in data-driven setting}\label{sec:ddcm_multibeam}

In this section, we explicitly state the discrete-continuous nonlinear optimization formulation (DCNLP) for static structural analysis of multi-beam structures. 
For the sake of clarity, we first recall the general formulation in Section \ref{sec:formulations} as follows:

\noindent
Find $\xF \in \xSpace$ such that:
\begin{equation}\label{eq:dcnlpCopy}
  \inf_{\xF,\,\ytilde} \; \sup_{\lF \in \lSpace} \;
        \globObjFuncP \left(\yF ,\, \ytilde\right)
        \, + \, \map\left( \lF,\, \xF; \, \vect{f} \right)
        \text{ s.t. } 
        \ytilde \in \datafieldset \,.
\end{equation}

\noindent
Here, we employ the modified global objective function, $\globObjFuncP$, proposed in this work (see also Equation \eqref{eq:objFuncModi}). 
For a multi-beam structure consisting of $\numBeam$ beam members, $\globObjFuncP$ and the enforcement of the constraint set, $\map$, take the following explicit form:

\noindent
\begin{equation}
    \begin{aligned}
        \globObjFuncP & \left(\yF ,\, \ytilde\right) = \sum_{i=1}^{\numBeam} \, \globObjFuncP^i \left(\yF^i ,\, \ytilde^i \right) \,,\\
        & \text{with }  
        \globObjFuncP^i \left(\yF^i ,\, \ytilde^i \right) = 
        \frac{1}{2} \, \langle\eV^i-\eTilV^i,\,\eV^i-\eTilV^i\rangle_{\mat{C}^i} + \frac{1}{2} \, \langle\sV^i-\sTilV^i,\,\sV^i-\sTilV^i\rangle_{(\mat{C}^i)^{-1}} + 
        \penFunc (\eV^i,\,\sV^i)\,,
    \end{aligned}
\end{equation}

\noindent
\begin{equation}
    \begin{aligned}
        \map & \left( \lF,\, \xF; \, \vect{f} \right) = \sum_{i=1}^{\numBeam} \, \map^i \left( \lF^i,\, \xF^i; \, \vect{f}^i \right) \,,\\
        & \text{with }  
        \map^i \left( \lF^i,\, \xF^i; \, \vect{f}^i \right) = 
        \langle \muV^i, \,\mathcal{B}^{i,T} \sV^i + \mathcal{H}^{i,T} \chiV^i - \vect{f}^i\rangle_{\mathcal{V}^i} 
        + \langle \lambV^i,\, \vect{\epsilon}(\qV^i) - \eV^i\rangle_{\mathcal{S}^i} + \langle \nuV^i,\, \hV^i(\qV^i)\rangle_{\mathcal{V}^i} \\
        & \qquad \qquad \qquad \qquad \quad
        + \langle \muV^i, \,\mathcal{H}_c^{i,T} \etaV\rangle_{\mathcal{V}^i} 
        + \langle \tauV, \,\hV_c\rangle_{\mathcal{T}} 
        \,,
    \end{aligned}
\end{equation}
respectively. 
Here, the superscript $i$ denotes the $i$-th beam member, 
$\etaV$ the Lagrange multipliers for enforcing the coupling constraints between beam members at the structural level, 
$\tauV$ the Lagrange multipliers for enforcing these constraints at the DCNLP level, 
$\hV_c$ is the collection of all coupling constraints of the multi-beam structure, i.e. of $\hV_{\text{hinged}}$ and/or $\hV_{\text{rigid}}$ (see also Equation \eqref{eq:jointCons}),  
$\mathcal{H}_c^i$ is the operator regarding the variation of $\hV_c$ with respect to $\qV^i$ of the $i$-th beam member that is:
\begin{equation*}
    \mathcal{H}_c^i \, \delta \qV^i = \frac{\partial}{\partial \, \qV^i} \,\hV_c \; \delta \qV^i \,, 
    \text{ and } \mathcal{T} := L^2(\domain;\mathbb{R}^{n_c}) \,,
\end{equation*}
where $n_c$ is the dimension of the coupling constraint vector $\hV_c$. 
We note that $\etaV \in \mathcal{T}$ and $\tauV \in \mathcal{T}$.

The first-order necessary optimality conditions (Karush-Kuhn-Tucker (KKT) conditions) are then:

\noindent
Given a fixed $\ytilde \in \datafieldset$:
\begin{equation}\label{eq:1stOptCondEq}
    \begin{aligned}
        0 = \, \delta\left( \globObjFuncP \left(\yF ,\, \ytilde\right)
        + \map\left( \lF,\, \xF; \, \vect{f} \right)
        \right) =
        & \sum_{i=1}^{\numBeam} \, \delta\left( \globObjFuncP^i + \map^i \right) \\
      + & \sum_{i=1}^{\numBeam} \, \langle \delta \qV^i, \,\sum_{j=1}^{\numBeam} \mathcal{K}_c^{ij} (\etaV) \, \muV^j + \mathcal{H}_c^{i,T} \, \tauV \rangle_{\mathcal{V}^i} + 
          \sum_{i=1}^{\numBeam} \, \langle \muV^i, \,\mathcal{H}_c^{i,T} \etaV\rangle_{\mathcal{V}^i} \\
      + & \langle \delta \etaV, \,\sum_{i=1}^{\numBeam} \, \mathcal{H}_c^{i,T} \muV^i\rangle_{\mathcal{T}} 
      +   \langle \delta \tauV, \,\hV_c\rangle_{\mathcal{T}}
        \,,
    \end{aligned}
\end{equation}
where the operator $\mathcal{K}_c^{ij}$ corresponds to the variation of the product $\left(\mathcal{H}_c^{j,T} \, \etaV \right)$ with respect to $\qV^i$ that is:
\begin{equation*}
    \left(\mathcal{K}_c^{ij}\right)^T \, \delta \qV^i = \frac{\partial}{\partial \, \qV^i} \,\left(\mathcal{H}_c^{j,T} \, \etaV \right) \; \delta \qV^i \,.
\end{equation*}

\noindent
For the explicit formulation of the KKT conditions of each beam member, $\delta\left( \globObjFuncP^i + \map^i \right)$, we refer to \cite{Gebhardtddcmstatic2020}. 
We note that Equation \eqref{eq:1stOptCondEq} leads to a nonlinear system of equations. 
One can solve this using the standard finite element method and Newton-Raphson method.

\section{Constitutive data generation}\label{sec:datageneration}

In this section, we describe how we generate synthetic datasets for our computations in Section \ref{sec:results} using virtual material tests. 
We also discuss an approach to rotate the resulting discrete stress and strain fields, to achieve the required limit values for the data-driven computations and enrich the dataset.

\section*{Virtual material tests}

\begin{table}[htb]
    \centering
    \renewcommand{\arraystretch}{1.4} 
    \setlength{\tabcolsep}{4pt} 
{\footnotesize
    \begin{tabularx}{1\textwidth}{A |
    B|
    B|
    >{\raggedright\arraybackslash}X|
    C}
        \textbf{Example} & \textbf{Number of elements} & \textbf{Number of load steps} & \centering{\textbf{Loading}}  & \textbf{Number of obtained states} \\
        \hline
        Cantilever beam with follower end force, Figure \ref{fig:simobeamSnapNcost}
        & 16 & 10 & 4 tests with constant end force:
        \begin{itemize}
            \item Tests 1 and 2: $\vect{F}_{\text{end}} = \pm [500 ,\,0,\,0]$ kN
            \item Tests 3 and 4: $\vect{F}_{\text{end}} = \pm [0 ,\,0,\,200]$ kN
        \end{itemize}        
        & 640 \\
        \hline
        Clamped-clamped curved beam, Figure \ref{fig:curvedbeam3Ddeformed} 
        & 20 & 1 & 2 tests:
        \begin{itemize}
            \item Test 1: $\vect{F} = [25 ,\,-25,\,50]$ N at nodes 10, 11, and 12
            \item Test 2: $\vect{F} = [0 ,\,0,\,-20]$ N at nodes 2, and 20 and $\vect{F} = [0 ,\,0,\,20]$ N at node 11
        \end{itemize}
         & 40  \\
        \hline
        Three-dimension frame, Figure \ref{fig:frame3dSketch} 
         & 8 (each beam member) & 5 & 3 tests with single force at point A:
         \begin{itemize}
            \item Test 1: $\vect{F}_{A} = [1 ,\,0,\,0]$ N
            \item Test 2: $\vect{F}_{A} = [0,\,2,\,0]$ N
            \item Test 3: $\vect{F}_{A} = [0 ,\,0,\,1]$ N
         \end{itemize} 
         & 120 (each beam member) 
    \end{tabularx}}

    \caption{Virtual material tests for generating synthetic datasets of numerical examples studied in Section \ref{sec:results}.}
    \label{tab:matTest}
\end{table}

In this work, we generate our synthetic datasets of stress and strain states using the discrete stress and strain fields resulting from the conventional finite element analysis (FEA) of the same structure of interest subjected to different load scenarios. 
We refer to this as virtual material tests since it is equivalent to numerical simulations of the laboratory material tests. 
We note that the FEA of such virtual tests, however, requires a prescribed constitutive model for the computation, while laboratory tests aim to identify this. 
In this work, we assume homogeneous isotropic materials and a linear constitutive relation to generate synthetic datasets. 
For our numerical examples in Section \ref{sec:results}, we perform the FEA of the same structure and boundary conditions, using the material properties given in the corresponding references (see also discussions in Section \ref{sec:results}), 
and choose different loads and numbers of load steps. 
Table \ref{tab:matTest} 
summarizes the chosen loading and number of obtained stress and strain states 
for each relevant example studied in Section \ref{sec:results}. 
We note that our choices of forces are to obtain a certain limit of the stress and strain which does not require too many rotations to achieve the required limits of the reference solution of each example, avoiding an overly dense dataset.

\section*{Rotation of discrete stress and strain fields}

\begin{table}[htb]
    \centering
    \renewcommand{\arraystretch}{1.4} 
    \setlength{\tabcolsep}{4pt} 
{\footnotesize
    \begin{tabularx}{0.9\textwidth}{Z | Y | Y| Y}
        \textbf{Example} & \textbf{Number of rotation angles} & \textbf{Total number of generated states} & \textbf{Number of selected states for a subset} \\
        \hline
        Cantilever beam with follower end force, Figure \ref{fig:simobeamSnapNcost}
        & 100 & 62,735 & 5,000  \\
        \hline
        Clamped-clamped curved beam, Figure \ref{fig:curvedbeam3Ddeformed}
        & 125 & 5,000 & $[-]$ \\
        \hline
        Three-dimensional frame, Figure \ref{fig:frame3dSketch}
        & 21 & 5,000 & $[-]$ \\
    \end{tabularx}}

    \caption{Number of rotation angles, $\theta$, and the number of resulting generated stress and strain states of the synthetic datasets employed in Section \ref{sec:results}.}
    \label{tab:numRot4data}
\end{table}

To obtain a dataset that covers the required range of stress and strain values for the loading scenarios of interest, we first rotate the deformation vectors, 
$\gammaV$ and $\omegaV$ (see also Equation \eqref{eq:deformVec}), around the beam axis, and then mirror the third component, i.e. the axial and torsional responses, of both the original and rotated vectors. 
In particular, we first multiply, for instance, $\gammaV$, by the following rotation matrix:
\begin{equation}
    \gammaV_r = \begin{bmatrix}
        \cos(\theta) & -\sin(\theta) & 0 \\
        \sin(\theta) &  \cos(\theta) & 0 \\
        0 & 0 & 1 \\
    \end{bmatrix} \, \gammaV \,,
\end{equation}
where $\theta \in [0,2\pi)$ is a rotation angle. We note that a zero rotation angle and an angle $\theta = 2\pi$ lead to the same rotation, resulting in the same vector as the original one. 
In the second step, we multiply both $\gammaV_r$ and $\gammaV$ with the following matrix to mirror the beam axis:
\begin{equation}
    \gammaV_{m} = \begin{bmatrix}
        1 & 0 & 0 \\
        0 & 1 & 0 \\
        0 & 0 & -1 \\
    \end{bmatrix} \, \gammaV \,, \qquad 
    \gammaV_{rm} = \begin{bmatrix}
        1 & 0 & 0 \\
        0 & 1 & 0 \\
        0 & 0 & -1 \\
    \end{bmatrix} \, \gammaV_r\,.
\end{equation}
We repeat these two steps with the second deformation vector, $\omegaV$, as well as the corresponding vectors of the generalized stress, $\vect{n}$ and $\vect{m}$. 
The synthetic dataset is then the collection of the original, rotated, and mirrored vectors, i.e. $\gammaV$, $\gammaV_r$, $\gammaV_{m}$, and $\gammaV_{rm}$, neglecting identical vectors. 
For our numerical examples in Section \ref{sec:results}, we rotate the obtained stress and strain vectors from the virtual material tests, described above, with uniform rotation angle, $\theta \in [0,2\pi)$. 
Table \ref{tab:numRot4data} gives an overview of the chosen number of rotation angles for each relevant example studied in Section \ref{sec:results}. 
We note that 
our choice is to achieve the required limits of the stress and strain components, without generating an overly dense dataset, while avoiding too coarse rotation. 
One can select a subset of the generated dataset, as we did for the studied cantilever beam subjected to a follower end force. 
We note that for this particular example, while we generated the dataset $\dataset_2$ of 5,000 stress and strain states, using the described procedure above and given parameters in Tables \ref{tab:matTest} and \ref{tab:numRot4data}, 
we obtain the dataset $\dataset_1$ by enriching $\dataset_2$ with the discrete solution of the corresponding FEA. 
The FEA results in 32,000 stress and strain states, leading to a dataset $\dataset_1$ with a total of 37,000 stress and strain states.

\bibliographystyle{elsarticle-num}
\bibliography{sections/refs}

\end{document}